\documentclass[12pt]{article}
\makeatletter \@addtoreset{equation}{section} \makeatother
\renewcommand{\theequation}{\thesection.\arabic{equation}}
\newcommand{\ba}{\begin{array}}
\newcommand{\ea}{\end{array}}
\newcommand{\beq}{\begin{equation}}
\newcommand{\eeq}{\end{equation}}
\newcommand{\bea}{\begin{eqnarray}}
\newcommand{\eea}{\end{eqnarray}}

\def\bce{\begin{center}}
\def\ece{\end{center}}

\def\nonu{\nonumber}

\def\pa{\partial}

\def\be{\beta}

\def\la{\lambda}

\def\eps6{{\displaystyle \mathop{\epsilon}^{6}}{}}
\def\g6{{\displaystyle \mathop{g}^{6}}{}}

\def\nab6{{\displaystyle \mathop{\nabla}^{6}}{}}

\def\0{{\sst{(0)}}}
\def\1{{\sst{(1)}}}
\def\2{{\sst{(2)}}}
\def\3{{\sst{(3)}}}
\def\4{{\sst{(4)}}}
\def\5{{\sst{(5)}}}
\def\6{{\sst{(6)}}}
\def\7{{\sst{(7)}}}
\def\8{{\sst{(8)}}}

\def\ba{\begin{array}}
\def\ea{\end{array}}
\def\beq{\begin{equation}}
\def\eeq{\end{equation}}
\def\be{\begin{equation}}
\def\ee{\end{equation}}

\def\la{\lambda}
\def\eps{\epsilon}

\def\ba{\begin{array}}
\def\ea{\end{array}}
\def\beq{\begin{equation}}
\def\eeq{\end{equation}}
\def\be{\begin{equation}}
\def\ee{\end{equation}}

\def\la{\lambda}
\def\eps{\epsilon}

\def\eps6{{\displaystyle \mathop{\epsilon}^{6}}{}}

\def\nab6{{\displaystyle \mathop{\nabla}^{6}}{}}

\newcommand{\bean}{\begin{eqnarray*}}
\newcommand{\eean}{\end{eqnarray*}}

\begin{document}
\thispagestyle{empty} \addtocounter{page}{-1}
   \begin{flushright}
\end{flushright}

\vspace*{1.3cm}
 
\centerline{ \Large \bf   
Higher Spin Currents with Arbitrary $N$ }
\vspace*{0.5cm}
\centerline{ \Large \bf  
in the ${\cal N}=1$ Stringy Coset Minimal Model }
\vspace*{1.5cm}
\centerline{{\bf Changhyun Ahn 
}} 
\vspace*{1.0cm} 
\centerline{\it 
Department of Physics, Kyungpook National University, Taegu
702-701, Korea} 
\vspace*{0.8cm} 
\centerline{\tt ahn@knu.ac.kr 
} 
\vskip2cm

\centerline{\bf Abstract}
\vspace*{0.5cm}

In the ${\cal N}=1$ supersymmetric coset model based on 
$(A_{N-1}^{(1)} \oplus A_{N-1}^{(1)}, A_{N-1}^{(1)})$ at level $(k, N)$,
the lowest ${\cal N}=1$ 
higher spin supercurrent with spins-$(\frac{5}{2}, 3)$, in terms of
two independent numerator WZW currents, is reviewed. 
By calculating the operator product expansions (OPE) between this 
${\cal N}=1$ higher spin supercurrent and itself, the next two 
${\cal N}=1$ higher spin 
supercurrents can be generated 
with spins-$(\frac{7}{2}, 4)$ and $(4, \frac{9}{2})$.
These four currents 
are polynomials of degree $3, 4, 4, 4$ in the first numerator 
WZW currents with level $k$.  
The complete nonlinear OPE of the lowest ${\cal N}=1$ 
higher spin supercurrent 
for general $N$ is obtained.
The three-point functions involving  two scalar primaries with one 
spin-$2$ current or spin-$3$ current 
are calculated in the large $N$ limit for all values of 
the 't Hooft coupling.  
In particular, 
the light states that appeared in the case when the second level was fixed by
$1$ are no longer light ones because the eigenvalues are finite in the large
$N$ limit.

\baselineskip=18pt
\newpage
\renewcommand{\theequation}
{\arabic{section}\mbox{.}\arabic{equation}}

\section{Introduction}

The $W_N(=W A_{N-1})$ 
minimal model conformal field theory (CFT) 
is dual, in the 't Hooft $\frac{1}{N}$ expansion,
to the higher spin theory of Vasiliev on the $AdS_3$ coupled 
to one complex scalar \cite{GG,GG1}.
See the recent review papers \cite{GG2,AGKP} for this duality.
The CFTs have the following coset form 
\bea
\frac{\widehat{SU}(N)_k \oplus \widehat{SU}(N)_1}
{\widehat{SU}(N)_{k+1}}.
\label{cosetexp}
\eea
The higher spin-$s$ currents from polynomial (of degree $s$ with $s=2, 3, 
\cdots, N$) combinations 
of the individual numerator $SU(N)$ currents of spin-$1$ 
can be constructed \cite{BS}. 
The diagonal denominator 
$SU(N)$ currents commute with these higher spin currents.
One of the levels for the spin-$1$ current is fixed by the positive 
integer $k$
and the other is fixed by $1$  in the numerator 
of the coset CFT (\ref{cosetexp}).
Using the isomorphism \cite{BS} between the coset construction and the 
Drinfeld-Sokolov reduction, the extension of above duality 
to the triality is described in \cite{GG1}. 
One of the specialty on the above coset model (\ref{cosetexp})
is that the additional currents that are generated in the OPEs for general
level $l$ in the second numerator $SU(N)$ current 
become null fields for the level $l=1$ and hence decouple \cite{BBSS2}. 

A more general diagonal coset with levels $(k,l)$ is expected to 
have many additional extra currents.   
The corresponding 
algebra should be  larger than the conventional $W_N$ algebra.
Recently, the two dimensional gauge theory coupled to  the adjoint fermions
for $k=l=N$ is described in \cite{GHKSS}. 
This coset has ${\cal N}=(2,2)$ supersymmetry \cite{BFK}.
It is known in \cite{BS} that 
the Viraosoro primary field, corresponding to 
the higest weight of a highest weight module of the 
affine Kac Moody algebra, has conformal dimension-$\frac{1}{2}$
for the adjoint representation at level $N$
which is a dual Coxeter number of $SU(N)$. 
By Sugawara construction, 
the spin-$\frac{3}{2}$ current, a superpartner of spin-$2$ 
stress energy tensor,  
can be obtained from the above adjoint free fermions of 
spin-$\frac{1}{2}$. 
Similarly, the second spin-$\frac{3}{2}$ current 
can be obtained from the second adjoint free fermions 
of spin-$\frac{1}{2}$ for the coset with the levels $(k=N,l=N)$. 

For the levels $(k,l=N)$ when the second level is fixed by 
$N$, the ${\cal N}=(1,1)$ supersymmetry is preserved
\cite{GKO,Douglas}.
The coset CFTs are described as
\bea
\frac{\widehat{SU}(N)_k \oplus \widehat{SU}(N)_N}
{\widehat{SU}(N)_{k+N}}.
\label{coset}
\eea
The coset central charge $c(N,k)$ can be calculated as 
\bea
c(N,k) = \frac{(N^2-1)}{2} \left[ 1- \frac{2N^2}{(k+N)(k+2N)} \right] < 
 \frac{(N^2-1)}{2}, \qquad k=1, 2, \cdots.
\label{centralc1}
\eea 
In \cite{Ahn1211}, the higher spin currents with spins
\bea
(\frac{5}{2}, 3), \,\,(\frac{7}{2}, 4), \,\, (4, 
\frac{9}{2}), \,\, (4, \frac{9}{2}), \,\, (\frac{9}{2}, 5), 
\,\, (\frac{11}{2}, 6),
\,\, (6, \frac{13}{2}) 
\label{list}
\eea
besides the ${\cal N}=1$ super stress energy 
tensor with spins-$(\frac{3}{2}, 2)$ are constructed for $N=3$
with levels $(k, l=3)$ by reconsidering the previous works in \cite{ASS}. 
The six additional super currents (or twelve currents in components) arise.
Furthermore, 
the general $N$ expressions for the currents with spins-$(\frac{3}{2}, 2)$
and $(\frac{5}{2},3)$ are obtained explicitly. 
The lowest model $(k=1)$ with $c=\frac{(3N+1)(N-1)}{2(2N+1)}$ 
in the series of coset models (\ref{centralc1}) has ``minimal'' 
${\cal N}=1$ supersymmetric $W_N$ algebra where there are currents 
of spins-$(\frac{3}{2}, 2), (\frac{5}{2}, 3), \cdots, 
(N-\frac{1}{2}, N)$ \cite{SS} \footnote{Recently, in \cite{BCGG}, 
the ${\cal N}=1$ minimal model holography corresponding to 
the above  ``minimal'' ${\cal N}=1$ supersymmetric $W_N$ algebra
is found.}. 
As the $k$ increases, the additional currents start to appear 
and the final algebra is associative for all values of $c$.
It was expected \cite{SS} that 
this algebra is determined by considering the limit model of the series
which is a simple model with $c=\frac{N^2-1}{2}$ describing 
$(N^2-1)$ fermions in the adjoint repesentation of $SU(N)$.
In other words, $k \rightarrow \infty$ limit model with fixed $N$.

In this paper,
the generalization to $N$ for the first three  super currents in the list
(\ref{list}) is constructed (to describe the three-point functions 
with scalars in the large $N$ limit as one of the reasons).
Actually, the first higher spin super current 
in (\ref{list}) was already given in 
\cite{Ahn1211}.
The explicit calculations of OPEs between the currents 
with spins-$(\frac{5}{2}, 3)$ are rather involved.
Among three OPEs, the OPE between spin-$3$ current and itself can be obtained 
from the previous results in \cite{Ahn1111}. 
Therefore, the second- and first-order poles of the OPE between 
spin-$\frac{5}{2}$ current and spin-$3$ current and the first-order pole of the 
OPE between spin-$\frac{5}{2}$ current and itself should be 
determined.
In doing these calculations, the fully normal ordering products 
are needed to describe the zero mode calculations.

In section $2$, the two ${\cal N}=1$ higher spin super currents with 
spins-$(\frac{7}{2}, 4)$ and $(4, \frac{9}{2})$
are constructed 
by calculating the OPEs between the ${\cal N}=1$ higher spin 
super current with spins-$(\frac{5}{2}, 3)$
and itself.
The OPE of the ${\cal N}=1$ 
lowest higher spin super current can be obtained for general
$N$ up to the normalizations of above two higher spin super currents.
Note the presence of the ${\cal N}=1$ higher spin super current of 
spins-$(4, \frac{9}{2})$, in the right hand side of OPE, which is one of
the  additional 
currents compared to the above minimal ${\cal N}=1$ 
supersymmetric $W_N$ algebra.

In section $3$, the three-point functions with scalars including the 
spins-$2$ or $3$ current are described. The conformal dimension for 
$(f;f)$ \cite{GG} becomes nonzero even in the large $N$ limit.

In section $4$, the summary of this work is presented and 
some discussions on the future directions are given.

In Appendices, the detailed descriptions appearing in sections 
$2$, $3$ and $4$ are presented.
       
\section{The three ${\cal N}=1$ higher spin (Casimir) super currents  }

In this section, 
the higher spin currents will be  constructed for general $N$ explicitly.
The fermion fields $\psi^a(z)$ of spin-$\frac{1}{2}$, 
where the $SU(N)$ adjoint index $a$ runs from $1$ to
$(N^2-1)$, are used 
for the construction of spin-$1$ current with level $N$ 
of the coset model (\ref{coset}). 
The fundamental OPE of this fermion field is expressed as
\bea
\psi^a(z) \; \psi^b(w) =-\frac{1}{(z-w)} \, \frac{1}{2} \, \delta^{ab} +\cdots.
\label{psipsi}
\eea
The Kac-Moody current of spin-$1$, $J^a(z)$,  is defined as follows:
\bea
J^a(z) \equiv f^{abc} \, \psi^b \psi^c(z).
\label{Jdef}
\eea
The standard OPE of this spin-$1$ current with level $N$ is obtained
from (\ref{psipsi}) and (\ref{Jdef})
\bea
J^a (z) \; J^b(w) = -\frac{1}{(z-w)^2} \, N \, \delta^{ab} + 
\frac{1}{(z-w)} \, f^{abc} \, J^c(w) +\cdots.
\label{JJ}
\eea
To match with the convention of \cite{Ahn1111}, 
the different normalization from that of \cite{Ahn1211} is used 
in this paper.
The following OPE is obtained from (\ref{psipsi}) and (\ref{Jdef})
\bea
\psi^a(z) \; J^b(w) = \frac{1}{(z-w)} \, f^{abc} \, \psi^c(w) +\cdots.
\label{psiJ}
\eea
The other spin-$1$ current of level $k$ living in the other 
$SU(N)$ factor of our coset model (\ref{coset}) satisfies
the following OPE
\bea
K^a (z) \; K^b(w) = -\frac{1}{(z-w)^2} \, k \, \delta^{ab} + 
\frac{1}{(z-w)} \, f^{abc} \, K^c(w) +\cdots.
\label{KK}
\eea
The normalization is corrected appropriately and  
matches with that in \cite{Ahn1111}.
The diagonal spin-$1$ current, $(J^a + K^a)(z)$,
living in the denominator of the coset (\ref{coset}),
satisfies similar OPE with the level $(N+k)$ from (\ref{JJ}) and 
(\ref{KK}). 
The OPE between $J^a(z)$ and $K^b(w)$ contains only regular terms.

The spin-$2$ stress energy tensor, via the Sugawara
construction, is 
expressed as 
\bea
T(z) =-\frac{1}{4N} \, J^a J^a(z) -\frac{1}{2(k+N)} \, K^a K^a(z) +
\frac{1}{2(k+2N)} \,
(J^a +K^a) (J^a +K^a)(z), 
\label{cosett}
\eea
and the central charge (\ref{centralc1}) 
depends on $N$ and $k$ as follows:
\bea
c = \frac{(N^2-1)}{2} \left[ 1- \frac{2N^2}{(k+N)(k+2N)} \right] < 
 \frac{(N^2-1)}{2}, \qquad
k =1, 2, \cdots.
\label{centralc}
\eea
The fourth-order pole of $T(z) \; T(w)$ with (\ref{cosett}) provides 
$\frac{c}{2}$ with (\ref{centralc}).
The superpartner of the spin-$2$ current (\ref{cosett}) of 
spin-$\frac{3}{2}$ can be described as follows \cite{Ahn1211}:
\bea
G(z) = -\frac{\sqrt{2} k}{3\sqrt{N(k+N)(k+2N)}} 
\, \psi^a \, \left( J^a -\frac{3N}{ k} K^a \right)(z).
\label{cosetg}
\eea
The relative coefficient in (\ref{cosetg}) can be fixed  using 
either  the OPE of the diagonal current $(J^a +K^a)(z)$ and 
$G(w)$ does not contain any singular terms
or this spin-$\frac{3}{2}$ current should transform as a primary field under 
the stress energy tensor (\ref{cosett}).
See also the description of \cite{BS} on the construction of the coset  
spin-$\frac{3}{2}$ current (in the subsection $7.3.2$ of \cite{BS}). 
For $N=2$, the form of spin-$\frac{3}{2}$ current was obtained in \cite{GKO}.
The overall normalization in (\ref{cosetg}) is fixed by 
the highest singular term $\frac{1}{(z-w)^3}$ in the OPE $G(z) \; G(w)$.
The central term here is given by $\frac{2c}{3}$, where $c$ is given by 
(\ref{centralc}).
In this analysis, the OPEs, (\ref{psipsi}), (\ref{JJ}), (\ref{psiJ}), and 
(\ref{KK}), are used.

How should the other higher spin currents be determined?
The spin-$3$ current can be described as the cubic terms in the  combination 
of spin-$1$ currents, $J^a(z)$ and $K^a(z)$, 
in the numerator of our coset model (\ref{coset}).
The explicit coefficients which depend on $N$ and $k$
were fixed previously in \cite{BBSS2}. The primary spin-$3$ current is,
by plugging the values $k_1=N$ and $k_2=k$ into $(2.8)$ of \cite{BBSS2}, 
\bea
W(z) & = &  \frac{i \, d^{abc}}{
6N(N+k)(2N+k) \sqrt{6(N+2k)(5N+2k)(N^2-4)}} \left[
  k(k+N)(2k+N)  J^a J^b J^c \right. \nonu \\
&& \left. -6N(k+N)(2k+N) J^a J^b K^c 
+ 18N^2(k+N) J^a K^b K^c -6N^3 K^a K^b K^c \right](z).
\label{Wk}
\eea

What about the superpartner of this spin-$3$ current? 
This spin-$\frac{5}{2}$ current is obtained from the method used in the 
construction of $W(z)$ (\ref{Wk}).  
Or from the ${\cal N}=1$ supersymmetry and the 
OPE $G(z) \; W(w)$ together with (\ref{cosetg}) and (\ref{Wk}) 
(the second-order singular term of this OPE),
the following spin-$\frac{5}{2}$ current
is completely determined \cite{Ahn1211}
\bea
U(z) & = &
\frac{i \, d^{abc}}{\sqrt{
50N(k+N)(2k+N) (k+2 N) (2 k+5 N) (-4+N^2)}} 
\left[  k (2 k+N)  
\psi^a J^b J^c \right. \nonu \\
&-&  \left.  5N (2 k+N) \psi^a J^b K^c  
 +   10 N^2
\psi^a K^b K^c  \right](z).
\label{Uk}
\eea
Equivalently, the coefficients in (\ref{Uk}) are 
fixed using the OPE $G(z) \; U(w)$ and reading off the first-order singular 
term, which is nothing but the spin-$3$ current $W(w)$ \footnote{In this 
calculation, the following identities can be derived 
$d^{abc} J^a J^b J^c(z)= -6N d^{abc} \psi^a \pa \psi^b J^c(z)$ 
and $d^{abc} J^a J^b K^c(z) =
f^{abc} d^{cde} \psi^a \psi^b J^d  K^e(z) -2N d^{abc} \psi^a \pa \psi^b K^c(z)$.
In the realization (\ref{Jdef}), the number of independent terms for 
$d^{abc} J^a J^b K^c(z)$ is two.  }. 

The other higher spin currents, by calculating the 
OPEs between the currrents $W(z)$ and $U(z)$, are determined.
Consider the OPE $W(z) \; W(w)$. This OPE for $N=3$ was obtained previously
and was expected in the more general coset model (\ref{coset}) 
in \cite{Ahn1211}:
\bea
W(z) \; W(w) 
& = & \frac{1}{(z-w)^6} \, \frac{c}{3} +\frac{1}{(z-w)^4} \, 2 T(w)
+\frac{1}{(z-w)^3} \, \pa T(w) \nonu \\
& + & \frac{1}{(z-w)^2} \left[ (\frac{3}{10}) \pa^2 T +\frac{32}{22+5c} 
\left( T^2 -\frac{3}{10} \pa^2 T 
\right) +P_{4}^{ww} +P_{4'}^{ww} \right](w)
\nonu \\
& + & \frac{1}{(z-w)} \left[ (\frac{1}{15}) 
\pa^3 T +(\frac{1}{2})   \frac{32}{22+5c} 
\pa \left( T^2 -\frac{3}{10} \pa^2 T 
\right)\right. \nonu \\
& + & \left. (\frac{1}{2}) 
\pa P_{4}^{ww} + (\frac{1}{2}) \pa P_{4'}^{ww} \right](w)+  \cdots,
\label{wwexp}
\eea
where the spin-$4$ current consisting of $T(z)$ and 
$G(z)$ (and their derivative terms) and transforming as a primary field 
under the stress energy tensor (\ref{cosett}) is given by \cite{HR}
\bea
 P_4^{ww}(z) & = &  \frac{8(10-7c)}
{(4c+21)(10c-7)} 
 \left[ -\frac{7}{10} \pa^2 T + \frac{17}{22+5c} 
\left( T^2 -\frac{3}{10} \pa^2 T 
\right) + G \pa G \right](z).
\label{expexp}
\eea
Furthermore, the other primary spin-$4$ current, which is located 
at the second-order pole of the OPE (\ref{wwexp}), 
can be described as
\bea
 P_{4'}^{ww}(z) & = &  \mbox{second-order pole}(z) 
-\frac{3}{10} \pa^2 T -\frac{32}{22+5c} 
\left( T^2 -\frac{3}{10} \pa^2 T 
\right)(z) -P_{4}^{ww}(z)
\nonu \\
& = & W^{(4)}(z) -P_{4}^{ww}(z).
\label{above}
\eea

In the last line of (\ref{above}), the previous result found 
in \cite{Ahn1111} for the primary spin-$4$ field with the values 
$k_1 =N$ and $k_2=k$
was used as follows \footnote{
\label{dabcd}
The completely symmetric traceless $d$ symbol of rank-$4$ \cite{Ahn1111} 
is introduced as 
follows: $
d^{abcd} = d^{abe} d^{ecd} +d^{ace} d^{ebd} +d^{ade} d^{ebc}
-\frac{4(N^2-4)}{N(N^2+1)} \left( \delta^{ab} \delta^{cd} +\delta^{ac}
\delta^{bd} +\delta^{ad} \delta^{bc} \right)$.
If this relation is used in (\ref{W4}), then 
each five term in the first line of (\ref{W4})
can be distributed to other independent terms.
See also (\ref{fiveid}).}:
\bea
W^{(4)}(z) & = &   d^{abcd} \left[   
c_1 J^a J^b J^c J^d +  c_2 J^a J^b J^c K^d  + c_3  J^a J^b K^c K^d
+ c_4  J^a K^b K^c K^d + 
c_5  K^a K^b K^c K^d \right] \nonu \\ 
 & + &    d^{abe} d^{cde} \left[ c_6 J^a J^b J^c J^d+  
c_7   J^a J^b J^c K^d + 
c_8   J^a J^b K^c K^d
+   c_9  J^a K^b K^c K^d + 
c_{10}   K^a K^b K^c K^d \right]
\nonu \\
& + &  c_{11} \, J^a J^a J^b J^b +
c_{12} \, J^a J^a J^b K^b 
 + c_{13} \, J^a J^a K^b K^b  + 
c_{14}  \, J^a K^a K^b K^b 
+
c_{15} \, K^a K^a K^b K^b 
\nonu \\
& + &
c_{16} \, \pa^2 J^a K^a  +  c_{17} \, \pa J^a \pa K^a   
 +  c_{18}  \, J^a \pa^2 K^a +
c_{19} \, J^a J^b K^a K^b.
\label{W4}
\eea
The explicit coefficient functions in (\ref{W4}) are 
presented in (\ref{coeffWW}) of Appendix $A$.
The composite fields 
$T^2(z)$ and $G\, \pa G(z)$
are expressed as (\ref{Tsquare}) and (\ref{GG'1}), respectively.
Therefore, the first nontrivial primary spin-$4$ current, 
which will belong to the ${\cal N}=1$ supermultiplet (together with other 
primary spin-$4$ current), 
is obtained from the formula (\ref{above}) together with (\ref{W4}), 
(\ref{coeffWW}), (\ref{expexp}), (\ref{cosett}) and (\ref{cosetg}).

Consider the OPE $U(z) \; W(w)$ which
was expected in the more general coset model (\ref{coset}) 
in \cite{Ahn1211} \footnote{ When the order of the OPE (\ref{wuexp1})
is changed, the explict form for the OPE can be derived as follows:
$W(z) \; U(w) 
  =  \frac{1}{(z-w)^4} \, \frac{3}{\sqrt{6}} G(w)  +
\frac{1}{(z-w)^3} \, (\frac{2}{3}) \frac{3}{\sqrt{6}} \pa G(w) 
 +  \frac{1}{(z-w)^2}
\left[ (\frac{1}{4}) \frac{3}{\sqrt{6}} \pa^2 G +
\frac{11\sqrt{6}}{(4c+21)} 
\left( G T -\frac{1}{8} \pa^2 G \right) +O_{\frac{7}{2}}  \right](w)
 +  \frac{1}{(z-w)} \left[ 
 (\frac{1}{15}) \frac{3}{\sqrt{6}} \pa^3 G + (\frac{4}{7})
\frac{11\sqrt{6}}{(4c+21)} 
\pa \left( G T -\frac{1}{8} \pa^2 G \right) + 
(\frac{4}{7}) \pa O_{\frac{7}{2}}
+  
\frac{12\sqrt{6}}{7(10c-7)} 
\left( \frac{4}{3} T \pa G - G \pa T -\frac{4}{15} \pa^3 G 
\right)  + O_{\frac{9}{2}} 
\right](w) +\cdots$. When the component OPEs are combined into a single 
${\cal N}=1$ super OPE, the OPE $W(z) \; U(w)$ is also used. See Appendix $E$.}:
\bea
U(z) \; W(w) 
 & = & \frac{1}{(z-w)^4} \, \frac{3}{\sqrt{6}} G(w)  +
\frac{1}{(z-w)^3} \, (\frac{1}{3}) \frac{3}{\sqrt{6}} \pa G(w) \nonu \\
& + & \frac{1}{(z-w)^2}
\left[ (\frac{1}{12}) \frac{3}{\sqrt{6}} \pa^2 G +
\frac{11\sqrt{6}}{(4c+21)} 
\left( G T -\frac{1}{8} \pa^2 G \right) +O_{\frac{7}{2}}  \right](w)
\nonu \\
& + & \frac{1}{(z-w)} \left[ 
 (\frac{1}{60}) \frac{3}{\sqrt{6}} \pa^3 G + (\frac{3}{7})
\frac{11\sqrt{6}}{(4c+21)} 
\pa \left( G T -\frac{1}{8} \pa^2 G \right) + 
(\frac{3}{7}) \pa O_{\frac{7}{2}} \right. \nonu \\
& - &  \left.  
\frac{12\sqrt{6}}{7(10c-7)} 
\left( \frac{4}{3} T \pa G - G \pa T -\frac{4}{15} \pa^3 G 
\right)  - O_{\frac{9}{2}} 
\right](w) +\cdots.
\label{wuexp1}
\eea
From the second order pole of (\ref{wuexp1}), 
the spin-$\frac{7}{2}$ current can be expressed as
\bea
O_{\frac{7}{2}}(z) = \mbox{second-order pole}(z) 
- \frac{1}{12} \frac{3}{\sqrt{6}} \pa^2 G(z) -
\frac{11\sqrt{6}}{(4c+21)} 
\left( G T -\frac{1}{8} \pa^2 G \right)(z).
\label{7half}
\eea
The following quantities are introduced to express concisely
\bea
Q^a(z)  & \equiv & d^{abc} J^b J^c(z),
\qquad
R^a(z)  \equiv   d^{abc} J^b K^c(z), \qquad
S^a(z)  \equiv  d^{abc} K^b K^c(z),
\nonu \\
Q(z) & \equiv &  d^{abc} J^a J^b J^c(z),
\qquad 
 S(z) \equiv d^{abc} K^a K^b K^c(z).
\label{newdef} 
\eea
As noted in \cite{HR}, 
for infinite limit of $k$, $Q^a(z)$ plays the role of 
superpartner of spin-$\frac{3}{2}$ field, $\phi^a(z) \equiv 
\frac{2}{3} d^{abc} \psi^b J^c(z)$ 
and spin-$\frac{5}{2}$ field, $\Phi^a(z) \equiv 
\frac{2}{3} d^{abc} (2 \pa \psi^b J^c - \psi^b 
\pa J^c)(z)$.
In other words, the OPE $G(z) \; Q^a(w)$ for infinite 
$k$
contains the second-order pole with this spin-$\frac{3}{2}$ field, $\phi^a(z)$,
and the first-order pole with the combination of 
the spin-$\frac{5}{2}$ field, $\Phi^a(z)$, 
and the derivative of spin-$\frac{3}{2}$ field, $\phi^a(z)$.

The next step is to determine how to obtain 
the second-order pole of OPE $U(z) \; W(w)$ together with (\ref{Uk}) 
and (\ref{Wk}).
At first, the OPE between the first term of $U(z)$,
$d^{abc} \psi^a J^b J^c(z) = \psi^a Q^a(z)$, 
and the first term of $W(w)$, $d^{def} J^d J^e J^f(w)=J^d Q^d(w)$
should be calculated.  
The OPE consists of two parts.
One is to calculate the OPE between $\psi^a Q^a(z)$ and $J^d(x)$
and the other is to calculate the OPE between $\psi^a Q^a(z)$ and $Q^d(w)$. 
The OPE of $J^d(x)$ and $\psi^a(z)$
can be derived from (\ref{psiJ}) and the OPE 
between $J^d(x)$ and $Q^a(z)$ is needed.
This was obtained in  \cite{Ahn1111}, where
the level $N$ is inserted,
\bea
J^a(z) \; Q^b(w)  & = &  -\frac{1}{(z-w)^2} \, 3N d^{abc} J^c(w)
+\frac{1}{(z-w)} \, f^{abc} Q^c(w) +\cdots.
\label{JaQb}
\eea
It is straightforward to 
obtain the second-order pole, $-5N d^{abc} (\psi^a J^b) Q^c(w)$, 
from the first part.
The following fully normal ordered product 
can be obtained as follows:
\bea
d^{abc} (\psi^a J^b) Q^c(w) & = & 
d^{abc} \psi^a J^b Q^c(w) -\frac{5}{2}(N^2-4) \psi^a \pa^2 J^a(w) \nonu \\
& + &
\frac{7}{2} (N^2-4) \pa \psi^a \pa J^a(w).
\label{exp2}
\eea

Moreover, 
to obtain the second part of the OPE,
the following OPE from (\ref{psiJ})  should be used
\bea
Q^c(z) \; \psi^a(w) & = & 
-\frac{1}{(z-w)^2} \, N d^{acd} \psi^d(w) \nonu 
\\
& + &
\frac{1}{(z-w)} \left[ -2 N d^{acd} \pa \psi^d - 2 d^{cde} f^{adf} \psi^f J^e 
\right](w)
+  \cdots.
\label{Qcpsia}
\eea 
The following OPE, which appeared in \cite{Ahn1111},  
should be used
\bea
Q^a (z) \; Q^b (w) & = & \frac{1}{(z-w)^4} \, 6N(N^2-4) \delta^{ab}
- \frac{1}{(z-w)^3} \, 6(N^2-4) f^{abc} J^c(w)
\nonu \\
&-& \frac{1}{(z-w)^2} \left[ N d^{abc} Q^c +6N d^{ace} d^{bde}
  J^c J^d 
\right](w) \nonu \\
& + & \frac{1}{(z-w)} 
\left[ -3N d^{ace} d^{bde} \pa J^c J^d + f^{ace} d^{bcd}
  Q^e J^d + f^{ade} d^{bcd} J^c Q^e \right. \nonu \\
& - & \left.  3N d^{ace} d^{bde} J^d \pa J^c \right](w) 
+ \cdots. 
\label{QaQb}
\eea
The second-order pole contributed from the second part of the OPE leads to
the intermediate 
result, $-5N d^{abc} d^{cde} J^a ( \psi^d J^e) J^b(w)-5N d^{abc} 
d^{cde} J^a J^b \psi^d J^e(w)$.  
The expression 
$d^{abc} d^{cde} J^a (\psi^d J^e)J^b(w)$ can be simplified further. 
Using the identity \cite{Ahn1111}
\bea
d^{adb} d^{bec} f^{cfa} & = & \frac{(N^2-4)}{N} f^{def},
\label{id1}
\eea
one can rearrange this normal ordered product by moving the field $J^b$ to 
the left. The identities in \cite{Ahn1111}
\bea
f^{abc} f^{dbc} & = & 2N \delta^{ad},
\qquad
d^{abc} d^{dbc}  =  \frac{2}{N}(N^2-4) \delta^{ad},
\label{id2}
\eea
are used.
Therefore, the final result can be expressed as
\bea
d^{abc} d^{cde} J^a (\psi^d J^e) J^b(w) &=&
d^{abc} \psi^a J^b Q^c(w) -(N^2-4) \pa^2 \psi^a J^a(w) 
-(N^2-4) \psi^a \pa^2 J^a(w) \nonu \\
&+ & 2(N^2-4) \pa \psi^a \pa J^a(w). 
\label{exp1}
\eea

Using (\ref{exp1}), (\ref{exp2}) and the 
identity
$d^{abc} d^{cde} J^a J^b \psi^d J^e(w) = d^{abc} d^{cde} \psi^a J^d J^e J^b(w) +
2(N^2-4) \pa \psi^a \pa J^a(w) -(N^2-4) \pa^2 \psi^a J^a(w)$,
which can be checked by moving the field $\psi^d$ to the left, 
the final second-order pole of OPE $ \psi^a Q^a(z)  \; J^d Q^d(w) $ 
can be summarized as
\bea
\{ \psi^a Q^a  \; J^d Q^d \}_{-2}(w) & = & 
-15 N d^{abc} \psi^a J^b Q^c(w) +\frac{45}{2} N(N^2-4) \psi^a \pa^2 J^a(w) 
\nonu \\
& - & \frac{75}{2} N(N^2-4) \pa \psi^a \pa J^a(w). 
\label{firstfirst}
\eea
The other OPEs can be obtained similarly.
In Appendix $B$, 
the eleven contributions including (\ref{firstfirst}) 
are presented explicitly.
Obviously, the OPE between the first term of $U(z)$ and the last term of
$W(w)$ does not contain the singular term. 
In particular, the OPE  $\psi^a R^a (z) \; Q(w)$
found in (\ref{4}) is most involved.
The normal ordered products in (\ref{1})-(\ref{11}) are not fully 
normal ordered.
Some nestings of these higher ordered products are 
presented in Appendix $B$ using (\ref{id1}) and (\ref{id2}).
Furthemore, 
the field $\psi^a(z)$ should be placed to the left of the field 
$J^b(z)$ and the field $K^c(z)$ is located at the right hand side of $J^d(z)$.

Then the final second-order pole, which consists of 
thirteen terms, is presented as follows: 
\bea
\{ U \; W \}_{-2}(w) & = &
c_1 \, d^{abc} \psi^a J^b Q^c(w) + c_2 \, d^{abc} \psi^a J^b S^c(w) +
c_3 \, d^{abc} \psi^a J^b R^c(w) + c_4 \, d^{abc} \psi^a K^b S^c(w) 
\nonu \\
& + & c_5 \, d^{abc} d^{cde} \psi^d J^a J^b K^e(w) +c_6 \,
d^{abc} d^{cde} \psi^d J^a K^b K^e(w)
+ c_7 \, \psi^a \pa^2 J^a(w) \nonu \\
& + &  c_8 \, \pa \psi^a \pa J^a(w) +
c_9 \, f^{abc} \psi^a K^b \pa K^c(w) 
+ c_{10} \, f^{abc} \pa \psi^a J^b K^c(w)
 +  c_{11} \, \pa^2 \psi^a K^a(w) \nonu \\
& + & c_{12} \, \pa \psi^a \pa K^a(w) +
c_{13} \, \psi^a \pa^2 K^a(w),
\label{UW2}
\eea
where the coefficient functions are determined as
\bea
c_1 &=& -15 B C k N (k+N) (2 k+N)^2 (k+2 N), 
\nonu \\
c_2  & = &   -60 B C N^3 (k+N) (k+2 N) (6 k+5 N),
\nonu \\
c_3 & = & 75 B C N^2 (k+N) (2 k+N)^2 (k+2 N),
\nonu \\
c_4  & = &  480 B C N^4 (k+N) (k+2 N),
\nonu \\
c_5 & = & 30 B C N^2 (k+N) (2 k+N)^2 (k+2 N), 
\nonu \\
c_6  & = &  -240 B C N^3 (k+N) (2 k+N) (k+2 N),
\nonu \\
c_7 & = & \frac{15}{2} B C k (-4+N^2) N  (k+N) (2 k+N) (k+2 N) (6 k+11 N),
\nonu \\
c_8 & = & -\frac{15}{2} B C k (-4+N^2) N  (k+N) (2 k+N) (k+2 N) (10 k+13 N),
\nonu \\
c_9 & = & 240 B C (-4+N^2) N^2  (k+N) (2 k+N) (k+2 N),
\nonu \\
c_{10} & = & -30 B C (-4+N^2) N  (k+N) (2 k+N) (k+2 N) (10 k+13 N),
\nonu \\
c_{11} & = & -60 B C (-4+N^2) N^2  (k+N) (2 k+N) (k+2 N) (2 k+3 N),
\nonu \\
c_{12} & = & 60 B C (-4+N^2) N^2  (k+N) (2 k+N)^2 (k+2 N),
\nonu \\
c_{13} & = & -30 B C (-4+N^2) N^2  (k+N) (2 k+N) (k+2 N) (2 k+9 N).
\label{UW2coeff}
\eea
Furthermore the following $(N,k)$ dependent functions, 
which are overall nomalization constants of $W(z)$ (\ref{Wk}) and 
$U(z)$ (\ref{Uk}), 
respectively,  are introduced
\bea
B(N,k) & \equiv & \frac{i}{6 \sqrt{6} N (k+N) (k+2 N) 
\sqrt{(2 k+N) (2 k+5 N)(-4+N^2)}},
\nonu \\
C(N,k) & \equiv & \frac{i}{5 \sqrt{2  
N (k+N) (2 k+N) (k+2 N) (2 k+5 N) (-4+N^2)}}. 
\label{BC}
\eea
The presence of the first nonderivative terms of (\ref{UW2})
was expected in \cite{Ahn1211}.
Each term contains $\psi^a(z)$ or its derivative term.
The $c_1$-term in (\ref{UW2}) can be seen from the result of \cite{HR}.
The $c_1$-, $c_7$- and $c_8$-terms do not contain the current $K^a(z)$.

Therefore, the spin-$\frac{7}{2}$ current 
is obtained from the formula (\ref{7half}), (\ref{newdef}), 
(\ref{UW2}), (\ref{UW2coeff}), 
(\ref{cosett}), (\ref{cosetg}) and (\ref{BC}).
Some identities in (\ref{GTcomp}) of Appendix $B$ can be used to
obtain the composite field $G \, T(z)$ that appears in (\ref{7half}). 

What happens in next order pole?
The explicit form for the spin-$\frac{9}{2}$ current 
can be derived as follows:
\bea
O_{\frac{9}{2}}(w) & = &  -\mbox{first-order pole}(w) + 
 (\frac{1}{60}) \frac{3}{\sqrt{6}} \pa^3 G(w) + (\frac{3}{7})
\frac{11\sqrt{6}}{(4c+21)} 
\pa \left( G T -\frac{1}{8} \pa^2 G \right)(w) \nonu \\
& + &  
(\frac{3}{7}) \pa O_{\frac{7}{2}}(w)  -    
\frac{12\sqrt{6}}{7(10c-7)} 
\left( \frac{4}{3} T \pa G - G \pa T -\frac{4}{15} \pa^3 G 
\right)(w).
\label{9half}
\eea
The next step is to determine how to obtain 
the first-order pole of OPE $U(z) \; W(w)$.
As done in previous second-order pole, the first-order pole from 
the OPE 
between $\psi^a Q^a(z)$ and $J^d Q^d(w)$
can be obtained.
Using (\ref{JaQb}), (\ref{Qcpsia}) and (\ref{QaQb}),
the first-order pole consists of
$-5N d^{abc} d^{cde} \pa (\psi^a J^b) (J^d J^e)(w)
 -  5N d^{abc} d^{cde} J^a \pa (\psi^d J^e) J^b(w)
 -5N d^{abc} d^{cde} J^a J^b \pa (\psi^d J^e)(w)$.
In (\ref{1-1}), the following nontrivial normal ordered products
are obtained
\bea
d^{abc} d^{cde} \pa (\psi^a J^b) (J^d J^e)(w)
& = &  d^{abc} \pa \psi^a J^b Q^c(w) + d^{abc} \psi^a  \pa J^b Q^c(w)
\nonu \\
& + & 
(N^2-4) \left( -\pa^3 \psi^a J^a +\pa \psi^a \pa^2 J^a -\frac{1}{3} \psi^a
\pa^3 J^a \right)(w),
\nonu \\
d^{abc} d^{cde} J^a \pa (\psi^d J^e) J^b(w) & = &
-\frac{2}{3} (N^2-4) J^a \pa^3 \psi^a(w) + d^{abc} d^{cde} J^a J^b \pa 
(\psi^d J^e)(w),
\nonu \\
 d^{abc} d^{cde} J^a J^b \pa (\psi^d J^e)(w)
& = & 
d^{abc} d^{cde} \psi^d J^a J^b \pa J^e(w) +\frac{2(N^2-4)}{N} f^{abc} \pa \psi^a
J^b \pa J^c(w) \nonu \\
& +& d^{abc} d^{cde} \pa \psi^d J^a J^b  J^e(w) +\frac{(N^2-4)}{N} 
f^{abc} \pa^2 \psi^a
J^b J^c(w) \nonu \\
& - & (N^2-4) \pa^2 \psi^a \pa J^a(w) -\frac{1}{3} (N^2-4) \pa^3 \psi^a J^a(w),
\label{above1}
\eea
where the OPEs of (\ref{JaQb}) and (\ref{Qcpsia})
are used. Some properties between the $d$ symbol and $f$ symbol are used.
In Appendix $C$, 
the  contributions are presented explicitly.
The normal ordered products in (\ref{1-1})-(\ref{1-11}) are not fully 
normal ordered.
Some nestings of these higher ordered products are 
presented in Appendix $C$.

Then the final first-order pole, which consists of 
thirty nine terms, is presented as follows: 
\bea
\{ U \; W\}_{-1}(w) & = &  c_1 \, f^{abc} d^{cde} \psi^a J^d Q^b K^e +
c_2 \, f^{abc} d^{cde} \psi^b J^d J^e R^a + c_3 \, d^{abc} d^{def} f^{ebh} \psi^a
J^d J^f J^h K^c 
\nonu \\
&+& c_4 \, f^{abc} d^{cde} \psi^d J^a Q^b K^e 
+ c_6 \, d^{abc} d^{def} f^{eah} \psi^h J^f J^b K^d K^c
\nonu \\
& + & 
c_7 \, d^{abc} d^{def} f^{ebh} \psi^a J^f J^h K^d K^c +
c_8 \, f^{abc} d^{cde} \psi^a J^d K^e S^b 
+c_9 \, f^{abc} d^{cde} \psi^d J^a K^e S^b
\nonu \\
&+& c_{10} f^{abc} d^{cde} \psi^b J^d J^e S^a 
+ c_{13} d^{abc} d^{def} f^{ecg} \psi^a J^b K^d K^f K^g
+ c_{14} f^{abc} d^{cde} \psi^a K^d K^e S^b
\nonu \\
&+& e_1 \, d^{abc} \pa \psi^a J^b Q^c + e_2 \, d^{abc} \psi^a \pa J^b Q^c +e_3 
\, \pa \psi^a \pa^2 J^a +e_4 \, \psi^a \pa^3 J^a + e_6 \, \pa^2 \psi^a \pa J^a
\nonu \\
&+& e_7 \, \pa^3 \psi^a K^a + e_8 \, d^{abc} d^{cde} \psi^d J^e \pa J^a K^b
+ e_9 \, f^{abc} \pa \psi^a J^b \pa K^c   
+ e_{10} \, f^{abc} \pa^2 \psi^a J^b K^c 
\nonu \\
&+& e_{11} \, f^{abc} \pa \psi^a \pa J^b K^c  +
e_{13} \, \pa \psi^a \pa^2 K^a + e_{14} \, d^{abc} \psi^a J^b \pa S^c 
+  e_{15} \, \psi^a \pa^3 K^a \nonu \\
& + & 
e_{16} \, d^{abc} d^{cde} \psi^d J^a J^b \pa K^e + e_{17} \,
d^{abc} d^{cde} \pa \psi^d
J^a J^b K^e 
 +  e_{18} \, d^{abc} \pa \psi^a J^b R^c \nonu \\
& + & e_{19} \, d^{abc} \psi^a 
\pa J^b R^c 
 +  e_{20} \, d^{abc} d^{cde} \psi^a \pa J^b J^d \pa K^e 
 +  e_{22} \, d^{abc} \pa \psi^a J^b S^c 
\nonu \\
& + & e_{23} \, d^{abc} \psi^a \pa J^b S^c
+e_{24} \, d^{abc}  d^{cde} \psi^d \pa J^a J^b K^e 
 +   e_{26} \, d^{abc} d^{cde} \pa \psi^d J^a K^b K^e 
\nonu \\
& + & 
e_{27} \, d^{abc} d^{cde} \psi^d J^a K^b \pa K^e +
e_{28} \, d^{abc} \psi^a \pa K^b S^c 
+ e_{29} \, f^{abc} \psi^a K^b \pa^2 K^c  \nonu \\
& + & e_{30} \,
d^{abc} \pa \psi^a K^b S^c 
 + e_{32} \, f^{abc} \psi^a \pa K^b \pa K^c 
+ e_{33} \, d^{abc} \psi^a K^b \pa S^c,
\label{UW1}
\eea
where the coefficient functions are presented in 
(\ref{coeffUWpole1}) of Appendix $C$.
Compared to the previous spin-$\frac{7}{2}$ current, where 
the field $d^{abc} \psi^a J^b Q^c(z)$ appears in (\ref{UW2}),
the nonderivative term among $c_1$-$c_{14}$ terms 
in (\ref{UW1}) contains $K^a(z)$ term.
Furthermore, the nonderivative terms 
contain a single $f$ symbol.
Because the spin-$1$ currents in (\ref{JJ}) and (\ref{KK}) 
has the first-order pole 
with the structure constant $f$ symbol, 
this reflects the final first-order pole in the OPE $U(z) \; W(w)$.
Note that the OPE (\ref{psiJ}) contains a $f$ symbol on the right hand side.
The $K^a(z)$ independent terms are given
by 
$e_1$-$e_6$ terms.
The last twenty eight derivative terms in (\ref{UW1}) ($e_1$-$e_{33}$ terms)
can be seen from the $\pa \{ U \; W\}_{-2}(w)$ with (\ref{UW2}).
This fact is consistent with the result of \cite{HR} where 
the field $\rho_{\frac{9}{2}}(z)$ in $(3.22)$  of \cite{HR} 
has no nonderivative term.

Therefore, the spin-$\frac{9}{2}$ current 
is obtained from the formula (\ref{9half}), (\ref{7half}), (\ref{newdef}), 
(\ref{UW1}), (\ref{coeffUWpole1}), 
(\ref{cosett}), (\ref{cosetg}) and (\ref{BC}).
Some identities in (\ref{TG'}) and (\ref{GT'}) of Appendix $C$ can be used to
obtain the composite field $ T \, \pa G(z)$ and $G \, \pa T(z)$, 
respectively.

Now move the following OPE 
\bea
U(z) \; U(w) 
 & = & \frac{1}{(z-w)^5} \, \frac{2c}{5}  
+ \frac{1}{(z-w)^3} \, 2T(w) +
\frac{1}{(z-w)^2} \, \pa T(w) \nonu \\
&+&\frac{1}{(z-w)} \, 
\left[ (\frac{3}{10}) \pa^2 T 
+\frac{27}{22+5c} \left( T^2 -\frac{3}{10} \pa^2 T 
\right) +  P_{4}^{uu} + P_{4'}^{uu} \right](w) \nonu \\
& + & \cdots,
\label{uulast}
\eea
where the primary spin-$4$ field is
\bea
 P_{4}^{uu}(z) & = &  
-\frac{3(2c-83)}{(4c+21)(10c-7)} 
 \left[ -\frac{7}{10} \pa^2 T + 
\frac{17}{22+5c} \left( T^2 -\frac{3}{10} \pa^2 T 
\right) + G \pa G \right](z).
\label{puuexp}
\eea
Then the other spin-$4$ field belonging to 
${\cal N}=1$ supermultiplet
is expressed as
\bea
 P_{4'}^{uu}(z)=
 \mbox{first-order pole}(z) 
-\frac{3}{10} \pa^2 T(z) -\frac{27}{22+5c} 
\left( T^2 -\frac{3}{10} \pa^2 T 
\right)(z) -P_{4}^{uu}(z).
\label{spin4}
\eea
The OPE of the first-term of (\ref{Uk}) and itself
can be obtained.
For the OPE $\psi^a Q^a(z) \; \psi^b Q^b(w)$,
the following OPE is needed.
\bea
\psi^a Q^a(z) \; \psi^b(w) =\frac{1}{(z-w)} \left[ -\frac{1}{2} Q^b
- 2 d^{acd} f^{bce} \psi^a \psi^e J^d - N d^{abc} \psi^a \pa \psi^c
 \right](w) +\cdots.
\label{idid1}
\eea
The second part of this OPE leads to 
the following first-order pole
\bea
\{ \psi^a Q^a \; Q^b \}_{-1}(w) =  \left[ -5 N d^{bcd} d^{cef} \pa (\psi^e J^f) J^d
-5N d^{bcd} d^{def} J^c \pa (\psi^e J^f) \right](w).
\label{idid}
\eea
In the normal ordering of $(\psi^a J^b)(\psi^d J^e)$,
one should be careful about the signs due to the fermionic property 
\cite{Fuchs}.
That is,
\bea
(\psi^a J^b)( \psi^d J^e) = \psi^a J^b \psi^d J^e + \{ (\psi^d J^e), 
(\psi^a J^b) \}
-\{ (\psi^d J^e), \psi^a \} J^b + \psi^a [(\psi^d J^e), J^b ].
\label{fermionic}
\eea

Furthermore, the following nontrivial normal ordered products 
 in (\ref{2-1}) with (\ref{idid1}) and (\ref{idid}) can be obtained
\bea
Q^a Q^a(w) & = &  d^{abc} d^{ade} J^b J^c J^d J^e(w) -
6(N^2-4) \pa^2 J^a J^a(w), \nonu \\ 
d^{abc} d^{cde} \psi^a \pa (\psi^d J^e) J^b(w) & = & d^{abc} d^{cde}
\psi^a J^b \pa (\psi^d J^e)(w) -\frac{2}{3}(N^2-4) \psi^a \pa^3 \psi^a(w),
\nonu \\
d^{abc} (\psi^a \pa \psi^b) Q^c(w) 
& = & d^{abc} d^{cde} \psi^a \pa \psi^b J^d J^e(w) + 
3(N^2-4) \pa \psi^a \pa^2 \psi^a(w)
\nonu \\
& -& \frac{5}{3} (N^2-4) \psi^a \pa^3 \psi^a(w), 
\nonu \\
d^{abc} d^{cde} \pa (\psi^a J^b) (\psi^d J^e)(w)
& = & d^{abc} d^{cde} \pa \psi^a J^b \psi^d J^e(w) 
+ d^{abc} d^{cde} \psi^a \pa J^b \psi^d J^e(w) 
\nonu \\
& + & \frac{N^2-4}{N} \pa^2 J^a J^a(w) + 
\frac{2}{N} (N^2-4) \pa J^a \pa J^a(w) 
\nonu \\
& + & 4(N^2-4) \pa \psi^a \pa^2 \psi^a(w) 
+\frac{14}{3} (N^2-4) \psi^a \pa^3 \psi^a(w),
\nonu \\
f^{abc} d^{cde} (\psi^d \psi^a J^e ) Q^b(w)
& = & - f^{abc} d^{cde} d^{bfg} \psi^a (\psi^d J^e) J^f J^g(w)
+ N d^{abc} d^{cde} \pa \psi^a \psi^d J^e J^b(w) 
\nonu \\
& + & N d^{abc} d^{cde} \pa \psi^a \psi^d J^b J^e(w) 
+ 2N d^{abc} d^{cde} \pa (\psi^a J^b)(\psi^d J^e)(w)
\nonu \\
& + & \frac{4}{3} N (N^2-4) \psi^a \pa^3 \psi^a(w) 
\nonu \\
& + & 7(N^2-4) f^{abc} \psi^a \pa (\pa \psi^b J^c)(w).
\label{express1}
\eea
Note that the first term of the fifth equation 
in (\ref{express1}) should be rearranged to obtain the final result. 
The two nonderivative terms will appear in the final results below 
\footnote{ The following simplification can be made
$f^{abc} d^{cde} d^{bfg} \psi^a (\psi^d J^e) J^f J^g(w)= 
f^{abc} d^{cde} d^{bfg} \psi^a \psi^d J^e J^f J^g(w)+ 2(N^2-4) f^{abc} \psi^a \pa
(\psi^b J^c)(w)-3(N^2-4) f^{abc} \psi^a \pa (\psi^b \pa J^c)(w)+ N d^{abc} d^{cde}
\psi^a \psi^b J^e \pa J^d(w)-N d^{abc} d^{cde} \psi^a \psi^e J^b \pa J^d(w) -N 
d^{abc} d^{cde} \psi^a \psi^d J^e \pa J^b(w)-N d^{abc} \psi^a \pa \psi^b Q^c(w) +
\frac{3}{2}(N^2-4) f^{abc} \psi^a \psi^b \pa^2 J^c(w)$.}.

The final first-order pole, which consists of 
thirty eight terms,
can be constructed 
\bea
\{ U \; U \}_{-1}(w) & = &
c_1 \, d^{abc} J^a J^b Q^c 
+ c_2 \, f^{abc} d^{cde} \psi^a \psi^d J^e Q^b
+c_3  \, d^{abc} J^a J^b R^c \nonu \\
& + & 
c_4 \, d^{abc} d^{def} f^{ebg} \psi^d \psi^a J^g K^f K^c
+  c_5 \, d^{abc} J^a K^b S^c +c_6 \, d^{abc} d^{def} f^{ecg} \psi^d \psi^a 
J^b K^f K^g \nonu \\
& + &  c_7 \, d^{abc} J^a K^b R^c   
+ c_9 \, d^{abc} d^{def} f^{fcg}
\psi^d \psi^a J^e J^b K^g 
 +  c_{10}  \, f^{abc} d^{cde}  \psi^d \psi^a J^e R^b
\nonu \\
& + & c_{11} Q^a S^a +c_{12} f^{abc} d^{cde} \psi^d \psi^a J^e S^b
+
 c_{13} f^{abc} d^{cde} \psi^d \psi^a Q^b K^e 
 +  c_{16} f^{abc} d^{cde} \psi^d \psi^b K^e S^a
\nonu \\
&+& c_{17} d^{abc} K^a K^b S^c
+ e_1 d^{abc} d^{cde} \pa \psi^a \psi^d J^e J^b
+ e_3 d^{abc} d^{cde} \psi^a \psi^d J^e \pa J^b
+ 
 e_6 d^{abc} \psi^a \pa \psi^b Q^c  
\nonu \\
& + & e_7 \, d^{abc} d^{cde} \psi^d \psi^a K^e \pa K^b 
+ e_8 \, f^{abc} \pa \psi^a \pa \psi^b J^c 
+ e_9 \, \pa J^a \pa J^a 
+ e_{10} \, \pa^2 J^a K^a \nonu \\
& + & e_{11} \, J^a \pa^2 J^a + e_{12} \,
f^{abc} \pa J^a J^b K^c 
+ e_{14} \, d^{abc} d^{cde} \psi^a \pa \psi^b J^d K^e 
+e_{16} \, \pa^2 K^a K^a 
\nonu \\
&+& e_{17} \, \pa \psi^a \pa^2 \psi^a + e_{18} \, \psi^a \pa^3 \psi^a 
+e_{20} \,
f^{abc} J^a K^b \pa K^c 
+ e_{21} \, d^{abc} d^{cde} \pa \psi^d \psi^a K^e K^b
\nonu \\
& + & e_{22} d^{abc} d^{cde} \psi^d \pa \psi^a  K^e K^b
+e_{23} f^{abc} \psi^a \pa \psi^b \pa K^c 
+ e_{24} J^a \pa^2 K^a  
+  e_{25} d^{abc} d^{cde} \psi^a \pa \psi^d J^b K^e
\nonu \\
&+& 
e_{28} \, d^{abc} d^{cde} \psi^d \psi^a \pa J^e K^b
+e_{29} \, d^{abc} d^{cde} \psi^d \psi^a J^e \pa K^b 
+ e_{30} \, d^{abc} \psi^a \pa \psi^b S^c
\nonu \\
&+& 
 e_{34} \, f^{abc} \pa \psi^a \pa \psi^b K^c
+ 
e_{36} \, d^{abc} d^{cde} \psi^d \pa \psi^a J^b K^e, 
\label{UUpole1}
\eea
where the coefficient functions are presented in 
(\ref{coeffUU}) of Appendix $D$.
In particular, the OPE  $\psi^a R^a (z) \; \psi^b Q^b(w)$
found in (\ref{2-4}) is most involved.
The nonderivative $c_1$-term in (\ref{UUpole1}) appears also in \cite{HR}. 
The expression which does not contain the current $K^a(z)$
consists of $c_1$-, $c_2$-, $e_1$-, $e_3$-, $e_6$-, $e_8$-, $e_9$-, 
$e_{11}$-, $e_{17}$- and $e_{18}$-terms.
The nonderivative terms with a single $f$ symbol can be 
obtained from the contractions between the spin-$1$ currents.  
As observed previously, the first-order pole between these currents 
has a $f$ symbol. Also the quadratic expression in $\psi^a$ appears because  
they do not play the role of the contraction and remain unchanged. 
On the other hand, 
the nonderivative terms without $f$ symbol can be 
obtained from the contractions between the spin-$\frac{1}{2}$ currents
and therefore there is no $\psi^a$ factor in their expressions.  

Therefore, the spin-$4$ current 
is obtained from the formula (\ref{spin4}), (\ref{newdef}), 
(\ref{UUpole1}), (\ref{puuexp}), (\ref{coeffUU}), 
(\ref{cosett}), (\ref{cosetg}) and (\ref{BC}).
Some identities in (\ref{GG'})  of Appendix $A$ can be used to
obtain the composite field $ G \, \pa G(z)$ that appears in (\ref{puuexp}).

Eventually, the correct spin-$4$ currents, which are the elements of 
${\cal N}=1$ super currents, are 
\bea
O_4(w) = \left( -\frac{1}{\sqrt{6}} P_{4'}^{uu} +\sqrt{6} P_{4'}^{ww} \right)(w),
\qquad
O_{4'}(w) = \frac{1}{8} \left( \frac{16}{7} \sqrt{\frac{2}{3}} P_{4'}^{uu} 
-\frac{4}{7} \sqrt{6} P_{4'}^{ww}\right)(w),
\label{twofour}
\eea
where the equations (\ref{above}) and (\ref{spin4})
are needed.

The main observation in this section is that the pole structures
(\ref{UW2}), (\ref{UW1}), and (\ref{UUpole1}) are obtained explicitly.
Together with the pole structure (\ref{W4}) found in \cite{Ahn1111}, 
they provide the WZW 
field contents for  $O_{\frac{7}{2}}(z), O_4(z), O_{4'}(z)$ and 
$O_{\frac{9}{2}}(z)$ given in (\ref{7half}), (\ref{9half}) and (\ref{twofour}).
Because the spin-$\frac{3}{2}$ current $G(z)$ in (\ref{cosetg})
determines the superpartner for component field in any primary superfield,
the OPE $G(z) \; O_{\frac{7}{2}}(w)$ should determine the current 
$O_4(w)$. Similarly, the OPE $G(z) O_{4'}(w)$ should lead to the current 
$O_{\frac{9}{2}}(w)$. 
It would be interesting to check these two facts explicitly for consistency.
As described in the begining of this section, 
the three ${\cal N}=1$ higher spin super 
currents with spins-$(\frac{5}{2}, 3), (\frac{7}{2},4)$ 
and $(4, \frac{9}{2})$ are constructed.

\section{The three-point functions with scalars in the large $N$ limit}

In this section, the three-point functions with scalars 
will be determined for the spins $s=2, 3$
because the normalization for these currents are known completely. 
For $s=4, 4'$, 
the normalization for these currents are not fixed because 
the OPEs between each current and itself are not constructed
and the highest singular terms, $\frac{1}{(z-8)^8}$ and $ 
\frac{1}{(z-w)^8}$, are not determined.
  
The conformal dimension $h(0;f)$ \cite{GG} and its large $N$ limit 
can be obtained as follows (the formula for the dimension can be found in 
\cite{BG,BG1}): 
\bea
h(0;f) = \frac{N^2-1}{2N} \left[ \frac{1}{N+N} -\frac{1}{N+N+k} \right]
= \frac{N^2-1}{4N^2} \left[ 1-\frac{2N}{2N+k} \right]
\rightarrow \frac{(1-\la)}{4(1+\la)},
\label{form}
\eea
where the eigenvalue of the quadratic Casimir operator of 
$SU(N)$, $\frac{N^2-1}{2N}$, is used. 
The 't Hooft coupling constant is
\bea
\la =\frac{N}{k+N}.
\label{lambda}
\eea
The level $N$ appears in the two denominators of (\ref{form}), whileas
the level $k$ appears in the second denominator of (\ref{form}).
At the final stage of (\ref{form}), after substituting $k = 
\frac{1-\la}{\la} N$, the large $N$ limit was taken.
There is no contribution from the trivial representation.

On the other hand, the conformal dimension $h(f;0)$ and its 
large $N$ limit can be derived as follows:
\bea
h(f;0) = \frac{N^2-1}{2N} \left[ \frac{1}{N+k} +\frac{1}{N+N} \right]
= \frac{N^2-1}{4N^2} \left[ 1+\frac{2N}{N+k} \right]
\rightarrow \frac{1+2\la}{4},
\label{form1}
\eea
where 
the level $k$ appears in the first denominator of (\ref{form1}), whileas
the level $N$ appears in the second denominator of (\ref{form1}).
At the final step, the large $N$ limit was taken similarly.
The trivial representation does not contribute to the conformal dimension.

The Virasoro zero mode (acting on the above primaries) 
eigenvalue is determined as follows. 
For the primary $(0;f)$, the zero mode $K_0^a$ corresponding to 
the numerator current with level $k$ of the coset model vanishes \cite{GH}.
That is,
the eigenvalue is zero when this zero mode acts on the state corresponding to 
this primary field.
On the other hand,
the ground state transforms as a fundamental representation with respect to
the zero mode 
$J_0^a$ corresponding to the numerator current with the level $N$
of the coset.
The nonzero contribution arises from the first and third terms of 
spin $2$ stress energy tensor (\ref{cosett}).
The result can be expressed as
\bea
\left(-\frac{1}{4N} J_0^a J_0^a + \frac{1}{2(k+2N)} J_0^a J_0^a \right) |f>
& = & \left(-\frac{1}{4N}  + \frac{1}{2(k+2N)}  \right)(-N) |f>
\nonu \\
& = &  \frac{(1-\la)}{4(1+\la)} | f> = h(0;f) | f>.
\label{res}
\eea
Note that because $J_0^a J_0^a = \delta^{ab} \mbox{Tr} (T^a T^b) =-\delta^{aa}=
-(N^2-1)$, the large $N$ limit for the eigenvalue equation 
leads to $-N$. Furthermore, the extra 
$\frac{1}{N}$ factor arises due to the fact that the eigenvalue 
(not a trace) is needed. Therefore, the final contribution to the fundamental
representation provides $-N$ \cite{CY,Ahn1111}.
At the final stage of (\ref{res}), the identity (\ref{form}) is used.
The result (\ref{res}) indicates that the Virasoro zero mode eigenvalue 
can be fixed by the conformal dimension of the scalar primary operator.

On the other hand, for the primary $(f;0)$, the zero mode
$K_0^a$ equals to $-J_0^a$.
Or the eigenvalue equation of the zero mode of 
the diagonal denominator current in the coset 
has zero eigenvalue.  
Furthermore,
the ground state transforms as a fundamental representation with respsect to 
$K_0^a$ and as an antifundamental representation with respect to 
$J_0^a$. 
The nonzero contribution arises from the first and second terms of 
spin $2$ Virasoro current (\ref{cosett}).
The result can be derived 
\bea
\left(-\frac{1}{4N} J_0^a J_0^a - \frac{1}{2(k+N)} J_0^a J_0^a \right) |f>
& = & \left(-\frac{1}{4N}  - \frac{1}{2(k+N)}  \right)(-N) |f>
\nonu \\
& = &  \frac{1}{4}(1+ 2\la) | f> = h(f;0) | f>.
\label{resultone}
\eea
Although the generators 
in the antifundamental representation have the extra minus signs,
compared to those in the fundamental representation,
the final result 
has no extra minus sign because the number of power of 
the $SU(N)$ generator $T^a (=K_0^a)$ is even.
The identity (\ref{form1}) is used.
Again, the Virasoro zero mode eigenvalue is fixed by
the conformal dimension of the scalar primary operator. 

The three-point functions with two real scalars, from (\ref{resultone})
and (\ref{res}), are 
\bea
< \overline{\cal{O}}_{+} {\cal{O}}_{+} T > =   \frac{1}{4}(1+ 2\la),
\qquad
< \overline{\cal{O}}_{-} {\cal{O}}_{-} T > =  \frac{(1-\la)}{4(1+\la)},
\label{spin2threepoint}
\eea
where 
the scalar primaries 
correspond to 
${\cal{O}}_{+} = (f;0) \otimes (f;0)$, $ \overline{\cal{O}}_{+}=
(\overline{f};0) \otimes (\overline{f};0)$,
${\cal{O}}_{-} = (0;f) \otimes (0;f)$ and $ \overline{\cal{O}}_{-}=
(0;\overline{f}) \otimes (0;\overline{f})$, respectively.
The normalization for the spin $2$ current is described as
$<T(z) \; T(w) > =\frac{1}{(z-w)^4} \left[ \frac{1}{2} 
N^2\frac{(1-\lambda ) (1+2\lambda )}{2 (1+\lambda )} \right]$ 
in the large $N$ 
limit.
By absorbing the $(N,\la)$ dependent term into 
the  spin $2$ current-spin $2$ current correlator 
(or equivalently dividing this two point function by $ \frac{1}{2} 
N^2\frac{(1-\lambda ) (1+2\lambda )}{2 (1+\lambda )}$),
\bea
< \overline{\cal{O}}_{+} {\cal{O}}_{+} T > & = &   \frac{1}{2N}
\sqrt{\frac{(1+\la)(1+2\la)}{(1-\la)}},
\qquad
< \overline{\cal{O}}_{-} {\cal{O}}_{-} T > =  \frac{1}{2N} 
\sqrt{\frac{(1-\la)}{(1+\la)(1+2\la)}},
\label{threepoint2}
\eea
where 
$<T(z) \; T(w) > =\frac{1}{(z-w)^4}$. 
Purely $\la$-dependent parts in each case of (\ref{threepoint2}) 
are proportional to each other 
inversely. In other words, the product of these leads to
$\frac{1}{4N^2}$.

Furthermore, the conformal dimension $h(f;f)$ \cite{GG}, which goes to zero
in the large $N$ limit when the second level is $1$
and forms a continuum of light  states near the vacuum,
and its large $N$ limit can be determined similarly.
\bea
h(f;f) = \frac{N^2-1}{2N} \left[ \frac{1}{N+k} -\frac{1}{N+N+k} \right]
= \frac{(N^2-1)}{2(N+k)(2N+k)}
\rightarrow \frac{\la^2}{2(1+\la)}.
\label{resultff}
\eea
The trivial representation corresponds to 
the numerator current with the level $N$ in the coset.
The appearance of $N$ (from the level) 
in the second denominator of (\ref{resultff})
gives rise to a nontrivial $N$ factor in the numerator 
and this cancels the same factor in the denominator.  
Therefore, in the large $N$ limit, 
the conformal dimension 
is nonzero.
The states are no longer light.
The above can be expressed as
$
h(f;f) = h(f;0) + h(0;f)-\frac{N^2-1}{2N^2}$
with (\ref{form}) and (\ref{form1}).

The Virasoro zero mode  
eigenvalue can be determined similarly. 
For the primary $(f;f)$, the zero mode $J_0^a$ vanishes.
More precisely, the eigenvalue becomes zero.
The ground state transforms as a fundamental representation with respect to
$K_0^a$.
The nonzero contribution arises from the second and third terms of 
(\ref{cosett}).
The result can be expressed as
\bea
\left(-\frac{1}{2(k+N)} K_0^a K_0^a + \frac{1}{2(k+2N)} K_0^a K_0^a \right) |f>
& = & \left(-\frac{1}{2(k+N)}  + \frac{1}{2(k+2N)}  \right)(-N) |f>
\nonu \\
& = &  \frac{\la^2}{2(1+\la)} | f> = h(f;f) | f>.
\label{resultresult}
\eea
In the last line of (\ref{resultresult}), the relation (\ref{resultff}) is 
used.

One expects that there is a nonzero three-point function.
The three-point function with two real scalars, from (\ref{resultresult}), 
are 
\bea
< \overline{\cal{O}} \, {\cal{O}} \, T > =   \frac{\la^2}{2(1+\la)},
\label{aboveresult}
\eea
where 
the scalar primaries correspond to 
${\cal{O}} = (f;f) \otimes (f;f)$ and $ \overline{\cal{O}}=
(\overline{f};\overline{f}) \otimes (\overline{f};\overline{f})$, respectively.
By rescaling the stress energy tensor as done before,
the above  (\ref{aboveresult}) becomes
\bea
< \overline{\cal{O}} \, {\cal{O}} \, T > & = &   \frac{\la^2}{N 
\sqrt{(1+\la)(1-\la)(1+2\la)}},
\label{oot}
\eea
where 
$<T(z) \; T(w) > =\frac{1}{(z-w)^4}$.

The spin $3$ zero mode eigenvalue can be determined 
similarly.
For the primary $(0;f)$ with vanishing $K_0^a$ (i.e. the eigenvalue is zero), 
the 
nontrivial contribution from the spin $3$ current 
(\ref{Wk}) together with 
(\ref{BC})
can be obtained 
\bea
B(N,k) k (k+N) (2k+N) d^{abc} J_0^a J_0^b J_0^c | f > & = &
\frac{i}{6 \sqrt{6}N^2} \frac{(1-\lambda )}{(1+\lambda )} 
\sqrt{\frac{(2-\lambda ) }{(2+3 \lambda )}}
(i N^2) | f >
\nonu \\
& = & -\frac{1}{6 \sqrt{6}} \frac{(1-\lambda )}{(1+\lambda )} 
\sqrt{\frac{(2-\lambda ) }{(2+3 \lambda )}} | f >.
\label{eigen}
\eea
The eigenvalue $i N^2$ in (\ref{eigen}) can be determined 
using $d^{abc} \mbox{Tr}(T^a T^b T^c)=-\frac{i}{2} d^{abd} d^{abc} \mbox{Tr}
(T^d T^c) \rightarrow i N^2$ after the extra $\frac{1}{N}$ is multiplied.

On the other hand, for the primary 
$(f;0)$ with $K_0^a =-J_0^a$ (i.e. the eigenvalue for the diagonal current is 
zero),
the following result can be derived
\bea
B(N,k) 
(k+2N) (k+3N) (2k+5N) 
d^{abc} J_0^a J_0^b J_0^c | f >  
& = & 
\frac{i}{6 \sqrt{6}N^2}
(1+2\lambda ) \sqrt{\frac{(2+3\lambda ) }{(2-\lambda )}}
(-i N^2) | f >  \nonu \\
& = & \frac{1}{6 \sqrt{6}}
(1+2\lambda ) \sqrt{\frac{(2+3\lambda ) }{(2-\lambda )}}
 | f >.
\label{eigen1}
\eea
In this case, each term in (\ref{Wk}) contributes to the 
eigenvalue equation.
The extra minus sign in the eigenvalue of the first line of 
(\ref{eigen1}) is due to the fact that the ground state transforms as 
an antifundamental representation with respect to $J_0^a$. 

Then  the three-point functions with scalars 
can be expressed as, from (\ref{eigen}) and (\ref{eigen1}),
\bea
< \overline{\cal{O}}_{+} {\cal{O}}_{+} W >  & = & \frac{1}{6 \sqrt{6}}
(1+2\lambda ) \sqrt{\frac{(2+3\lambda ) }{(2-\lambda )}},
\nonu \\
< \overline{\cal{O}}_{-} {\cal{O}}_{-} W >  & = & 
-\frac{1}{6 \sqrt{6}} \frac{(1-\lambda )}{(1+\lambda )} 
\sqrt{\frac{(2-\lambda ) }{(2+3 \lambda )}}.
\label{spin3threepoint}
\eea
Each three-point function in (\ref{spin3threepoint}) contains 
the corresponding two point function 
in (\ref{spin2threepoint}), respectively. 
The other $\la$ dependent part of one three-point function 
appears in the other three-point function inversely.  
The normalization for the spin $3$ current is as follows:
$<W(z) \; W(w) > =\frac{1}{(z-w)^6} \left[ \frac{1}{3} 
N^2\frac{(1-\lambda ) (1+2\lambda )}{2 (1+\lambda )} \right]$.
By different normalization,
the above three-point function become
\bea
< \overline{\cal{O}}_{+} {\cal{O}}_{+} W >   & = &  \frac{1}{6N}
\sqrt{\frac{(1+\la)(1+2\la)(2+3\la) }{(1-\lambda )(2-\la)}},
\nonu \\
 < \overline{\cal{O}}_{-} {\cal{O}}_{-} W >  & = &  
-\frac{1}{6 N}  
\sqrt{\frac{(1-\la)(2-\lambda ) }{(1+\la)(1+2 \lambda )(2+3\la)}}.
\label{threepoint3} 
\eea
The $\la$-dependent parts in each case of (\ref{threepoint3}) 
are proportional to each other 
inversely.

For the primary $(f;f)$ with vanishing $J_0^a$ (i.e. the eigenvalue is zero), 
the 
nontrivial contribution in the spin $3$ zero mode
from (\ref{Wk}) together with 
(\ref{BC})
can be obtained 
\bea
B(N,k) (-1) 6N^3 d^{abc} K_0^a K_0^b K_0^c | f > & = &
-\frac{i \la^3 }{N^2(1+\la)
\sqrt{6(2-\lambda )(2+3 \lambda)}}
(i N^2) | f >
\nonu \\
& = & \frac{\la^3}{(1+\lambda )  
\sqrt{6 (2-\lambda )(2+3 \lambda)}} | f >.
\label{resultabove}
\eea
From (\ref{resultabove}), the three-point functions with scalars 
can be expressed as
\bea
< \overline{\cal{O}} \, {\cal{O}} \, W >   & = &
 \frac{\la^3}{(1+\lambda )  
\sqrt{6 (2-\lambda )(2+3 \lambda)}}. 
\label{resresres}
\eea
By different normalization with $<W(z) \; W(w) > =\frac{1}{(z-w)^6}$,
the above three-point function (\ref{resresres}) becomes
\bea
< \overline{\cal{O}} \, {\cal{O}} \, W >   & = &  \frac{\la^3}{N
\sqrt{(1+\la)(1-\la)(1+2\la)(2+3\la)(2-\la)}}.
\label{aboveaboveabove}
\eea
The factor in (\ref{oot}) appears in (\ref{aboveaboveabove}). In other words, 
$< \overline{\cal{O}} \, {\cal{O}} \, W >  = 
 \frac{\la}{\sqrt{(2+3\la)(2-\la)}} < \overline{\cal{O}} \, {\cal{O}} \, T > $.

Therefore, the three-point functions with scalars are completely determined
for $s=2, 3$ in the large $N$ limit. 
How should the other three-point functions corresponding to 
$s=4, 4'$?
The unnormalized three-point functions can be obtained using the prescriptions
in this section. As described in the begining of this section,
because the normalizations for $O_4(z)$ and $O_{4'}(z)$ 
are not determined in this paper, the right hand side of
the three-point functions are not fixed.
Moreover, 
compared to the $W_N$ minimal model, the result (\ref{resultff}) for the 
conformal dimension (and corresponding three-point function in (\ref{oot})) 
indicates that the corresponding states can survive in the
large $N$ limit. The states are no longer light.
 
\section{Conclusions and outlook }

In this paper, 
the explict expressions for the ${\cal N}=1$ higher spin currents 
of spins-$(\frac{7}{2}, 4)$ and $(4, \frac{9}{2})$,
which are the second and third elements in (\ref{list}), are constructed.
These are obtained from the OPE between the ${\cal N}=1$ 
lowest higher spin super 
currrent of spins-$(\frac{5}{2},3)$, which is the first element in 
(\ref{list}), and itself. 

Furthermore, the three OPEs (\ref{wwexp}), (\ref{wuexp1}) 
and (\ref{uulast}) can be expressed as a single ${\cal N}=1$ super OPE, via 
Appendix $E$, as follows:
\bea
\hat{W}(Z_1) \; \hat{W}(Z_2) &= &
\frac{1}{z_{12}^5} \, \frac{c}{15} +
\frac{\theta_{12}}{z_{12}^4} \, \hat{T}(Z_2)+
\frac{1}{z_{12}^3} \, \frac{1}{3} 
D \hat{T}(Z_2) +\frac{\theta_{12}}{z_{12}^3} \, \frac{2}{3} \pa \hat{T}(Z_2)
 +  \frac{1}{z_{12}^2} \, \frac{2}{3} \hat{T}^2(Z_2) \nonu \\
& + & \frac{\theta_{12}}{z_{12}^2} \left[ \frac{(2c+5)}{2(4c+21)} 
\pa^2 \hat{T} + \frac{22}{(4c+21)} \hat{T} D \hat{T} +
\frac{1}{\sqrt{6}} \hat{O}_{\frac{7}{2}}\right](Z_2)
\nonu \\
&+& \frac{1}{z_{12}} \left[ \frac{2(18c+1)}{(4c+21)(10c-7)} 
D \hat{T} D \hat{T}+ \frac{(2c^2-c-37)}{(4c+21)(10c-7)} D \pa^2 \hat{T} 
\right. \nonu \\
& - & \left.  
\frac{2(2c-83)}{(4c+21)(10c-7)} \hat{T} \pa \hat{T} +
\frac{1}{7\sqrt{6}} D \hat{O}_{\frac{7}{2}} + \frac{2}{\sqrt{6}} 
\hat{O}_4 \right](Z_2)
\nonu \\
&+& \frac{\theta_{12}}{z_{12}} \left[  \frac{16(7c-10)}{(4c+21)(10c-7)} 
\hat{T} D \pa \hat{T} +  
\frac{4(2c^2-29c+3)}{3(4c+21)(10c-7)} \pa^3 \hat{T} \right. \nonu \\
& + & \left. \frac{8(18c+1)}{(4c+21)(10c-7)}
 D \hat{T} \pa \hat{T}  + 
\frac{4}{7 \sqrt{6}} \pa \hat{O}_{\frac{7}{2}} + \frac{1}{\sqrt{6}} 
D \hat{O}_4 \right](Z_2) +\cdots,
\label{singleOPE}
\eea
where the ${\cal N}=1$ super currents have the following component fields 
\cite{Ahn1211}
\bea
\hat{W} (Z)  & = & \frac{1}{\sqrt{6}} \, U(z) + \theta \; W(z), 
\nonu \\ 
\hat{O}_{\frac{7}{2}}(Z) & = &  O_{\frac{7}{2}} (z) +\theta \; O_4 (z),
\nonu \\
\hat{O}_4 (Z) & = &  O_{4'} (z) +\theta \; O_{\frac{9}{2}} (z),
\qquad
Z=(z,\theta),
\label{highersome} 
\eea
and
$z_{12} = z_1 -z_2-\theta_1 \theta_2, 
\theta_{12} = \theta_1 -\theta_2, D = \pa_{\theta} + \theta
\pa_z$ and $\pa =\pa_z$.
Because the $c$-dependent coefficients in the linear superfield 
terms of (\ref{singleOPE})
approach to constant value and those in the quadratic superfield terms 
approach to $\frac{1}{c}$ when $c \rightarrow \infty$, 
the corresponding classical algebra can be obtained from (\ref{singleOPE})
with appropriate coefficients found in this limit. 
This will correspond to the asymptotic symmetry algebra of $AdS_3$ bulk theory.

The ${\cal N}=1$ super fusion rule of (\ref{singleOPE}) 
can be summarized as
\bea
\left[\hat{W} \right] \left[\hat{W} \right] & = &  \left[\hat{I} \right] + \left[\hat{O}_{\frac{7}{2}}
\right] +\left[\hat{O}_4 \right].
\label{fusion}
\eea
The coupling contants, $c_{\frac{5}{2} \frac{5}{2}}^{\frac{7}{2}}$ and 
 $c_{\frac{5}{2} \frac{5}{2}}^{4}$, 
appearing in front of $\left[\hat{O}_{\frac{7}{2}}\right]$ and 
$\left[\hat{O}_4\right]$, respectively
are not determined in this expression (\ref{fusion}) because 
the normalizations of $\hat{O}_{\frac{7}{2}}(Z)$ and $\hat{O}_4(Z)$
(\ref{highersome}) are not fixed.
These normalizations can be fixed 
only if the OPEs $\hat{O}_{\frac{7}{2}}(Z_1) \; \hat{O}_{\frac{7}{2}}(Z_2)$
and $\hat{O}_4(Z_1) \; \hat{O}_4(Z_2)$ are calculated.
A further study should determine these OPEs.
These coupling constants should vanish when $N=3$ and $c=\frac{10}{7}$ (or 
$k=1$).
See also the recent developements 
\cite{CGKV,CG} where the approach $1$ of \cite{BS} is used to fix the 
normalizations.
They make the most general ansatz for the OPEs of the currents with each 
other and solve the Jacobi identities at the level of OPEs rather than 
modes.

The three-point functions with two scalars and a single conserved current
of spins $2$, $3$ are obtained. The new observation is 
that the conformal dimension and the three-point function including the state 
$(f;f)$ or its complex conjugate representation
behave nonzero eigenvalue equation in the large $N$ limit.
This arises from the fact that the difference between  
the level $N+k$ for the diagonal WZW current and the level $k$ for the first
numerator WZW current 
is given by $N(= N+k -k)$. Recall that when the level for the second 
numerator WZW current is equal to $1$, the above difference leads to 
$1(=1+k-k)$.   
Therefore, the previous light states are no longer light and they
participate in the nonzero states even in the large $N$ limit.

It is not known what the spin contents are in the coset model (\ref{coset}). 
How do the extra spin contents arise in addition to the spin contents in 
the minimal ${\cal N}=1$
super $W_N$ algebra? 
In \cite{ASS}, the character technique was used to extract the spin contents 
for $c=4$ eight fermion model (i.e. $k \rightarrow \infty$ limit with 
$N=3$ in (\ref{coset})). As a warmup, the analysis for $N=4$ with 
$k \rightarrow \infty$ can be done.  
Finding the general $N$ character technique will be an open problem.
This will be a generalization of the recent construction found 
in \cite{BCGG} because the spin contents for the minimal ${\cal N}=1$
supersymmetric $W_N$ algebra are contained.

Compared to the ${\cal N}=2$ ${\cal W}_{N+1}$ algebra studied in 
\cite{Ahn1206,CG,Ahn1208},
the OPE of the ${\cal N}=1$ 
lowest higher spin super current looks similar 
but the right hand side of (\ref{fusion}) in the present case
does not contain the ${\cal N}=1$ lowest higher spin super current. 
Suppose that $\hat{W}(Z_2)$ can arise in the right hand side of (\ref{fusion}).
Then the super current 
$\hat{W}(Z_2)$ can appear in $\frac{\theta_{12}}{z_{12}^3}$-term. However, 
due to the symmetry in $\hat{W}(Z_1) \; \hat{W}(Z_2)$(same fermionic 
super currents), after reversing 
the arguments $Z_1$ and $Z_2$ and expanding around $Z_2$, the same 
expression $\hat{W}(Z_2)$ will appear with an opposite sign. This suggests
that the super current $\hat{W}(Z_2)$ should vanish.

What about the OPEs $\hat{W}(Z_1) \; \hat{O}_{\frac{7}{2}}(Z_2)$
and $\hat{W}(Z_1) \; \hat{O}_{4'}(Z_2)$?
At the moment, one expects that these OPEs can be expressed in terms of
other ${\cal N}=1$ 
super currents (the explicit calculations will be very complicated). 
In Appendix $E$, the two ${\cal N}=1$ super OPEs are presented 
by replacing the $k$-dependent coefficient functions with 
the central charge when $N=3$. 
Note that the structure constant $c_{\frac{5}{2} \frac{7}{2}}^{\frac{5}{2}}$
has a factor $(7c-10)$ in the present case and those constant for 
the minimal ${\cal N}=1$ $W_N$ algebra was studied in \cite{BCGG}.
One expects that this structure constant will depend on both the 
central charge $c$ and $N$
and have still a factor $(7c-10)$ for general $N$.
Furthermore, the new ${\cal N}=1$ 
higher spin currents should appear when the spins of two
super 
currents in the super 
OPE increase. It would be interesting to determine the complete
set of (super) currents for the coset (\ref{coset}). At the moment, it will be 
very difficult to determine the field contents in terms of WZW currents 
but the generalization of the character technique for general $N$ can be 
used to determine the spin contents at least.   

As described in the introduction, 
the ${\cal N}=2$ supersymmetric extension can arise for $k=N$
where
the coset is given by
\bea
\frac{\widehat{SU}(N)_N \oplus \widehat{SU}(N)_N}
{\widehat{SU}(N)_{2N}}.
\label{othercoset}
\eea
It would be interesting to determine the ${\cal N}=2$ higher spin super 
currents
as well as the ${\cal N}=2$ super 
stress energy tensor with spins-$(1, \frac{3}{2}, \frac{3}{2}, 2)$.  
In this case, it is not clear how to take the large $N$ limit with fixed 
't Hooft coupling constant because 
the $k$ dependence disappears and the 't Hooft coupling constant becomes 
$\la =\frac{1}{2}$ \cite{CY}.
In doing this, the first step is to construct the ${\cal N}=2$
superconformal algebra realized by the WZW currents living in the coset
(\ref{coset}) with the substitution of $k=N$ where $c=\frac{N^2-1}{3}$. 
The two independent spin-$\frac{3}{2}$ currents 
can be obtained from two independent free fermions of spin-$\frac{1}{2}$.
Obviously, the $U(1)$ spin $1$ current in the ${\cal N}=2$ 
superconformal algebra can be constructed from the product 
of these two fermions.  
The next step is to determine the higher spin currents, $(2,\frac{5}{2}, 
\frac{5}{2}, 3), \cdots, (s, s+\frac{1}{2}, s+\frac{1}{2}, s+1), \cdots$
in terms of WZW currents. 
The structure for $N=3$ will be useful to determine the lower ${\cal N}=2$ 
higher spin 
super currents in terms of the WZW currents living in this specific coset. 
It would be interesting to see whether the recent findings 
with ${\cal N}=1$ supersymmetry in \cite{BCGG}
can arise in the coset model (\ref{othercoset}) or not. 
Although they have different supersymmetries, they share the common 't Hooft
coupling constant $\la =\frac{1}{2}$.
The first step in this direction is to understand the precise relation 
between ${\cal N}=2$ 
${\cal W}_3$ algebra and the ${\cal N}=1$ ${\cal W}_3$ algebra and to see 
how the 
latter can be embedded in the former.

As described in \cite{BS}, one of the levels is equal to $1$ 
for the equivalence between the 
coset construction and Drinfeld-Sokolov reduction and the triality in 
\cite{GG1} is based on this particular case. 
Because the coset model (\ref{coset}) in this paper has the second level $N$
which is not equal to $1$, it is not clear how to 
observe the isomorphism between these two constructions.
It would be interesting to study Drinfeld-Sokolov reduction for the more 
general 
coset model.  

One application of the result of this paper is to analyze the ${\cal N}=1$ 
higher spin 
super currents living in the coset model with orthogonal group. In particular, 
the holographic minimal model with ${\cal N}=1$ supersymmetry is 
described in \cite{CHR,CV} where the coset is expressed as the orthogonal 
groups. In this case, although the second level of the numerator WZW current
is equal to $1$, the spin-$\frac{3}{2}$ current is obtained.
It would be interesting to construct the ${\cal N}=1$ higher spin 
super currents 
when the second level is given by the dual Coxeter number of 
$SO(2N)$.
Similar analysis can be done in the context of 
\cite{AP,Ahn1202,Ahn1106,GV}. By applying to the present methods, 
the construction of ${\cal N}=1$ 
higher spin currents when the second level is 
given by the dual Coxeter number of $SO(2N)$. 
Furthermore, 
it would be interesting to 
determine the ${\cal N}=2$ 
higher spin currents by restricting the first level to
the dual Coxeter number of $SO(2N)$ also.
 
The asymptotic symmetry (for the lowest super current) of the ${\cal N}=1$ 
higher 
spin $AdS_3$ gravity  
can be read off from the two dimensional CFT results
obtained thus far.
The three-point functions from the CFT computations  
should correspond to the three-point functions in the 
$AdS_3$ bulk theory. 
The bulk theory would have  higher spin gauge symmetry
in $AdS_3$ string theory because  the central charge 
 in 
this coset model is proportional to $N^2$ rather than $N$. 
Furthermore, the light states in \cite{GG,GG1,GG2} are no longer light ones 
because the conformal dimension of spin $2$ for the state $(f;f)$
has finite value in the large $N$ limit. 
It would be highly nontrivial to find the $AdS_3$ bulk string theory.
One direction to find the $AdS_3$ string theory was described in the context of
large ${\cal N}=4$ holography in \cite{ggi,GG1305}. 
The two-dimensional CFT has more supersymmetry and the extra transverse space 
(in type IIB string theory) has a ${\bf S}^1$ factor.
It would be interesting to see any relations between the coset model in this 
paper and the coset model in \cite{ggi}.

Because there are fermionic currents including the spin-$\frac{3}{2}$ current
(\ref{cosetg}), the three-point functions with a bosonic operator, 
a fermionic operator (superpartner of scalar operator) 
and a fermionic current
can be constructed. 
It would be interesting to describe
the fermionic operators explicitly
and see how to construct the corresponding three-point functions 
with two fermionic operators and a single higher spin current 
with integer spin \cite{CHR1,MZ}. 

\vspace{.7cm}

\centerline{\bf Acknowledgments}

CA would like to thank P. Bowcock, J. Fuchs, M.R. Gaberdiel, R. Gopakumar, 
A. Jevicki, E. Joung, H. Kim  and M. Vasiliev 
for discussions. 
This work was supported by the Mid-career Researcher Program through
the National Research Foundation of Korea (NRF) grant 
funded by the Korean government (MEST) (No. 2012-045385).
We thank the Galileo Galilei Institute for Theoretical Physics 
for the hospitality and the INFN for partial support during 
the completion of this work. 
CA acknowledges warm hospitality from 
the School of  Liberal Arts (and Institute of Convergence Fundamental
Studies), Seoul National University of Science and Technology.

\newpage

\appendix

\renewcommand{\thesection}{\large \bf \mbox{Appendix~}\Alph{section}}
\renewcommand{\theequation}{\Alph{section}\mbox{.}\arabic{equation}}

\section{The second-order pole in the OPE $W(z) \; W(w)$ 
and some expressions relevant to spin $4$ field 
}

In (\ref{wwexp}), to extract the primary spin $4$ current, 
the second-order pole is needed to calculate. 
Because this second order pole of $W(z) \; W(w)$ 
for the arbitrary levels $(k_1, k_2)$ in the coset model
(\ref{coset}) was found in \cite{Ahn1111},
the corresponding expression can be obtained  by simply substituting 
$k_1 =N$ and $k_2=k$ into the primary spin $4$ current in \cite{Ahn1111}. 
Each fourteen term with (\ref{newdef}) is presented as follows:
\bea
&& \{ Q\; Q\}_{-2} 
 =  -18N Q^a Q^a 
 -  54N(N^2-4)J^a \pa^2
J^a, \nonu \\
&& \{ Q\;  Q^a K^a \}_{-2}  = 
-6N(2 d^{abc} J^a Q^b K^c 
 -3(N^2-4) \pa^2 J^a
K^a), \nonu \\
&& \{ Q\; J^a S^a\}_{-2}  = 
-6N Q^a S^a, \nonu \\
&& \{ Q^a  K^a
\; Q \}_{-2}  =   
-6N(2 d^{abc} J^a Q^b K^c 
-3
(N^2-4) J^a \pa^2 K^a), \nonu \\
&& \{ Q^a K^a \;
Q^b K^b \}_{-2}  = 
  d^{abcd} \left[ 
\frac{2}{3}  N  J^a J^b J^c K^d 
- 3N  J^a J^b K^c K^d
 \right] \nonu \\
&&+      d^{abe} d^{cde} \left[ -  k  J^a J^b J^c J^d+
2  N    J^a J^b K^c K^d \right] 
+  
\frac{8(N^2-4)}{N^2+1}  J^a J^a J^b K^b 
\nonu \\
&& -\frac{12(N^2-4)}{(N^2+1)} J^a J^a K^b K^b  
+  6k (N^2-4)   \pa^2 J^a J^a
+\frac{8(N^2-4)(N^2+2)}{N^2+1}
f^{abc} J^a \pa
J^b K^c
\nonu \\
&&  -  \frac{2N(N^2-4)(11N^2+15)}{3(N^2+1)}
 \pa^2 J^a K^a 
+  
3N(N^2-4)
 \pa^2 K^a K^a 
 +   6(N^2-4) 
f^{abc} J^a K^b \pa K^c
\nonu \\
&& -   \frac{3N(N^2-4)(N^2-3)}{N^2+1}
 \pa J^a \pa K^a 
-
  6N(N^2-4)
 J^a \pa^2 K^a 
 - 
 \frac{24(N^2-4)}{(N^2+1)}
 J^a J^b K^a K^b,
\nonu 
\\
&& \{ Q^a K^a \;  
J^b S^b\} _{-2}  = 
-(N+2k) d^{abc} J^a Q^b K^c 
 + 2N Q^a S^a \nonu \\
&&- 3N d^{abc} J^a K^b
S^c   + 
3(N+2k)(N^2-4) \pa^2 J^a  K^a,\nonu \\
&& \{ Q^a K^a \; S  \}_{-2}  =  -3(N+k) Q^a S^a,
\nonu \\
&& \{ J^a S^a \;  
Q \}_{-2}  =  -6N Q^a S^a, \nonu \\
&& \{ J^a S^a \;  Q^b K^b \}_{-2}  =     
-3N d^{abc} K^a S^b J^c 
 + 2N Q^a S^a  - (N+2k) d^{abc} J^a K^b
Q^c   \nonu \\
&& + 
3(N+2k)(N^2-4) \pa^2 K^a  J^a, \nonu \\
&& \{ J^a S^a \; J^a S^a \}_{-2}  =   
  d^{abcd} \left[ 
\frac{2}{3}  N  K^a K^b K^c J^d 
- (N+2k)  K^a K^b J^c J^d
 \right] 
 \nonu \\
&& +      d^{abe} d^{cde} \left[ -  N  K^a K^b K^c K^d+
2  k    K^a K^b J^c J^d \right] 
\nonu \\
&& +  
\frac{8(N^2-4)}{N^2+1}  K^a K^a K^b J^b 
-\frac{4(N^2-4)(N+2k)}{N(N^2+1)} K^a K^a J^b J^b  
 +  2 (N^2-4)(N+2k)   \pa^2 K^a K^a
\nonu \\
&&  -  \frac{2(N^2-4)(6k+9N+6k N^2 +5N^3)}{3(N^2+1)}
 \pa^2 K^a J^a 
\nonu \\
&& +  \frac{1}{N}
(N^2-4)k(N+2k)
 \pa^2 J^a J^a  -   \frac{(N+2k)(N^2-4)(N^2-3)}{N^2+1}
 \pa K^a \pa J^a 
\nonu \\
&& -
  2(N+2k)(N^2-4)
 K^a \pa^2 J^a 
- 
 \frac{8(N^2-4)(N+2k)}{N(N^2+1)}
 K^a K^b J^a J^b  \nonu \\
&& +   \frac{2(N^2-4)(N+2k)}{N} 
f^{abc} K^a J^b \pa J^c
+\frac{4(N^2-4)(k+3N+k N^2+N^3)}{N(N^2+1)}
f^{abc} K^a \pa
K^b J^c, \nonu \\
&& \{ J^a S^a \;  S \}_{-2}  =   
-3(N+k)(2 d^{abc} K^a S^b J^c 
-\frac{1}{N} (N+2k)
(N^2-4) K^a \pa^2 J^a), \nonu \\
&& \{ S  \; Q^a K^a \}_{-2}  =   
-3(N+k) Q^a S^a, \nonu \\
&& \{ S  \; J^a S^a \}_{-2}  =  
-3(N+k)(2 d^{abc} K^a S^b J^c 
 -\frac{1}{N} (N+2k)(N^2-4) \pa^2 K^a
J^a), \nonu \\
&& \{ S \;  S \}_{-2}  = 
-9(N+k) S^a S^a 
 -  \frac{9}{N}(N+k)(N+2k)(N^2-4)K^a \pa^2
K^a. 
\label{pole2WW}
\eea
Considering the above fourteen terms (\ref{pole2WW}) with 
the relative coefficients in (\ref{Wk}) appropriately
provides the complete expression for the second order pole 
of $W(z) \; W(w)$. 

For the calculation of $W^{(4)}(z)$ in (\ref{W4}), 
the following expression for 
$T^2(z)$ (by inserting $k_1=N$ and $k_2=k$ into Appendix $(G.6)$ 
in \cite{Ahn1111}) is used 
\bea
T^2(z) & = & \frac{k^2}{16 N^2(2N+k)^2}\left[- 4 N
\pa^2 J^a J^a(z) + 
J^a J^a J^b J^b(z)\right] \nonu \\
&+&   \frac{N^2}{4(N+k)^2(2N+k)^2}\left[- 2 (k+N)
\pa^2 K^a K^a(z) + 
K^a K^a K^b K^b(z)\right] \nonu \\
& +&  \frac{N k}{4N(N+k)(2N+k)^2} J^a J^a K^b K^b(z)
\nonu \\
& + & \frac{1}{(2N+k)^2} \left[
J^a K^a J^b K^b(z) + N \pa^2 J^a K^a-\frac{N}{2}
\pa^2 K^a K^a(z) + f^{abc} J^a \pa K^b K^c(z) \right. \nonu \\
& - & \left. \frac{k}{2} J^a \pa^2 J^a(z)
-  f^{abc}
J^a \pa J^b K^c(z)\right]
\nonu \\
& - & \frac{N}{2(N+k)(2N+k)^2} \left[J^a K^a K^b K^b(z) 
-(N+k) \pa^2 J^a K^a(z)
\right. \nonu \\
&+ & \left. J^b K^a K^a K^b(z) -(2N+k) J^a
\pa^2 K^a(z) + 
2 f^{abc} J^a K^b \pa K^c(z) \right]
\nonu \\
& - & \frac{k}{4N(2N+k)^2} \left[J^a K^a J^b J^b(z) -2N 
J^a \pa^2 K^a(z) 
\right. \nonu \\
& + & \left. J^a J^a J^b K^b(z) -3N
\pa^2 J^a K^a(z) + 
2 f^{abc} J^a \pa J^b K^c(z)\right].
\label{Tsquare}
\eea

The field $G \pa G(z)$ can be obtained from the following identities
(again the rearrangement lemma that appeared in (\ref{fermionic}) should be
used) 
\bea
(\psi^a J^a) \pa (\psi^b J^b) & = &
\psi^a J^a \pa (\psi^b J^b) -\frac{N}{2} \psi^a \pa^3 \psi^a 
+\frac{3}{4} \pa^2 J^a J^a -\frac{9}{4} \pa (J^a \pa J^a),
\nonu \\
(\psi^a J^a) \pa (\psi^b K^b) & = &
\psi^a \pa \psi^b J^a K^b + \psi^a \psi^b J^a \pa K^b-
\frac{3}{4} \pa^2 J^a K^a -\frac{3}{2} \pa J^a \pa K^a,
\nonu \\
(\psi^a K^a) \pa (\psi^b J^b)
& = & \psi^a \pa \psi^b J^b K^a + \psi^a \psi^b \pa J^b K^a -
\frac{3}{2} \pa J^a \pa K^a -\frac{3}{4} J^a \pa^2 K^a,  
\nonu \\
(\psi^a K^a) \pa (\psi^b K^b) & = &
\psi^a \psi^b K^a \pa K^b + \psi^a \pa \psi^b K^a K^b +\frac{1}{3} k \psi^a 
\pa^3 \psi^a -\frac{1}{2} f^{abc} \psi^a \pa^2 \psi^b K^c \nonu \\
&+ & \frac{1}{2} \pa J^a \pa K^a -\frac{1}{4} \pa^2 K^a K^a -\frac{1}{2}
\pa K^a \pa K^a + \frac{1}{2} \pa^2 J^a K^a \nonu \\
& + & \frac{k}{2} \pa \psi^a \pa^2 
\psi^a, 
\label{GG'}
\eea
and it is expressed as
\bea
G \pa G(z) & = &
 \frac{2 k^2}{9N(k+N)(k+2N)} \left[
\psi^a J^a \pa (\psi^b J^b) -\frac{N}{2} \psi^a \pa^3 \psi^a 
+\frac{3}{4} \pa^2 J^a J^a -\frac{9}{4} \pa (J^a \pa J^a)
\right.
\nonu \\
& - & \frac{3N}{k} \left( \psi^a \pa \psi^b J^a K^b + \psi^a \psi^b J^a \pa K^b-
\frac{3}{4} \pa^2 J^a K^a -\frac{3}{2} \pa J^a \pa K^a \right.
\nonu \\
& + &  \left. \psi^a \pa \psi^b J^b K^a + \psi^a \psi^b \pa J^b K^a -
\frac{3}{2} \pa J^a \pa K^a -\frac{3}{4} J^a \pa^2 K^a \right)
+ \frac{9N^2}{k^2} \left( \psi^a \psi^b K^a \pa K^b \right. \nonu \\
&+ &    \psi^a \pa \psi^b K^a K^b +\frac{1}{3} k \psi^a 
\pa^3 \psi^a -\frac{1}{2} f^{abc} \psi^a \pa^2 \psi^b K^c + 
 \frac{1}{2} \pa J^a \pa K^a \nonu \\
 &-  &   \left. \left. \frac{1}{4} \pa^2 K^a K^a -\frac{1}{2}
\pa K^a \pa K^a + \frac{1}{2} \pa^2 J^a K^a  +  
\frac{k}{2} \pa \psi^a \pa^2 
\psi^a \right) \right].
\label{GG'1}
\eea

As noticed in the footnote \ref{dabcd}, the first five terms in (\ref{W4})
can be expressed as \cite{Ahn1211}
\bea
&& d^{abcd} J^a J^b J^c J^d(z)   =  3 d^{abe} d^{cde} J^a J^b J^c J^d(z)
-\frac{12(N^2-4)}{N(N^2+1)}
J^a J^a J^b J^b(z) \nonu \\
&& -  \frac{3(N^2-4)(N^2-3)}{N^2+1}  
\pa J^a \pa J^a(z)
+\frac{2(N^2-4)(N^2-3)}{N^2+1} \pa^2 J^a J^a(z), \nonu \\
&& d^{abcd} J^a J^b J^c K^d(z) = 
3 d^{abc} J^a Q^b K^c(z)
 -\frac{12(N^2-4)}{N(N^2+1)} J^a J^b J^b K^a(z) \nonu \\
&& +
 \frac{2(N^2-4)(N^2-3)}{N(N^2+1)}
f^{abc} \pa J^a J^b K^c(z) 
- \frac{(N^2-4)(N^2-3)}{N(N^2+1)} f^{abc} J^a \pa J^b K^c(z),
\nonu \\
&& d^{abcd} J^a J^b K^c K^d(z)  =  
 2 d^{ace} d^{bde} J^a J^b K^c K^d(z) 
+ d^{abc} J^a J^b S^c(z) \nonu \\
&& -
\frac{(N^2-4)(N^2-3)}{N(N^2+1)} f^{abc} J^a J^b \pa K^c(z)
 - \frac{4(N^2-4)}{N(N^2+1)} \left[ 2 J^a J^b K^a K^b(z) + J^a J^a K^b
  K^b(z) \right], 
\nonu \\
&& d^{abcd} J^a K^b K^c K^d(z)  =  
 3 d^{abc} J^a K^b S^c(z)
-\frac{12(N^2-4)}{N(N^2+1)} J^a K^a K^b K^b(z) 
\nonu \\
&& +
 \frac{2(N^2-4)(N^2-3)}{N(N^2+1)}
f^{abc} J^a \pa K^b K^c(z) 
 - \frac{(N^2-4)(N^2-3)}{N(N^2+1)} f^{abc} J^a K^b \pa K^c(z),
\nonu \\
&& d^{abcd} K^a K^b K^c K^d(z)   =  3 d^{abe} d^{cde} K^a K^b K^c K^d(z)
-\frac{12(N^2-4)}{N(N^2+1)}
K^a K^a K^b K^b(z) \nonu \\
&& -  \frac{3(N^2-4)(N^2-3)}{N^2+1}  
\pa K^a \pa K^a(z)
+\frac{2(N^2-4)(N^2-3)}{N^2+1} \pa^2 K^a K^a(z).
\label{fiveid}
\eea

The explicit coefficient functions, which depend on $N$ and $k$,
in (\ref{W4})
can be read off from the results \cite{Ahn1111}
\bea
c_1 & =& -\frac{k (1 + N^2) (-5 k^2 - 15 k N - 8 N^2 + k^2 N^2 + 3 k N^3)}
{4 (-4 + N^2) N  (k + 2 N) (-3 + N^2) D(N,k)},
\nonu \\
c_2 &=& -\frac{2 (1 + N^2) (-36 k^3 - 201 k^2 N - 215 
k N^2 + 20 k^3 N^2 - 40 N^3 + 
   85 k^2 N^3 + 75 k N^4)}{9 (-4 + N^2) (k + 2 N) 
(2 k + 5 N) (-3 + N^2) D(N,k)},
\nonu \\
c_3 &= & \frac{N}{(-4+N^2) (k+2 N) (2 k+5 N)},
\nonu \\
c_4 & = & \frac{2 N (1 + N^2) (-25 k^2 - 107 k N - 72 N^2 + 5 k^2 N^2 + 
15 k N^3)}{3 (-4 + N^2)  (k + 2 N) (2 k + 5 N) (-3 + N^2) D(N,k)},
\nonu \\
c_5 &= &- \frac{N (1 + N^2) (-5 k^2 - 15 k N - 8 N^2 + k^2 N^2 + 3 k N^3)}
{2 (-4 + N^2)  (k + N) (k + 2 N) (-3 + N^2) D(N,k)},
\nonu \\
c_6 & \equiv & \frac{n_1}{d_1},
\nonu \\
  n_1  & = &
k (-162 k^3 - 657 k^2 N - 849 k N^2 - 12 k^3 N^2 - 312 N^3 - 
   114 k^2 N^3 - 218 k N^4 \nonu \\
& + &  14 k^3 N^4 - 136 N^5 + 67 k^2 N^5 + 
   75 k N^6), 
\nonu \\
 d_1 & = & 
 6 (-4 + N^2) N  (k + 2 N) (2 k + 5 N)  (-3 + N^2)  D(N,k),
\nonu \\
c_7 & \equiv & \frac{n_2}{d_2}, 
\nonu \\
  n_2  & = &
4 (99 k^3 + 309 k^2 N + 332 k N^2 - 32 k^3 N^2 + 112 N^3 - 
   142 k^2 N^3 - 194 k N^4 \nonu \\
& + &  5 k^3 N^4 - 64 N^5 + 25 k^2 N^5 + 
   30 k N^6), 
\nonu \\ 
d_2 & = & 3 (-4 + N^2) (k + 2 N) (2 k + 5 N)  (-3 + N^2) D(N,k),
\nonu \\
c_8 & = & \frac{2 N}{3 (-4+N^2) (2 k+N) (2 k+5 N)},
\nonu \\
c_9 & \equiv & \frac{n_3}{d_3}, \nonu \\
  n_3 & = & -4 N 
(-25 k^3 - 295 k^2 N - 652 k N^2 - 20 k^3 N^2 - 432 N^3 - 
   66 k^2 N^3 - 10 k N^4 \nonu \\
& + &  5 k^3 N^4 + 96 N^5 + 25 k^2 N^5 + 30 k N^6),
\nonu \\
 d_3 & = & 
(-4 + N^2) (2 k + N) (k + 2 N) (2 k + 5 N) (-3 + N^2) D(N,k),
\nonu \\
c_{10} & \equiv & \frac{n_4}{d_4}, \nonu \\
n_4  & = & 3 N (-10 k^4 - 60 k^3 N - 177 k^2 N^2 - 8 k^4 N^2 - 261 k N^3 - 
   48 k^3 N^3 - 152 N^4 \nonu \\
& - &  86 k^2 N^4 
 +   2 k^4 N^4 - 42 k N^5 + 
   12 k^3 N^5 + 24 N^6 + 23 k^2 N^6 + 15 k N^7), 
\nonu \\
d_4  & = & (-4 + N^2)  (k + N) (2 k + N) (k + 2 N) (2 k + 5 N)  (-3 + 
   N^2) D(N,k),
\nonu \\
c_{11} & = & -\frac{k (-27 k^2 - 57 k N - 24 N^2 + 7 k^2 N^2 + 13 k N^3)}
{N^2 (k + 2 N) (-3 + N^2) D(N,k)},
\nonu \\
c_{12} & = & -\frac{8 
(36 k^3 + 51 k^2 N - 35 k N^2 - 4 k^3 N^2 - 40 N^3 + k^2 N^3 + 
   15 k N^4)}{3 N (k + 2 N) (2 k + 5 N) (-3 + N^2) D(N,k)},
\nonu \\
c_{13} & = & -
\frac{4 (-31 k^2 - 97 k N - 88 N^2 + 3 k^2 N^2 + 5 k N^3)}
{(k + 2 N) (2 k + 5 N) (1 + N^2) D(N,k)},
\nonu \\
c_{14} & = & \frac{8 (-25 k^2 - 155 k N - 192 N^2 + 5 k^2 N^2 + 31 k 
N^3 + 40 N^4)}{(k + 2 N) (2 k + 5 N) (-3 + N^2) D(N,k)},
\nonu \\
c_{15} & = & -\frac{2 (-15 k^2 - 45 k N - 48 N^2 + 3 k^2 N^2 + 9 k N^3 + 
8 N^4)}{(k + N) (k + 2 N) (-3 + N^2) D(N,k)},
\nonu \\
c_{16} & = & \frac{2 (-63 k^3 - 348 k^2 N - 517 k N^2 + 19 k^3 N^2 - 200 N^3 + 
   92 k^2 N^3 + 105 k N^4)}{9 (k + 2 N) (2 k + 5 N) D(N,k)},
\nonu \\
c_{17} & \equiv &  \frac{n_5}{d_5}, \nonu \\
 n_5  & = &
 -2 (-15 k^3 - 24 k^2 N + 39 k N^2 - 12 k^3 N^2 + 72 N^3 
- 78 k^2 N^3 - 
   150 k N^4 \nonu \\
& + &  3 k^3 N^4 - 104 N^5 + 14 k^2 N^5 + 15 k N^6), 
\nonu \\
 d_5  & = &
(k + 2 N) (2 k + 5 N)  (1 + N^2) D(N,k),
\nonu \\
c_{18} & = & \frac{2 (k + 3 N) (-15 k^2 - 25 k N - 8 N^2 + 3 k^2 N^2 + 
5 k N^3)}{3 (k + 2 N) (2 k + 5 N) D(N,k)},
\nonu \\
c_{19} & = & -\frac{8 (-23 k^2 - 61 k N - 48 N^2 + 11 k^2 N^2 + 41 k N^3 + 
40 N^4)}{(k + 2 N) (2 k + 5 N) (1 + N^2) D(N,k)},
\nonu \\
D(N,k) & \equiv &  (39 k^2 + 117 k N + 88 N^2 + 
   5 k^2 N^2 + 15 k N^3). 
\label{coeffWW}
\eea

The large $N$ limit with fixed 't Hooft coupling constant 
$\la$ in (\ref{lambda})  provides the following limiting values
on the coefficient functions
\bea
c_1 & \rightarrow & \frac{(-1+\lambda) }{20N^3 (1+\lambda )},
\qquad 
c_2 \rightarrow -\frac{2 \lambda  (4+\lambda )}
{9 N^3 (1+\lambda ) (2+3 \lambda )},
\nonu \\
c_3 & \rightarrow & \frac{\lambda ^2}{N^3(1+\lambda ) (2+3 \lambda )},
\qquad 
c_4 \rightarrow \frac{2 \lambda ^2}{3N^3 (1+\lambda ) (2+3 \lambda )},
\nonu \\
c_5 & \rightarrow & -\frac{\lambda ^2}{10N^3 (1+\lambda )},
\qquad
c_6 \rightarrow -\frac{(-1+\lambda ) (14+11 \lambda )}
{30N^3 (1+\lambda ) (2+3 \lambda )},
\nonu \\
c_7 & \rightarrow &  \frac{4 \lambda }{3N^3 (2+3 \lambda )},
\qquad
c_8 \rightarrow -\frac{2 \lambda ^2}{3N^3 (-2+\lambda ) (2+3 \lambda )},
\nonu \\
c_9 & \rightarrow & \frac{4 \lambda ^2}{N^3(-2+\lambda ) (2+3 \lambda )},
\qquad
c_{10} \rightarrow -\frac{3 \lambda ^2 \left(2+2 \lambda +\lambda ^2\right)}
{5N^3 (-2+\lambda ) (1+\lambda ) (2+3 \lambda )},
\nonu \\
c_{11} & \rightarrow & \frac{(-1+\lambda ) (7+6 \lambda )}{5N^4 (1+\lambda ) 
(1+2 \lambda )},
\qquad
c_{12} \rightarrow -\frac{8 \lambda  \left(-4+9 \lambda +10 \lambda ^2\right)}
{15 N^4 (1+\lambda ) (1+2 \lambda ) (2+3 \lambda )},
\nonu \\
c_{13} & \rightarrow & -\frac{4 \lambda ^2 (3+2 \lambda )}{5N^4 (1+\lambda ) 
(1+2 \lambda ) (2+3 \lambda )},
\qquad
c_{14} \rightarrow -\frac{8 \lambda ^2 \left(5+21 \lambda +14 
\lambda ^2\right)}
{5N^4 (-1+\lambda ) (1+\lambda ) (1+2 \lambda ) (2+3 \lambda )},
 \nonu \\
c_{15} & \rightarrow & \frac{2 \lambda ^2 \left(3+3 \lambda +2 \lambda ^2\right)}
{5N^4 (-1+\lambda ) (1+\lambda ) (1+2 \lambda )},
\qquad
c_{16} \rightarrow \frac{2 \lambda  (19+16 \lambda )}
{45 N (1+\lambda ) (2+3 \lambda )},
\nonu \\
c_{17} & \rightarrow &  -\frac{2 \lambda  (3+2 \lambda )}{5N (1+\lambda ) 
(2+3 \lambda )},
\qquad
c_{18} \rightarrow  \frac{2 \lambda  (3+2 \lambda )}
{15N (1+\lambda ) (2+3 \lambda )},
\nonu \\
c_{19} & \rightarrow &  \frac{8 \lambda ^2 \left(11+19 \lambda 
+10 \lambda ^2\right)}{5N^4 
(-1+\lambda ) (1+\lambda ) (1+2 \lambda ) (2+3 \lambda )}.
\label{coefflarge}
\eea
Because the coset model (\ref{coset}) in this paper is different from 
that in \cite{Ahn1111}, some of the missing terms in \cite{Ahn1111} can 
survive in (\ref{coefflarge}): two terms with coefficients $c_5$ and 
$c_{10}$. 
The terms with the coefficients, $c_{11}, c_{12}, c_{13}, c_{14}, c_{15}$ and 
$c_{19}$ do not contribute to the final eigenvalue equation because the 
$\frac{1}{N}$ behavior of those terms is suppressed. 

\section{The second-order pole in the OPE $U(z) \; W(w)$ and 
other expression relevant to spin $\frac{7}{2}$ field  }

Let us describe the second-order singular terms in the OPE $U(z) \; W(w)$.
They come from the following $11$ OPEs.  
\bea
&& \{ \psi^a  Q^a \; Q \}_{-2}  = 
-5 N d^{abc} (\psi^a J^b) Q^c -5N d^{abc} d^{cde} J^a (\psi^d J^e) J^b
 \nonu \\
&& -   5N  d^{abc} d^{cde} J^a J^b \psi^d J^e,
\label{1} \\
&& \{ \psi^a  Q^a \; Q^b  K^b \}_{-2}  = 
-5N d^{abc} d^{cde} (\psi^a J^b ) J^d K^e -5N d^{abc} d^{cde} J^a \psi^d 
J^e K^b,  
\label{2} \\
&& \{ \psi^a  Q^a \; J^b  S^b \}_{-2}  = 
-5N d^{abc} \psi^a J^b S^c,
\label{3} \\
&& \{ \psi^a  R^a  \; Q \}_{-2}  = 
d^{abc} d^{def} \left[ f^{dag} f^{egh} (\psi^h J^b K^c) J^f  -  
 f^{dag} f^{egh} f^{hbi} \pa (\psi^i K^c) J^f \right. \nonu \\
&& +  f^{dag} f^{fgh} J^e \psi^h J^b K^c
-  f^{dag} f^{fgh} f^{hbi} J^e \pa (\psi^i K^c) 
+ f^{eag} f^{fgh} J^d \psi^h J^b K^c
\nonu \\
&&-  f^{eag} f^{fgh} f^{hbi} J^d \pa (\psi^i K^c)
+f^{eag} f^{gbh} J^d (\psi^h K^c) J^f
 + f^{fag} f^{gbh} J^d J^e \psi^h K^c 
\nonu \\
&& + f^{dag} f^{gbh} (\psi^h K^c)(J^e J^f)
+ f^{dag} f^{ebh} (\psi^g J^h K^c) J^f + N f^{dag} \delta^{eb} \pa (\psi^g K^c)
J^f  
\nonu \\
&& + f^{dag} f^{fbh} J^e (\psi^g J^h K^c) 
+N f^{dag} \delta^{fb} J^e \pa (\psi^g K^c) 
- N \delta^{bd} (\psi^a K^c)( J^e J^f) \nonu \\
&& -N \delta^{be} J^d (\psi^a K^c) 
J^f  
- N \delta^{bf} J^d J^e \psi^a K^c + N f^{eag} \delta^{bf} J^d \pa 
(\psi^g K^c)  
+ f^{eag} f^{fbh} J^d \psi^g J^h K^c \nonu \\
&& + f^{ebh} f^{fai}J^d \psi^i J^h K^c
  +  f^{ebh} f^{fhi} J^d \psi^a J^i K^c 
 + f^{dbg} f^{eah} (\psi^h J^g K^c) J^f 
\nonu \\
&& \left. + f^{dbg} f^{egh} (\psi^a J^h K^c) J^f 
 +f^{dbg} f^{fah} J^e \psi^h J^g K^c
+  f^{dbg} f^{fgh} J^e \psi^a J^h K^c \right],
\label{4} 
\\
&& \{ \psi^a  R^a   \; Q^b K^b \}_{-2}  = 
d^{abc} d^{def} \left[ 
- k \delta^{fc} (\psi^a J^b)( J^d J^e) 
 +  f^{fcg} f^{dah} (\psi^h J^b K^g) J^e \right.
\nonu \\
&& -f^{fcg} f^{dah} f^{hbi} (\pa (\psi^i 
K^g )) J^e
+ f^{fcg} f^{eah} J^d \psi^h J^b K^g -f^{fcg} f^{eah} f^{hbi} J^d \pa 
(\psi^i K^g)
\nonu \\
&& + f^{eag} f^{fgh} K^d \psi^h J^b K^c - f^{eag} f^{fgh} f^{hbi} K^d \pa (\psi^i 
K^c) 
 + f^{eag} f^{gbh} K^d (\psi^h K^c) J^f \nonu \\
&& + f^{fag} f^{gbh} K^d J^e \psi^h K^c
+ f^{fcg} f^{dbh} (\psi^a J^h K^g) J^e +N f^{fcg} \delta^{db} \pa 
(\psi^a K^g) J^e
\nonu \\
&& + f^{fcg} f^{ebh} J^d \psi^a J^h K^g  + N f^{fcg} \delta^{eb} J^d \pa 
(\psi^a K^g)  - N \delta^{be} K^d (\psi^a K^c) J^f 
 - N \delta^{bf} K^d J^e \psi^a K^c
\nonu \\
&&  + N f^{eag} \delta^{bf} K^d \pa (\psi^g K^c)
 + f^{eag} f^{fbh} K^d \psi^g J^h K^c
+  f^{ebh} f^{fai} K^d \psi^i J^h K^c 
  \nonu \\
&& \left. + f^{ebh} f^{fhi} K^d \psi^a J^i K^c \right],
\label{5} 
\\
&& \{ \psi^a  R^a   \; J^b  S^b \}_{-2}  = 
d^{abc} d^{def} \left[ f^{dag} f^{ech} (\psi^g J^b K^h) K^f
 + f^{dag} f^{fch} K^e \psi^g J^b K^h
\right. \nonu \\
&& - k \delta^{ec} J^d \psi^a J^b K^f
 +  f^{ecg} f^{fgh} J^d \psi^a J^b K^h
- k \delta^{fc} J^d K^e \psi^a J^b 
+ f^{dag} f^{gbh} (\psi^h K^c)(K^e K^f)
\nonu \\
&& \left. - N \delta^{bd} (\psi^a K^c)(K^e K^f)
 +  f^{dbg} f^{ech} (\psi^a J^g K^h) K^f  +
f^{dbg} f^{fch} K^e \psi^a J^g K^h \right], 
\label{6} \\
&& \{ \psi^a  R^a \; S \}_{-2}  = 
d^{abc} d^{def} \left[
- k \delta^{dc} (\psi^a J^b)(K^e K^f)
+  f^{dcg} f^{egh} (\psi^a J^b K^h) K^f
\right.
\nonu \\
&& \left. +f^{fgh} f^{dcg} K^e \psi^a J^b K^h
 -  k \delta^{ec} K^d \psi^a J^b K^f+ f^{ecg} f^{fgh} K^d \psi^a J^b 
K^h  
-  k \delta^{fc} K^d K^e \psi^a J^b \right],
\label{7} \\
&& \{ \psi^a  S^a \; Q \}_{-2}  = 
d^{abc} d^{def} \left[ f^{dag} f^{egh} (\psi^h K^b K^c) J^f 
+  f^{dag} f^{fgh} J^e \psi^h K^b K^c  
\right. \nonu \\
&& \left. + f^{eag} f^{fgh} J^d \psi^h K^b K^c \right], 
\label{8} \\
&& \{ \psi^a  S^a \; Q^b  K^b \}_{-2}  = 
d^{abc} d^{def} \left[ -(N+2k) \delta^{cd} (\psi^a K^b )( J^e J^f )
 + f^{dga} f^{egi} (\psi^i K^b K^c ) J^f
\right.
\nonu \\
&& \left. +f^{dga} f^{fgi} J^e \psi^i K^b K^c
+  f^{eag} f^{fgh} K^d \psi^h K^b K^c \right],
\label{9} \\
&& \{ \psi^a  S^a \; J^b  S^b \}_{-2}  = 
-d^{abc} d^{def} f^{dag} \left[ -f^{ebh} (\psi^g K^h K^c)K^f
- f^{ech} (\psi^g K^b K^h) K^f
\right. \nonu \\
&& - k \delta^{ec} \pa (\psi^g K^b) K^f
+ f^{ebh} f^{hci} \pa (\psi^g K^i) K^f
 - f^{fbh} K^e \psi^g K^h K^c
-f^{fch} K^e \psi^g K^b K^h
\nonu \\
&& \left.
- k \delta^{bf} K^e \pa (\psi^g K^c)
 - k \delta^{fc} K^e \pa (\psi^g K^b)
+  f^{fbh} f^{hci} K^e \pa (\psi^g K^i)
- k \delta^{be} \pa (\psi^g K^c) K^f
 \right]
\nonu \\
&& +  \frac{k}{N} (N^2-4)(N+2k) J^a \pa^2 \psi^a
+ \frac{2}{N} (N^2-4)(N+2k) f^{abc} J^a \pa (\psi^b K^c)
 \nonu \\
&& - N d^{abc} J^a \psi^b S^c
-2(N+2k) d^{abc} d^{cde} J^a \psi^d K^b K^e,
\label{10} \\
&& \{ \psi^a  S^a \; S \}_{-2}  = 
-(N+2k) d^{abc} d^{cde} (\psi^a K^b)(K^d K^e)
- d^{gbc} d^{def} f^{dag} \left[ 
- f^{ebh} (\psi^a K^h K^c) K^f
\right.
\nonu \\
&& - f^{ech} (\psi^a K^b K^h) K^f 
- k \delta^{be} \pa (\psi^a K^c)K^f 
 - k \delta^{ec} \pa (\psi^a K^b) K^f
+ f^{ebh} f^{hci} \pa (\psi^a K^i) K^f
\nonu \\
&&   - f^{fbh} K^e \psi^a K^h K^c
-f^{fch} K^e \psi^a K^b K^h 
 - k \delta^{bf} K^e \pa (\psi^a K^c)
- k \delta^{fc} K^e \pa (\psi^a K^b) 
 \nonu \\
&&  \left. + f^{fbh} f^{hci} K^e \pa (\psi^a K^i) \right]
\nonu \\
&& +  \frac{k}{N} (N^2-4)(N+2k) K^a \pa^2 \psi^a
 + \frac{2}{N}(N^2-4)(N+2k) f^{abc} K^a \pa (\psi^b K^c)
\nonu \\
&& - N d^{abc} K^a \psi^b S^c - 2(N+2k) d^{abc} d^{cde} K^d
\psi^a K^e K^b.
\label{11}
\eea
The following identity, which can be used in (\ref{2}), 
is obtained,  by moving the field $J^d$ to the left,
\bea
d^{abc} d^{cde} (\psi^a J^b) J^d K^e & = & 
d^{abc} d^{cde} J^a \psi^d J^e K^b - 2(N^2-4) \pa^2 \psi^a K^a. 
\label{abo}
\eea
The nonderivative first term of (\ref{abo}) 
will be contributed to the $c_3$-term in (\ref{UW2}) by moving $\psi^d$ 
to the left.
The second term of (\ref{2}) should be rearranged to obtain the final 
expression. 
In (\ref{3}), no rearrangement is needed because the field $S^c(z)$
does not contain $\psi^a(z)$ or $J^b(z)$ from (\ref{newdef}).
In order to simplify the expression (\ref{4}) further, one uses
the identity
\bea
d^{adb} f^{bec} f^{cfa} & = & - N d^{def}, \nonu \\
f^{hae} d^{ebf} f^{fcg} d^{gdh} & = & 
\frac{N}{2} ( d^{ace} d^{bde} - d^{ade} d^{bce}
)-\frac{N}{2} d^{abe} d^{cde},
\label{ididid}
\eea
which will be used continually 
in this paper.
The normal ordered products, using the previous method, are expressed as 
\bea
d^{abc}  (\psi^a R^b ) J^c & = & d^{abc} \psi^a J^b R^c -(N^2-4) \pa^2 
(\psi^a K^a), \nonu \\
f^{abc} \pa (\psi^a K^b) J^c & = & 
-N \pa^2 (\psi^a K^a) +f^{abc} J^a \pa (\psi^b K^c),
\nonu  \\
d^{def} d^{ceh} J^d (\psi^h K^c) J^f & = &
d^{def} d^{ceh} J^d J^f \psi^h K^c +\frac{N^2-4}{N} f^{abc} J^a \pa 
(\psi^b K^c),
\nonu \\
d^{abc} (\psi^a K^b) Q^c & = &
d^{abc}  \psi^a K^b Q^c +2(N^2-4) \pa \psi^a \pa K^a -(N^2-4)
\psi^a \pa^2 K^a,
\nonu \\
d^{abc} d^{def} f^{ebh} f^{dag} (\psi^g J^h K^c) J^f & = &
N(N^2-4) \pa^2 (\psi^a K^a)-\frac{N}{2} d^{abc} d^{cde} J^d J^b \psi^a K^e 
\nonu \\
& + & \frac{N}{2} d^{abc} d^{cde} J^b J^d \psi^a K^e 
+\frac{N}{2} d^{abc} d^{cde} J^d J^e \psi^a K^b,
\nonu \\
d^{abc} d^{def} f^{dbg} f^{eah} (\psi^h J^g K^c) J^f 
& = &
N(N^2-4) \pa^2 (\psi^a K^a) 
-\frac{N}{2} d^{abc} d^{cde} J^d J^a \psi^b K^e 
\nonu \\
& + & \frac{N}{2} d^{abc} d^{cde} J^d J^a \psi^e K^b 
+\frac{N}{2} d^{abc} d^{cde} J^a J^b \psi^d K^e.
\label{abo1}
\eea
The first, second, seventh, ninth, tenth and twenty first terms of
(\ref{4}) are related to the first, second, third, fourth, and fifth equation
of (\ref{abo1}), respectively.
Partial results are presented due to the space of this paper.
Other terms, which were not presented here should be further rearranged to 
arrive at the final expression.
As noticed previously, 
the nonderivative term of first equation in (\ref{abo1})
will contribute to $c_3$-term of (\ref{UW2}), 
the nonderivative term of third and fourth equations in (\ref{abo1})
will be combined to $c_5$-term of (\ref{UW2}).  
The $\psi^a$ appearing in the nonderivative terms in 
the fifth and sixth equations of (\ref{abo1}) should be moved to the left.

The nontrivial normal ordered products in (\ref{5}) can be expressed as 
\bea
d^{abc} d^{def} f^{fcg} f^{dah} (\psi^h J^b K^g) J^e 
& = &
-\frac{N}{2} d^{abc} d^{cde} J^d J^e \psi^a K^b +\frac{N}{2}
 d^{abc} d^{cde} J^b J^d \psi^e K^a 
\nonu \\
& + & \frac{N}{2} d^{abc} d^{cde} J^d J^a \psi^e K^b
+ 2N(N^2-4) \pa \psi^a \pa K^a 
\nonu \\
& - & (N^2-4) f^{abc} J^a \pa (\psi^b K^c)
-2N(N^2-4) \pa^2 (\psi^a K^a),
\nonu \\
d^{def} d^{ceh} K^d (\psi^h K^c) J^f 
& = &
d^{def} d^{ceh} K^d J^f \psi^h K^c +\frac{N^2-4}{N} f^{abc}
K^a \pa (\psi^b K^c),
\nonu \\
d^{abc} d^{def} f^{dbh} f^{fcg} (\psi^a J^h K^g) J^e 
& = & -2N (N^2-4) \pa \psi^a \pa K^a 
+(N^2-4) f^{abc} J^a \pa (\psi^b K^c)
\nonu \\
&-& \frac{N}{2} d^{abc} d^{cde} J^d J^a \psi^e K^b
+ \frac{N}{2} d^{abc} d^{cde} J^a J^d \psi^e K^b
\nonu \\
& +& \frac{N}{2}
 d^{abc} d^{cde} J^d J^e \psi^a K^b,
\nonu \\
d^{abc} d^{dbf} K^d (\psi^a K^c) J^f & = &
d^{abc} d^{cde} J^d \psi^a K^e K^b +\frac{N^2-4}{N} 
f^{abc} K^a \pa (\psi^b K^c).
\label{expexpexp}
\eea
The second, eighth, tenth, and fourteenth terms of
(\ref{5}) are related to the first, second, third, and fourth equation
of (\ref{expexpexp}), respectively.
The $c_1$-term of (\ref{UW2}) arises from the first term of (\ref{5})
and the second and fourth equations of (\ref{expexpexp}) contribute to the 
$c_6$-term of (\ref{UW2}). 
The $c_3$-term of (\ref{UW2}) can be obtained from the 
third equation of (\ref{expexpexp}).
The $c_2$-term can be seen from the last term of (\ref{5}).
All the $\psi^a$ field appearing to the right of $J^b$ in (\ref{expexpexp}) 
should be 
moved to the left.

Similarly, the normal ordered products in (\ref{6})
can be expressed as 
\bea
d^{abc} d^{def} f^{dag} f^{ech} 
(\psi^g J^b K^h) K^f 
& = & -\frac{N}{2} d^{abc} d^{cde} \psi^a J^d K^e K^b 
+\frac{N}{2}  d^{abc} d^{cde} \psi^a J^d K^b K^e 
\nonu \\
& + & \frac{N}{2}  d^{abc} d^{cde} \psi^a J^b K^d K^e, 
\nonu   \\
d^{abc} (\psi^a K^b) S^c & =
& d^{abc} \psi^a K^b S^c -\frac{1}{N} (N+2k)(N^2-4) \pa^2 \psi^a K^a, 
\nonu \\
d^{abc} d^{def} f^{dbg} f^{ech} (\psi^a J^g K^h) K^f 
& = & \frac{N}{2} d^{abc} d^{cde} \psi^d J^a K^e K^b 
-\frac{N}{2} d^{abc} d^{cde} \psi^d J^a K^b K^e
\nonu \\
&-& \frac{N}{2} d^{abc} d^{cde} \psi^a J^b K^d K^e.
\label{aboexp}
\eea
The first, sixth and eighth terms of
(\ref{6}) are related to the first, second, and third equation
of (\ref{aboexp}), respectively.
The nonderivative term of the first equation in (\ref{aboexp})
appears in the $c_2$-, $c_6$-terms of (\ref{UW2}) and the second equation of 
(\ref{aboexp}) contributes to the $c_4$-term of (\ref{UW2}). 

The normal ordered products in (\ref{7}) and (\ref{8})  
are expressed as 
\bea
d^{abc} d^{fch} (\psi^a J^b K^h) K^f & = & d^{abc} d^{cde} \psi^a J^b K^d K^e,
\label{bid0} 
\eea
coming from the second term of (\ref{7})
and 
\bea
d^{abc} d^{fah} (\psi^h K^b K^c) J^f & = &
d^{abc} d^{cde} \psi^d J^e K^a K^b,
\label{bid1}
\eea
coming from the first term of (\ref{8}), 
respectively.
These two expressions play the role of $c_2$-term in (\ref{UW2}).

The following OPEs can be seen from (\ref{9})
\bea
d^{abc} (\psi^a K^b) Q^c & = &
d^{abc} d^{cde} \psi^a J^d J^e K^b +2(N^2-4) \pa \psi^a \pa K^a 
\nonu \\
& - & (N^2-4) \psi^a \pa^2 K^a, 
\nonu \\
d^{fgj} d^{ghi} (\psi^j K^h K^i) J^f 
& = & d^{abc} d^{cde} J^a \psi^b K^d K^e.
\label{abbo} 
\eea
The first and second terms of (\ref{9})
are related to the first and second equations of (\ref{abbo}), 
respectively.
Note that the last expression of (\ref{abbo})
goes to the $c_2$-term of (\ref{UW2}) whileas the first relation of 
(\ref{abbo}) provides the $c_5$-term of (\ref{UW2}).
The third, and last terms of (\ref{9}) 
contribute to the $c_2$-, $c_4$-terms of (\ref{UW2}), respectively.
The $\psi^b$ in the last equation of (\ref{abbo}) should be moved to the 
left.

In (\ref{10}), the following OPE is used
\bea
S^a (z) S^b (w) & = & \frac{1}{(z-w)^4} \frac{2}{N}(N^2-4) k (N+2k) \delta^{ab}
- \frac{1}{(z-w)^3} \frac{2}{N}(N^2-4)(N+2k) f^{abc} K^c(w)
\nonu \\
&-& \frac{1}{(z-w)^2} \left[ N d^{abc} S^c(w) +2(N+2k) d^{ace} d^{bde}
  K^c K^d(w) 
\right] \nonu \\
& + & \frac{1}{(z-w)} 
\left[ -(N+2k) d^{ace} d^{bde} \pa K^c K^d(w) + f^{ace} d^{bcd}
  S^e K^d(w) + f^{ade} d^{bcd} K^c S^e(w) \right. \nonu \\
& - & \left.  (N+2k) d^{ace} d^{bde} K^d \pa K^c(w) \right] 
+ \cdots. 
\label{bid2}
\eea
The normal ordered products in (\ref{10})
are expressed as
\bea
d^{abc} d^{def} f^{dag} f^{ebh} (\psi^g K^h K^c) K^f
& = &
-(N+k) (N^2-4) \pa^2 (\psi^a K^a) + (N^2-4) f^{abc}
\pa (\psi^a K^b K^c)
\nonu \\
& -& \frac{N}{2} d^{abc} d^{cde} \psi^a K^d K^b K^e  
+ \frac{N}{2} d^{abc} d^{cde} \psi^a K^b K^d K^e
\nonu \\
& + & 
\frac{N}{2} d^{abc} d^{cde} \psi^a K^d K^e K^b,
\nonu \\
d^{abc} d^{def} f^{dag} f^{ech} (\psi^g K^b K^h) K^f 
& = & -(N^2-4) f^{abc} \pa (\psi^a K^b K^c) -k (N^2-4) \pa^2 (\psi^a K^a)
\nonu \\
&- & \frac{N}{2} d^{abc} d^{cde} \psi^a K^d K^e K^b 
+\frac{N}{2} d^{abc} d^{cde} \psi^a K^b K^d K^e 
\nonu \\
& +& \frac{N}{2} d^{abc} d^{cde} \psi^a K^d K^b K^e,
\nonu \\
f^{abc} \pa (\psi^a K^b) K^c
& = & N \pa^2 (\psi^a K^a) + f^{abc} K^a \pa (\psi^b K^c).
\label{abbbo}
\eea
The first, second, and third terms of
(\ref{10}) are related to the first, second and third equation
of (\ref{abbbo}), respectively.
The nonderivative term of the first equation in (\ref{abbbo})
appears in $c_4$-term of (\ref{UW2}).
The $c_6$-term of (\ref{UW2}) is obtained from the last term of (\ref{10})
by rearranging the fields.
Moreover, the $c_2$-term of (\ref{UW2}) can be obtained from the last line of 
(\ref{10}).

In (\ref{11}), the normal ordered products 
are simplified as follows: 
\bea
d^{def} d^{gbc} f^{dag} f^{ebh} (\psi^a K^h K^c) K^f 
& = & \frac{N}{2} d^{abc} d^{cde} \psi^a K^d K^b K^e -\frac{N}{2}
d^{abc} d^{cde} \psi^a  K^d K^e K^b 
\nonu \\
& -& \frac{N}{2} d^{abc} d^{cde} \psi^a K^b K^d K^e 
-(N^2-4) f^{abc} \pa (\psi^a K^b K^c) 
\nonu \\
& + & (N+k)(N^2-4) \pa^2 (\psi^a K^a), 
\nonu \\
d^{def} d^{gbc} f^{dag} f^{ech} (\psi^a K^b K^h) K^f 
& = &
k(N^2-4)  \pa^2 (\psi^a K^a) +\frac{N}{2} d^{abc} d^{cde} \psi^a K^d K^e K^b
\nonu \\
&-& \frac{N}{2} d^{abc} d^{cde} \psi^a K^d K^b K^e 
-\frac{N}{2} d^{abc} d^{cde} \psi^a K^b K^d K^e 
\nonu \\
& +& (N^2-4) f^{abc} \pa (\psi^a K^b K^c),
\nonu \\
d^{abc} (\psi^a K^b) S^c & = & d^{abc} d^{cde}
\psi^a K^b K^d K^e \nonu \\
& - & \frac{1}{N} (N^2-4)(N+2k) \pa^2 \psi^a  K^a.
\label{tod}
\eea
The second, third, and first terms of
(\ref{11}) are related to the first, second, and third equation
of (\ref{tod}), respectively.
The $c_4$-term of (\ref{UW2}) is obtained from the above 
nonderivative terms in (\ref{tod}).

Therefore, collecting all the nontrivial terms in (\ref{1})-(\ref{11})
using the extra rearragements for the composite fields described 
previously with (\ref{ididid}), (\ref{bid0}), (\ref{bid1}) and (\ref{bid2}), 
the final second-order pole  of the OPE $U(z) \; W(w)$
is presented as (\ref{UW2}).
The nonderivative terms can be seen easily and the derivative terms 
are collected from various places.

To express $G T(w)$ with (\ref{cosett}) and (\ref{cosetg}) 
explicitly, the following identities are needed
\bea
(\psi^a J^a)( J^b J^b) & = &
\psi^a J^a J^b J^b + N \psi^a \pa^2 J^a - 6N \pa^2 \psi^a J^a,
\nonu \\
(\psi^a J^a)( J^b K^b) & = &
\psi^a J^a J^b K^b + 2 f^{abc} \psi^a \pa J^b K^c - \frac{3}{2} N 
\pa^2 \psi^a K^a,
\nonu \\
(\psi^a K^a)( J^b J^b) & = &
\psi^a J^b J^b K^a + 2N \pa \psi^a \pa K^a - N \psi^a \pa^2  K^a,
\nonu \\
(\psi^a K^a)( J^b K^b) & = &
\psi^a J^b K^a K^b - f^{abc} \psi^a \pa J^b K^c -f^{abc} \psi^a K^b \pa K^c \nonu
\\
& - &  \frac{k}{2} \pa^2 \psi^a J^a + N \psi^a \pa^2 K^a,
\nonu \\
(\psi^a K^a)( K^b K^b) & = &
\psi^a K^a K^b K^b - (N+k) \pa^2 \psi^a  K^a,
\label{GTcomp}
\eea
where the rearrangement lemma $(A.15)$ of \cite{BBSS1}.

The large $N$ limit with fixed $\la$ in (\ref{lambda}) 
for the coefficient functions in (\ref{UW2}) 
leads to 
\bea
c_1 & \rightarrow & \frac{(-2+\lambda ) 
(-1+\lambda )}{4 \sqrt{3} N^{5/2} \sqrt{1+\lambda } (2+3 \lambda )},
\qquad
c_2   \rightarrow  \frac{(-6+\lambda ) 
\lambda ^2}{\sqrt{3} N^{5/2} (-2+\lambda ) \sqrt{1+\lambda } (2+3 \lambda )},
\nonu \\
c_3 &   \rightarrow & \frac{5 (-2+\lambda ) 
\lambda }{4 \sqrt{3} N^{5/2} \sqrt{1+\lambda } (2+3 \lambda )},
\qquad 
c_4  \rightarrow  \frac{8 \lambda ^3}{\sqrt{3} N^{5/2} (-2+\lambda ) 
\sqrt{1+\lambda } (2+3 \lambda )},
\nonu \\
c_5 &  \rightarrow  & \frac{(-2+\lambda ) \lambda }
{2 \sqrt{3} N^{5/2} \sqrt{1+\lambda } (2+3 \lambda )}, 
\qquad
c_6   \rightarrow 
\frac{4 \lambda ^2}{\sqrt{3} N^{5/2} \sqrt{1+\lambda } (2+3 \lambda )},
\nonu \\
c_7 &  \rightarrow  & \frac{(-1+\lambda ) 
(6+5 \lambda )}{8 \sqrt{3} \sqrt{N} \sqrt{1+\lambda } (2+3 \lambda )},
\qquad
c_8   \rightarrow \frac{(-1+\lambda ) (10+3 \lambda )}{8 \sqrt{3} 
\sqrt{N} \sqrt{1+\lambda } (2+3 \lambda )},
\nonu \\
c_9 &  \rightarrow  & \frac{4 \lambda ^2}{\sqrt{3} N^{3/2} \sqrt{1+\lambda } 
(2+3 \lambda )},
\qquad
c_{10}   \rightarrow \frac{\lambda  (10+3 \lambda )}{2 
\sqrt{3} N^{3/2} \sqrt{1+\lambda } (2+3 \lambda )},
\nonu \\
c_{11} &  \rightarrow  & \frac{\lambda  (2+\lambda )}{\sqrt{3} \sqrt{N} 
\sqrt{1+\lambda } (2+3 \lambda )},
\qquad
c_{12}  \rightarrow  \frac{(-2+\lambda ) 
\lambda }{\sqrt{3} \sqrt{N} \sqrt{1+\lambda } (2+3 \lambda )}, 
\nonu \\
c_{13} &  \rightarrow  & \frac{\lambda  (2+7 \lambda )}{2 \sqrt{3} \sqrt{N} 
\sqrt{1+\lambda } (2+3 \lambda )}.
\label{limivalue}
\eea
Then by counting the $N$ behavior from (\ref{limivalue}), 
the zero mode eigenvalue equation 
can be analyzed once the superpartners of scalar fields are found.

\section{The first-order pole in the OPE $U(z) \; W(w)$
and coefficient functions of spin $\frac{9}{2}$ field}

As in previous Appendix $B$, the first-order pole 
of $U(z) \; W(w)$ can be derived
\bea
&& \{ \psi^a  Q^a  \; Q \}_{-1}  = 
-5N d^{abc} d^{cde} \pa (\psi^a J^b) (J^d J^e)
 -  5N d^{abc} d^{cde} J^a \pa (\psi^d J^e) J^b
 \nonu \\
&& -5N d^{abc} d^{cde} J^a J^b \pa (\psi^d J^e),
\label{1-1} 
\\
&& \{ \psi^a  Q^a \; Q^b  K^b \}_{-1}  = 
-5N d^{abc} d^{cde} K^a \pa (\psi^d J^e) J^b
 -  5N d^{abc} d^{cde} K^a J^b \pa (\psi^d J^e),
\label{1-2} \\
&& \{ \psi^a  Q^a  \; J^b  S^b \}_{-1}  = 
-5N d^{abc} d^{cde} \pa (\psi^a J^b) K^d K^e, 
\label{1-3} \\
&& \{ \psi^a  R^a  \; Q \}_{-1}  = 
d^{abc} d^{def} \left[ f^{dag} f^{gbh} \pa (\psi^h K^c) (J^e J^f) 
 - N \delta^{bd} \pa (\psi^a K^c) (J^e J^f)
 \right. \nonu \\
&& -
f^{dag} (\psi^g J^b K^c)(J^e J^f) 
 - f^{dbg} (\psi^a J^g K^c)(J^e J^f)
-f^{eah} J^d ((\psi^h J^b K^c)J^f)
 \nonu \\
&& +  f^{eah} f^{hbi} J^d (\pa (\psi^i K^c) J^f )
-f^{ebh} J^d ((\psi^a J^h K^c) J^f)
 - N \delta^{eb} J^d (\pa (\psi^a K^c) J^f )
\nonu \\
&&   -f^{fah} J^d J^e \psi^h J^b K^c
 + f^{fah} f^{hbi} J^d J^e \pa (\psi^i K^c)
-f^{fbh} J^d J^e \psi^a J^h K^c 
\nonu \\
&& \left. -  N \delta^{fb} J^d J^e \pa (\psi^a K^c) \right],
\label{1-4} \\
&& \{ \psi^a  R^a \; Q^b  K^b \}_{-1}  = 
d^{abc} d^{def} \left[ - k \delta^{fc} \pa (\psi^a J^b)( J^d J^e) 
 - f^{fcg} (\psi^a J^b K^g)( J^d J^e) 
\right. \nonu \\
&& -f^{eah} K^d ((\psi^h J^b K^c) J^f) 
+ f^{eah} f^{hbi} K^d (\pa ( \psi^i K^c) J^f)
-f^{ebh} K^d ((\psi^a J^h K^c)J^f)
 \nonu \\
&& - N \delta^{eb} K^d (\pa (\psi^a K^c) J^f)
\nonu \\
&& -f^{fah} K^d J^e \psi^h J^b K^c
 + f^{fah} f^{hbi} K^d J^e \pa (\psi^i K^c)
 -f^{fbh} K^d J^e \psi^a J^h K^c
\nonu \\
&& \left.  -  N \delta^{fb} K^d J^e \pa (\psi^a K^c) \right],
\label{1-5} \\
&& \{ \psi^a  R^a  \; J^b  S^b \}_{-1}  = 
d^{abc} d^{def} \left[ -f^{dag} (\psi^g J^b K^c)(K^e K^f)  
+  f^{dag} f^{gbh} \pa (\psi^h  K^c)( K^e K^f) \right.
\nonu \\
&& - N \delta^{bd} \pa 
(\psi^a K^c) (K^e K^f)
- f^{dbg} (\psi^a J^g K^c)(K^e K^f) - k \delta^{ec} J^d \pa (\psi^a J^b) 
K^f \nonu \\
&& \left.
-  f^{ecg} J^d (\psi^a J^b K^g) K^f
- k \delta^{fc} J^d K^e \pa (\psi^a J^b)
 -  f^{fcg} J^d K^e \psi^a J^b K^g \right],
\label{1-6} \\
&& \{ \psi^a  R^a  \; S \}_{-1}  = 
d^{abc} d^{def} \left[ - k \delta^{dc} \pa (\psi^a J^b) (K^e K^f)
- f^{dcg} (\psi^a J^b K^g)(K^e K^f)
 \right. \nonu \\
&&  -k \delta^{ec} K^d \pa (\psi^a J^b) K^f
  - f^{ecg} K^d (\psi^a J^b K^g)K^f-
k\delta^{fc} K^d K^e \pa (\psi^a J^b)
 \nonu \\
&& \left. -  f^{fcg} K^d K^e \psi^a J^b K^g \right], 
\label{1-7} \\
&& \{ \psi^a  S^a  \; Q \}_{-1}  = 
d^{abc} d^{def} \left[ - f^{dag} (\psi^g K^b K^c)( J^e J^f)
 -  f^{eag} J^d (\psi^g K^b K^c)J^f  \right. 
\nonu \\
&& \left. - f^{fag} J^d J^e (\psi^g K^b K^c)
\right],
\label{1-8} \\
&& \{ \psi^a  S^a  \; Q^b  K^b \}_{-1}  = 
d^{abc} d^{def} \left[
-(N+2k) \pa (\psi^a K^b) ( J^d J^e)  
- f^{fga} (\psi^g K^b K^c)(J^d J^e) \right. \nonu \\
&& \left. -f^{dag} K^f (\psi^g K^b K^c)J^e
-  f^{eag} K^f J^d (\psi^g K^b K^c) \right],
\label{1-9} \\
&& \{ \psi^a  S^a  \; J^b  S^b \}_{-1}  = 
d^{abc} d^{def} \left[-f^{dag} (\psi^g K^b K^c)( K^e K^f)
- (N+2k) \delta^{cd} J^d \pa (\psi^a K^b) K^f
\right. \nonu \\
&& \left. -f^{ega} J^d (\psi^g K^b K^c) K^f
-  (N+2k) \delta^{cd} J^d K^e \pa (\psi^a K^b)
-f^{fga} J^d K^e (\psi^g K^b K^c) \right],
\label{1-10} \\
&& \{ \psi^a  S^a  \; S \}_{-1}  = 
d^{abc} d^{def} \left[ -(N+2k)\delta^{cd} \pa (\psi^a K^b) (K^e K^f) 
- f^{dga} (\psi^g K^b K^c)(K^e K^f) \right. \nonu \\
&& -(N+2k) \delta^{cd} K^d \pa (\psi^a K^b ) K^f
- f^{ega} K^d (\psi^g K^b K^c )K^f
-(N+2k) \delta^{cd} K^d K^e \pa (\psi^a K^b) 
\nonu \\
&& - \left. f^{fga} K^d K^e \psi^g K^b K^c \right].
\label{1-11}
\eea
The second relation of (\ref{above1})
determines the following normal ordered product in (\ref{1-2})
\bea
d^{abc} d^{cde} K^a \pa (\psi^d J^e) J^b
& = &  -\frac{2}{3} (N^2-4) K^a \pa^3 \psi^a + d^{abc} d^{cde} K^a J^b \pa 
(\psi^d J^e). 
\label{express} 
\eea
The $K^a$ in (\ref{express}) can be moved to the right easily.
No normal ordering procedure in (\ref{1-3}) is needed.

The normal ordered products appearing in (\ref{1-4}) 
are summarized as
\bea
d^{abc} (\pa (\psi^a K^b)) Q^c
& = &
d^{abc} \pa \psi^a K^b Q^c + d^{abc} \psi^a \pa K^b Q^c 
+(N^2-4) \pa \psi^a \pa^2 K^a \nonu \\
& - & \frac{1}{3} (N^2-4) \psi^a \pa^3 K^a,
\nonu \\
d^{abc} d^{def} f^{eah} J^d (\psi^h J^b K^c) J^f & = &
d^{abc} d^{def} f^{eah} J^d J^f \psi^h J^b K^c +(N^2-4) f^{abc} J^a \pa^2 (\psi^b K^c)
\nonu \\
&+ & \frac{1}{2} N d^{abc} d^{cde} J^d \pa (\psi^e J^a K^b)
-\frac{1}{2} N d^{abc} d^{cde} J^d \pa (\psi^a J^e K^b)
\nonu \\
&+& \frac{1}{2} N d^{abc} d^{cde} J^d \pa (\psi^a J^b K^e),
\nonu  \\
d^{abc} d^{def} f^{ebh} J^d (\psi^a J^h K^c ) J^f 
& = & \frac{1}{2} N d^{abc} d^{cde} J^d \pa (\psi^b J^a K^e)
+\frac{1}{2} N d^{abc} d^{cde} J^b \pa (\psi^d J^a K^e)
\nonu  \\
&-& \frac{1}{2} N d^{abc} d^{cde} J^d \pa (\psi^e J^a K^b)+
d^{abc} d^{def} f^{ebh} J^d J^f \psi^a J^h K^c,
\nonu \\
d^{abc} d^{def} f^{eah} f^{hbi} J^d \pa (\psi^i K^c) J^f
& = &
-\frac{1}{2} (N^2-4) f^{abc} J^a \pa^2 (\psi^b K^c)
-N d^{abc} d^{cde} J^a J^b \pa ( \psi^d K^e), 
\nonu \\
d^{abc} d^{def} \delta^{eb} J^d \pa (\psi^a K^c) J^f 
& = & \frac{N^2-4}{2N} f^{abc} J^a \pa^2 (\psi^b K^c)
+ d^{abc} d^{cde} J^d J^e \pa (\psi^a K^b),
\nonu \\
d^{abc} d^{def} f^{dag} (\psi^g J^b K^c) (J^e J^f)
& = &
5(N^2-4) f^{abc} \psi^a \pa J^b \pa K^c 
+ 2N d^{abc} d^{cde} \psi^d J^e J^a \pa K^b
\nonu \\
&-& 2(N^2-4) f^{abc} \pa \psi^a J^b \pa K^c - N 
d^{abc} \psi^a Q^b \pa K^c
\nonu \\
&-& \frac{4}{3} N(N^2-4) \psi^a \pa^3 K^a +
d^{abc} f^{dag} (\psi^g J^b) K^c Q^d,
\nonu \\
d^{abc} d^{cde} (\psi^b J^d) J^a \pa K^e
& = & d^{abc} d^{cde} \psi^b J^d J^a \pa K^e -\frac{N^2-4}{N} f^{abc}
\psi^a \pa J^b \pa K^c \nonu \\
& -& (N^2-4) \pa^2 \psi^a \pa K^a,
\nonu \\
d^{abc} d^{def} f^{dbg}
(\psi^a J^g K^c)( J^e J^f)
& = & 
d^{abc} f^{dag} (\psi^b J^g) K^c Q^d -2(N^2-4) f^{abc} \pa \psi^a 
J^b \pa K^c 
\nonu \\
&+& N d^{abc} d^{cde} \psi^b J^d J^a \pa K^e
- N d^{abc} d^{cde} \psi^d J^b J^a \pa K^e  
\nonu \\
& - & N d^{abc} d^{cde} \psi^d J^e J^a \pa K^b + 2 N d^{abc} \psi^a Q^b \pa K^c 
\nonu \\
&-& 2(N^2-4) f^{abc} \psi^a \pa J^b \pa K^c. 
\label{today} 
\eea
The first, fifth, seventh, sixth, eighth, third, and fourth terms of
(\ref{1-4}) are related to the first, second, third, fourth, fifth,  sixth, 
and eighth equation
of (\ref{today}), respectively.
The $c_1$-, $c_2$-, $c_3$-, and 
$c_ 4$-terms of (\ref{UW1}) can be obtained from the second, 
third, sixth, and last equations of
(\ref{today}), respectively.
Further rearrangement for the fields is needed. In particular, the last 
equation of (\ref{today}): $d^{abc} f^{dag} (\psi^b J^g) K^c Q^d(z)$ 
\footnote{This becomes $d^{abc} f^{dag} (\psi^b J^g) K^c Q^d(z)=
f^{dag} d^{abc} \psi^b J^g Q^d K^c(z)-2(N^2-4) f^{abc} \pa 
(\pa \psi^a J^b) K^c(z) +N d^{abc} d^{cde} \psi^b J^d \pa J^a K^e(z) - (N^2-4) 
f^{abc} \psi^a \pa^2 J^b K^c(z) - N d^{bc} d^{cde} \psi^d J^b \pa J^a K^e(z)-N 
d^{abc} d^{cde} \psi^a J^b \pa J^d K^e(z)+
2N d^{abc} \pa \psi^a Q^b K^c(z) + \frac{3}{2}
(N^2-4) f^{abc} \pa^2 \psi^a J^b K^c(z)$.}.

The following normal ordered products from (\ref{1-5})
are obtained 
\bea
d^{abc} d^{def} f^{fcg} (\psi^a J^b K^g)( J^d J^e)
& = & 
4(N^2-4) f^{abc} \pa \psi^a J^b \pa K^c -3(N^2-4)
f^{abc} \psi^a \pa J^b \pa K^c \nonu \\
& + &
\frac{4}{3} N (N^2-4) \psi^a \pa^3 K^a -
N d^{abc} d^{cde} \psi^d J^b J^e \pa K^a
\nonu \\
& -& N d^{abc} d^{cde} \psi^d J^e J^b \pa K^a 
+ d^{abc} d^{def} f^{fcg} (\psi^a J^b) K^g J^d J^e,
\nonu \\
d^{abc} d^{def} f^{eah} K^d (\psi^h J^b K^c) J^f
& = &
d^{abc} d^{def} f^{eah} J^f \psi^h J^b K^d K^c
+ (N^2-4) f^{abc} K^a \pa^2 (\psi^b K^c)
\nonu \\
&+& \frac{N}{2} d^{abc} d^{cde} K^d \pa (\psi^e J^a K^b)
-\frac{N}{2} d^{abc} d^{cde} K^d \pa (\psi^a J^e K^b)
\nonu \\
&+& \frac{N}{2} d^{abc} d^{cde} K^d \pa (\psi^a J^b K^e), 
\nonu \\
d^{abc} d^{def} f^{ebh} K^d (\psi^a J^h K^c) J^f 
& = &
d^{abc} d^{def} f^{ebh} J^f \psi^a J^h K^d K^c
+ \frac{N}{2} d^{abc} d^{cde} K^d \pa (\psi^b J^a K^e)
\nonu \\
&+& \frac{N}{2}  d^{abc} d^{cde} K^b \pa (\psi^d J^a K^e)
-\frac{N}{2}  d^{abc} d^{cde} K^d \pa (\psi^e J^a K^b),
\nonu  \\
d^{abc} d^{def} f^{eah} f^{hbi} K^d \pa (\psi^i K^c) J^f
& = & -\frac{1}{2} (N^2-4) f^{abc} K^a \pa^2 (\psi^b K^c)
- N d^{abc} d^{cde} K^a J^b \pa (\psi^d K^e),
\nonu \\
d^{abc} d^{def} \delta^{eb} K^d (\pa (\psi^a K^c) J^f)
& = &
\frac{N^2-4}{2N} f^{abc} K^a \pa^2 (\psi^b K^c) +
d^{abc} d^{cde} K^d J^e \pa (\psi^a K^b).
\label{today1}
\eea
The second, third, fifth,  fourth and sixth terms of
(\ref{1-5}) are related to the first, second, third, fourth, fifth equations
of (\ref{today1}), respectively.
The $c_6$-, and 
$c_7$-terms of (\ref{UW1}) can be obtained from the second, and 
third equations of
(\ref{today1}), respectively.
The product $f^{abc} d^{cde} \psi^d J^e Q^a K^b(z)$ can be expressed as
$c_1$- and $c_4$-terms of (\ref{UW1}) using the Jacobi identity.

The expression in (\ref{1-6}) can be simplified as
\bea
d^{abc} d^{def} f^{dag} (\psi^g J^b K^c) (K^e K^f)
& = & 
d^{abc} d^{def} f^{dag} \psi^g J^b K^c K^e K^f 
-N d^{abc} \pa (\psi^a J^b) S^c,
\nonu \\
d^{abc} d^{def} f^{dbg}
(\psi^a J^g K^c) (K^e K^f)
& = & 
d^{abc} d^{def} f^{dbg} \psi^a J^g K^c K^e K^f - N d^{abc} 
\pa (\psi^a J^b ) S^c, 
\nonu \\
d^{abc} d^{def} f^{gbh} f^{dag} \pa (\psi^h K^c) (K^e K^f)
& = & - N d^{abc} \pa \psi^a K^b S^c - N d^{abc} \psi^a \pa K^b S^c
\nonu \\
&+& \frac{1}{3} (N^2-4) (N+2k) \pa^3 \psi^a K^a,
\nonu \\
d^{abc} d^{def} f^{ecg} J^d (\psi^a J^b K^g) K^f 
& = &
d^{abc} d^{def} f^{ecg} J^d \psi^a J^b K^f K^g \nonu \\
& + & 
N d^{abc} d^{cde} J^d \pa (\psi^a J^b K^e).
\label{today2}
\eea
The first, fourth, second,  and sixth terms of
(\ref{1-6}) are related to the first, second, third and fourth equation
of (\ref{today2}), respectively.
The $c_8$- and $c_{10}$-terms of (\ref{UW1}) can be obtained from the first, 
second, and last  equations of
(\ref{today2}), respectively.
Further rearrangement should be done when $\psi^a$ is located to the right of
$J^b$ in (\ref{today2}). 
The product $d^{abc} d^{def} f^{ecg} \psi^a J^d J^b K^f K^g(z)$
can be expressed as $c_6$- and $c_7$-terms of (\ref{UW1}) using the Jacobi 
identity.

Similarly, the normal ordered products in (\ref{1-7})
can be reduced to  
\bea
d^{abc} d^{def} f^{dcg}
(\psi^a J^b K^g) (K^e K^f)
& = &
d^{abc} d^{def} f^{dcg} \psi^a J^b K^g K^e K^f + 2N d^{abc} \pa (\psi^a J^b) S^c,
\nonu \\
d^{abc} d^{def} f^{ecg}
K^d (\psi^a J^b K^g) K^f
& = & d^{abc} d^{def} f^{ecg} \psi^a J^b K^d K^f K^g 
\nonu \\
& + & N d^{abc} d^{cde} K^d \pa (\psi^a J^b K^e). 
\label{ex1}
\eea
The second and fourth terms of
(\ref{1-7}) are related to the first and second  equation
of (\ref{ex1}), respectively.
The $c_{8}$-, $c_{9}$-, and 
$c_{13}$-terms of (\ref{UW1}) can be obtained from the first and 
second equations of
(\ref{ex1}), respectively.
Note that the normal ordered product 
$d^{abc} d^{def} f^{dcg} \psi^a J^b K^g K^e K^f(z)$
can be expressed as $c_8$- and $c_9$-terms of (\ref{UW1}) using the Jacobi 
identity.

For the normal ordered products in (\ref{1-8}),
the rearrangement can be made as follows:
\bea
d^{abc} d^{def} f^{dag} (\psi^g K^b K^c)(J^e J^f)
 & = & d^{abc} d^{def} f^{dag} \psi^g K^b K^c J^e J^f 
+ 2N d^{abc} d^{cde} \psi^d J^e \pa (K^a K^b),
\nonu \\
d^{abc} d^{def} f^{eag} J^d (\psi^g K^b K^c) J^f
& = & d^{abc} d^{def} f^{eag} J^d J^f \psi^g K^b K^c +
N d^{abc} d^{cde} J^d \pa (\psi^e K^a K^b). 
\nonu
\eea
The 
$c_{10}$-term of (\ref{UW1}) can be obtained from these two equations by 
moving the field $\psi^g$ to the left.

The expression in (\ref{1-9})
leads to 
\bea
d^{abc} \pa (\psi^a K^b) Q^c & = &
d^{abc} \pa \psi^a K^b Q^c + d^{abc} \psi^a \pa K^b Q^c 
\label{cid} \\
& + & (N^2-4) \pa \psi^a \pa^2 K^a -\frac{1}{3} (N^2-4) \psi^a \pa^3 K^a, 
\nonu \\
d^{def} f^{fag} d^{ghi} (\psi^a K^h K^i)(J^d J^e)
& = &
d^{def} f^{fag} d^{ghi} \psi^a J^d J^e K^h K^i
-2N d^{abc} d^{cde} \psi^d J^e \pa (K^a K^b), 
\nonu \\
d^{abc} d^{def} f^{dag} K^f (\psi^g K^b K^c) J^e 
& = & d^{abc} d^{def} f^{dag} K^f J^e \psi^g K^b K^c 
+ N d^{abc} d^{cde} K^d \pa (\psi^e K^a K^b).
\nonu 
\eea
The second, and third  equations provide the $c_{8}$-, $c_{10}$-terms
 of (\ref{UW1}), respectively.

The following simplications are applied to the normal ordered products 
in (\ref{1-10})
\bea
d^{abc} d^{def} f^{dag} (\psi^g K^b K^c)(K^e K^f)
& = &
f^{dag} d^{abc} d^{def} \psi^g (K^b K^c) K^e K^f +
\frac{2}{3} (N^2-4)(N+2k) \pa^3 \psi^a K^a \nonu \\
&+ & (N+2k)(N^2-4) \pa^2 \psi^a \pa K^a 
\nonu \\
& + &
\frac{1}{N} (N+2k)(N^2-4) f^{abc} \pa \psi^a \pa K^b K^c
\nonu \\
& - & N d^{abc} \pa \psi^a S^b K^c
- N d^{abc} \pa \psi^a K^b S^c  
\nonu \\
&-& \frac{1}{N} (N+2k)(N^2-4) f^{abc} \pa \psi^a  K^b \pa K^c,
\nonu \\
d^{def} d^{eag} J^d \pa (\psi^a K^g) K^f 
& = &
d^{def} d^{eag} J^d K^f \pa (\psi^a K^g)-\frac{k}{3N} (N^2-4)
J^a \pa^3 \psi^a
\nonu \\
&-& \frac{N^2-4}{2N} f^{abc} J^a \pa^2 (\psi^b K^c),
\nonu \\
d^{def} f^{eag} d^{ghi} J^d (\psi^a K^h K^i) K^f
& = &
d^{def} f^{eag} d^{ghi} J^d \psi^a K^h K^i K^f 
\nonu \\
& - & \frac{1}{2N}(N^2-4)(N+2k) f^{abc} J^a \pa^2 (\psi^b K^c)
\nonu \\
&+ & N d^{abc} J^a \pa (\psi^b S^c). 
\label{yes}
\eea
The nonderivative terms of first and third equations 
provide the $c_{8}$-, $c_{14}$-terms of (\ref{UW1}), respectively.
Further rearrangement is needed. For example, the first equation of 
(\ref{yes}). All the $\psi^a$ located at the right hand side of $J^b$
should be moved to the left to obtain the final expression of (\ref{UW1}).
 
Finally,  the following normal ordered products in 
(\ref{1-11}) are obtained 
\bea
d^{def} d^{eag} K^d \pa (\psi^a K^g) K^f
& = &
d^{def} d^{eag} K^d K^f \pa (\psi^a K^g)
-\frac{k}{3N} (N^2-4) K^a \pa^3 \psi^a
\nonu \\
&- & \frac{N^2-4}{2N} f^{abc} K^a \pa^2 (\psi^b K^c),
\nonu \\
d^{def} f^{eag} d^{ghi} K^d (\psi^a  K^h K^i ) K^f
& = &
d^{def} f^{eag} d^{ghi} K^d \psi^a  K^h K^i  K^f
\nonu \\
& - & \frac{1}{2N} (N^2-4)(N+2k) f^{abc} K^a \pa^2 (\psi^b K^c) + N d^{abc} K^a 
\pa (\psi^b S^c),
\nonu \\
d^{def} d^{egh} K^d \pa (\psi^g J^h) J^f 
& = &-\frac{2}{3} (N^2-4) K^a \pa^3 \psi^a + d^{abc} d^{cde} K^a J^b \pa 
(\psi^d J^e).
\label{cid2} 
\eea
The nonderivative term contributes to the $c_{14}$ term of (\ref{UW1}). 

In summary, collecting all the nontrivial terms in (\ref{1-1})-(\ref{1-11})
using the extra rearragements for the composite fields described 
previously, the final first-order pole  of the OPE $U(z) \; U(w)$
is presented as (\ref{UW1}).
The following identities are used, together with (\ref{cid}) and 
(\ref{cid2}), 
\bea
f^{abc} \pa \psi^a J^b \pa J^c & = &
-\frac{N}{2} \pa^2 \psi^a \pa J^a + N \pa \psi^a \pa^2 J^a,
\nonu \\
f^{abc} \psi^a \pa^2 J^b K^c & = & 
-f^{abc} \pa^2 \psi^a J^b K^c - 2 f^{abc} \pa \psi^a \pa J^b K^c,
\nonu \\
d^{abc} d^{cde} \pa \psi^d J^a K^e K^b & = &
d^{abc} d^{cde} \pa \psi^d J^a K^b K^e +\frac{N^2-4}{N} f^{abc} \pa \psi^a J^b \pa 
K^c.
\nonu
\eea

The following results are used to express $T \pa G(w)$
\bea
(J^a J^a)\pa (\psi^b J^b)
& = & J^a J^a \pa (\psi^b J^b) -3N \pa \psi^a \pa^2 J^a -2N \psi^a \pa^3 J^a,  
\nonu \\
(J^a K^a)\pa (\psi^b J^b)
& = & J^a  \pa (\psi^b J^b) K^a -\frac{3}{2}N \pa \psi^a \pa^2 K^a -N 
\psi^a \pa^3 K^a,  
\nonu \\
(K^a K^a)\pa (\psi^b J^b)
& = & \pa (\psi^b J^b) K^a K^a,  
\nonu \\
(J^a J^a)\pa (\psi^b K^b)
& = & J^a J^a \pa (\psi^b K^b) + f^{abc} J^a \pa^2 (\psi^b K^c)
-2N \pa^3 \psi^a K^a \nonu \\
& - & 4N \pa^2 \psi^a \pa K^a -2N \pa \psi^a \pa^2 K^a,  
\nonu \\
(K^a K^a)\pa (\psi^b K^b)
& = & K^a K^a \pa (\psi^b K^b) -f^{abc}  K^a \pa^2 (\psi^b K^c) 
-N \pa^2 \psi^a \pa K^a \nonu \\
& -& (2N+k) \pa \psi^a \pa^2 K^a -(N+\frac{2}{3} k) \psi^a \pa^3 K^a,  
\nonu \\
(J^a K^a)\pa (\psi^b K^b)
& = & J^a K^a \pa (\psi^b K^b) -\frac{1}{2}f^{abc}  J^a \pa^2 (\psi^b K^c) 
+\frac{1}{2} f^{abc} K^a \pa^2 (\psi^b  K^c) -\frac{k}{2} \pa \psi^a 
\pa^2 J^a  \nonu \\
& -& \frac{k}{3} \psi^a \pa^3 J^a +N \pa^3 \psi^a  K^a +\frac{5}{2} N \pa^2 
\psi^a \pa K^a +\frac{N}{2} \psi^a \pa^3 K^a 
\nonu \\
& + &   2N \pa \psi^a \pa^2 K^a.
\label{TG'}
\eea

The composite field $G \pa T(w)$ can be obtained from  
\bea
(\psi^a J^a) \pa (J^b J^b) & = &
\psi^a J^a \pa (J^b J^b) +\frac{N}{3} \psi^a \pa^3 J^a -\frac{10N}{3} \pa^3 
\psi^a J^a - 3N \pa^2 \psi^a \pa J^a, 
\nonu \\
(\psi^a K^a) \pa (J^b J^b) & = &
\psi^a K^a \pa (J^b J^b) +2 N \pa^2 \psi^a \pa K^a -\frac{2}{3} N \psi^a 
\pa^3 K^a,
\nonu \\
(\psi^a J^a ) \pa (J^b K^b) & = &
\psi^a J^a \pa (J^b K^b) +
\frac{1}{2} f^{abc} \psi^a \pa^2 (J^b K^c) +\frac{1}{2} f^{abc} J^a \pa^2 
(\psi^b K^c)
\nonu \\
&-& \frac{4}{3} N \pa^3 \psi^a K^a -\frac{5}{2} N \pa^2 \psi^a \pa K^a -N
\pa \psi^a \pa^2 K^a, 
\nonu \\
(\psi^a K^a ) \pa (J^b K^b) & = &
\psi^a K^a \pa (J^b K^b) -\frac{1}{2} f^{abc} \psi^a \pa^2 (J^b K^c)
+\frac{1}{2} f^{abc} K^a \pa^2 (\psi^b K^c)
+  \frac{N}{2} \pa^2 \psi^a \pa K^a \nonu \\
& + &  N \pa \psi^a \pa^2 K^a +
\frac{N}{2} \psi^a \pa^3 K^a -\frac{k}{3} \pa^3 \psi^a J^a 
-\frac{k}{2} \pa^2 \psi^a \pa J^a,
\nonu \\
(\psi^a J^a ) \pa (K^b K^b) & = &
\psi^a J^a  \pa (K^b K^b),
\nonu \\
(\psi^a K^a) \pa (K^b K^b) & = &
\psi^a K^a \pa (K^b K^b) -\frac{2}{3} (N+k) \pa^3 \psi^a K^a
-(N+k) \pa^2 \psi^a \pa K^a.
\label{GT'}
\eea

The coefficient functions, which appear in (\ref{UW1})
are determined completely  
\bea
c_1 & = & 
5 B C N (k+N) (2 k+N)^2 (k+6 N),
\qquad
c_2  =  10 B C k N (k + N) (2 k + N)^2,
\nonu \\
c_3 & = & 10 B C k N (k + N) (2 k + N)^2,
\qquad
c_4  =  5 B C N (k+N) (2 k+N)^2 (k+6 N),
\nonu \\
\qquad
c_6  & = & -120 B C N^2 (k+N) (2 k+N) (k+2 N),
\nonu \\
c_7 & = &  -120 B C N^2 (k+N) (2 k+N) (k+2 N),
\qquad
c_{8}  =  -30 B C N^3 (2 k^2+13 k N+12 N^2),
\nonu \\
c_9 & = & 30 B C N^3 (2 k+N) (3 k+4 N),
\qquad
c_{10}  =  
-30 B C N^2 (k + N) (2 k + N) (k + 2 N),
\nonu \\
c_{13} & = & -60 B C N^4 (2 k + N),
\qquad
c_{14}  =  -60 B C N^4 (3 k + 2 N),
\nonu \\
e_1 & = & -15 B C k N (k + N) (2 k + N)^2 (k + 2 N),
e_2  =   -15 B C k N (k + N) (2 k + N)^2 (k + 2 N),
\nonu \\
e_3 & = & -15 B C k (-2 + N) N (2 + N) (k + N) (2 k + N)^2 (k + 2 N),
\nonu \\
e_4 & = & \frac{5}{2} B C 
k (-2 + N) N (2 + N) (k + N) (2 k + N) (k + 2 N) (6 k + 11 N),
\nonu \\
e_6 & = & -\frac{15}{2} B C  
k (-2 + N) N (2 + N) (k + N) (2 k + N) (k + 2 N) (2 k + 9 N),
\nonu \\
e_7 & = & -20 B C (-2 + N) N^2 (2 + N) (k + N) (2 k + N) (k + 2 N) (2 k + 3 N),
\nonu \\
e_8 & = & 5 B C N^2 (k + N) (2 k + N)^2 (k + 6 N),
\nonu \\
e_9 & = &  -30 B C (-2 + N) N (2 + N) (k + N) (2 k + N) (k + 2 N) (2 k + 11 N),
\nonu \\
e_{10} & = & -\frac{15}{2} B C (-2 + N) N (2 + N) (k + N) (2 k + N) (k + 2 N) 
(18 k + 25 N),
\nonu \\
e_{11} & = &  -90 B C (-2 + N) N (2 + N) (k + N) (2 k + N)^2 (k + 2 N),
\nonu \\
e_{13} & = &   30 B C (-2 + N) N^2 (2 + N) (k + N) (2 k + N)^2 (k + 2 N),
\nonu  \\
e_{14} & = &  -15 B C N^3 (8 k^3 + 30 k^2 N + 45 k N^2 + 22 N^3),
\nonu \\
e_{15} & = &  -10 B C 
(-2 + N) N^2 (2 + N) (2 k + N) (2 k^3 + 15 k^2 N + 37 k N^2 + 
   26 N^3),
\nonu \\
e_{16} & = & 
30 B C N^2 (k + N) (2 k + N)^2 (k + 2 N),
e_{17}  =  45 B C N^2 (k + N) (2 k + N)^2 (k + 2 N),
\nonu \\
e_{18} & = &  45 B C N^2 (k + N) (2 k + N)^2 (k + 2 N),
e_{19}  =  10 B C N^2 (k + N) (2 k + N)^2 (7 k + 12 N),
\nonu \\
e_{20} & = & 15 B C N^2 (k + N) (2 k + N) (k + 2 N) (2 k + 5 N),
\nonu \\
e_{22} & = &  -60 B C N^3 (k + N) (k + 2 N) (2 k + 3 N),
e_{23}  =   -150 B C N^3 (k + N) (2 k + N) (k + 2 N),
\nonu \\
e_{24} & = &  -\frac{5}{2} B C N^2 (k + N) (2 k + N)^2 (k + 6 N),
e_{26}  =  -240 B C N^3 (k + N) (2 k + N) (k + 2 N),
\nonu \\
e_{27} & = &  -240 B C N^3 (k + N) (2 k + N) (k + 2 N),
e_{28}  =  60 B C N^4 (6 k^2 + 21 k N + 14 N^2),
\nonu \\
e_{29} & = & 120 B C (-2 + N) N^2 (2 + N) (k + N) (2 k + N) (k + 2 N),
\nonu \\
e_{30} & = & 240 B C N^4 (k + N) (k + 2 N),
e_{32}  =    120 B C (-2 + N) N^3 (2 + N) (2 k + N) (3 k + 4 N),
\nonu \\
e_{33} & = &  60 B C N^4 (2 k^2 + 3 k N + 2 N^2),
\label{coeffUWpole1}
\eea
and their large $N$ limit with (\ref{lambda}) can be expressed as
\bea
c_1 & \rightarrow & 
\frac{(-2+\lambda ) 
\lambda  (1+5 \lambda )}{12 \sqrt{3} N^{7/2} (1+\lambda )^{3/2} (2+3 \lambda )},
\qquad 
c_2 \rightarrow \frac{(-2+\lambda ) 
(-1+\lambda ) \lambda }{6 \sqrt{3} N^{7/2} (1+\lambda )^{3/2} (2+3 \lambda )},
\nonu \\
c_3 & \rightarrow &  \frac{(-2+\lambda ) 
(-1+\lambda ) \lambda }{6 \sqrt{3} N^{7/2} (1+\lambda )^{3/2} (2+3 \lambda )},
\qquad  c_4 \rightarrow 
\frac{(-2+\lambda ) \lambda  (1+5 \lambda )}{12 \sqrt{3} N^{7/2} (1+\lambda )^{3/2} (2+3 \lambda )},
\nonu \\
c_6 & \rightarrow & 
\frac{2 \lambda ^2}{\sqrt{3} N^{7/2} \sqrt{1+\lambda } (2+3 \lambda )},
\qquad
c_7  \rightarrow   
\frac{2 \lambda ^2}{\sqrt{3} N^{7/2} \sqrt{1+\lambda } (2+3 \lambda )},
\nonu \\
c_8  & \rightarrow & 
\frac{\lambda ^3 \left(2+9 \lambda +\lambda ^2\right)}{2 \sqrt{3} N^{7/2} (-2+\lambda ) (1+\lambda )^{3/2} (2+3 \lambda )},
\qquad
c_9  \rightarrow  
\frac{\lambda ^3 (3+\lambda )}{2 \sqrt{3} N^{7/2} (1+\lambda )^{3/2} (2+3 \lambda )},
\nonu \\
c_{10} & \rightarrow & \frac{\lambda ^2}{2 \sqrt{3} N^{7/2} 
\sqrt{1+\lambda } (2+3 \lambda )},
\qquad
c_{13}  \rightarrow   \frac{\lambda ^4}{\sqrt{3} N^{7/2} 
(1+\lambda )^{3/2} (2+3 \lambda )},
\nonu \\ 
c_{14} & \rightarrow & \frac{(-3+\lambda ) \lambda ^4}{\sqrt{3} 
N^{7/2} (-2+\lambda ) (1+\lambda )^{3/2} (2+3 \lambda )},
\nonu \\
e_1 & \rightarrow & \frac{(-2+\lambda ) (-1+\lambda )}
{4 \sqrt{3} N^{5/2} \sqrt{1+\lambda } (2+3 \lambda )}, 
\qquad e_2 \rightarrow \frac{(-2+\lambda ) 
(-1+\lambda )}{4 \sqrt{3} N^{5/2} \sqrt{1+\lambda } (2+3 \lambda )},
\nonu \\
e_3 & \rightarrow &  \frac{(-2+\lambda ) (-1+\lambda )}
{4 \sqrt{3} \sqrt{N} \sqrt{1+\lambda } (2+3 \lambda )},
\qquad e_4 \rightarrow \frac{(-1+\lambda ) (6+5 \lambda )}{24 
\sqrt{3} \sqrt{N} \sqrt{1+\lambda } (2+3 \lambda )},
\nonu \\
e_6 & \rightarrow & \frac{(-1+\lambda ) (2+7 \lambda )}{8 \sqrt{3} 
\sqrt{N} \sqrt{1+\lambda } (2+3 \lambda )},
\qquad  e_7 \rightarrow \frac{\lambda  (2+\lambda )}{3 \sqrt{3} 
\sqrt{N} \sqrt{1+\lambda } (2+3 \lambda )},
\nonu \\
e_8 & \rightarrow & \frac{(-2+\lambda ) \lambda  
(1+5 \lambda )}{12 \sqrt{3} N^{5/2} (1+\lambda )^{3/2} (2+3 \lambda )},
\qquad e_9 
\rightarrow \frac{\lambda  (2+9 \lambda )}{2 \sqrt{3} N^{3/2} 
\sqrt{1+\lambda } (2+3 \lambda )},
\nonu \\
e_{10} & \rightarrow & \frac{\lambda  (18+7 \lambda )}
{8 \sqrt{3} N^{3/2} \sqrt{1+\lambda } (2+3 \lambda )},
\qquad e_{11} \rightarrow \frac{\sqrt{3} (-2+\lambda ) \lambda }
{2 N^{3/2} \sqrt{1+\lambda } (2+3 \lambda )},
\nonu \\
e_{13} & \rightarrow &  \frac{(-2+\lambda ) \lambda }{2 \sqrt{3} 
\sqrt{N} \sqrt{1+\lambda } (2+3 \lambda )},
\qquad  e_{14} \rightarrow
\frac{\lambda ^2 \left(-8-6 \lambda -9 \lambda ^2+\lambda ^3\right)}
{4 \sqrt{3} N^{5/2} (-2+\lambda ) (1+\lambda )^{3/2} (2+3 \lambda )},
\nonu \\
e_{15} & \rightarrow & \frac{\lambda  \left(2+9 \lambda +13 
\lambda ^2+2 \lambda ^3\right)}{6 \sqrt{3} \sqrt{N} 
(1+\lambda )^{3/2} (2+3 \lambda )},
\qquad
e_{16} 
\rightarrow \frac{(-2+\lambda ) \lambda }{2 \sqrt{3} N^{5/2} 
\sqrt{1+\lambda } (2+3 \lambda )},
\nonu \\
e_{17} & \rightarrow &
\frac{\sqrt{3} (-2+\lambda ) \lambda }{4 N^{5/2} \sqrt{1+\lambda } 
(2+3 \lambda )},
\qquad e_{18}
 \rightarrow \frac{\sqrt{3} (-2+\lambda ) \lambda }{4 N^{5/2} 
\sqrt{1+\lambda } (2+3 \lambda )},
\nonu \\
e_{19} & \rightarrow & \frac{(-2+\lambda ) \lambda  
(7+5 \lambda )}{6 \sqrt{3} N^{5/2} (1+\lambda )^{3/2} (2+3 \lambda )},
\qquad
e_{20} \rightarrow  \frac{\lambda }{4 \sqrt{3} N^{5/2} \sqrt{1+\lambda }},
\nonu \\
e_{22} & \rightarrow & \frac{\lambda ^2 (2+\lambda )}{\sqrt{3} 
N^{5/2} (-2+\lambda ) \sqrt{1+\lambda } (2+3 \lambda )},
\qquad
e_{23}  \rightarrow  \frac{5 \lambda ^2}{2 \sqrt{3} N^{5/2} 
\sqrt{1+\lambda } (2+3 \lambda )},
\nonu \\
e_{24} & \rightarrow & \frac{(-2+\lambda ) \lambda  
(1+5 \lambda )}{24 \sqrt{3} N^{5/2} (1+\lambda )^{3/2} (2+3 \lambda )},
\qquad
 e_{26} \rightarrow \frac{4 \lambda ^2}{\sqrt{3} N^{5/2} 
\sqrt{1+\lambda } (2+3 \lambda )},
\nonu \\
e_{27} & \rightarrow & \frac{4 \lambda ^2}{\sqrt{3} N^{5/2}
 \sqrt{1+\lambda } (2+3 \lambda )},
\qquad
e_{28} \rightarrow \frac{\lambda ^3 \left(-6-9 \lambda +
\lambda ^2\right)}{\sqrt{3} N^{5/2} (-2+\lambda ) 
(1+\lambda )^{3/2} (2+3 \lambda )},
\nonu \\
e_{29} & \rightarrow & \frac{2 \lambda ^2}{\sqrt{3} N^{3/2} 
\sqrt{1+\lambda } (2+3 \lambda )},
\qquad
e_{30} \rightarrow  
\frac{4 \lambda ^3}{\sqrt{3} N^{5/2} (-2+\lambda ) 
\sqrt{1+\lambda } (2+3 \lambda )},
\nonu \\
e_{32} & \rightarrow & \frac{2 \lambda ^3 (3+\lambda )}
{\sqrt{3} N^{3/2} (1+\lambda )^{3/2} (2+3 \lambda )},
\qquad
e_{33} \rightarrow 
\frac{\lambda ^3 \left(2-\lambda +\lambda ^2\right)}
{\sqrt{3} N^{5/2} (-2+\lambda ) (1+\lambda )^{3/2} (2+3 \lambda )}.
\label{fromabove}
\eea
From these results (\ref{fromabove}), it would be interesting to see
that the zero mode eigenvalue equation for the spin-$\frac{9}{2}$ 
current can be done similarly once the superpartners of 
scalar fields are determined. 

\section{The first-order pole in the OPE $U(z) \; U(w)$
and coefficient functions of spin $4$ field }

The first-order pole in $U(z) \; U(w)$, consisting of nine terms,
are summarized as 
\bea
&& \{ \psi^a  Q^a  \; \psi^b  Q^b \}_{-1}  = 
-\frac{1}{2} Q^a Q^a  -2 f^{abc} d^{cde} (\psi^d \psi^a J^e) Q^b
 - N d^{abc} (\psi^a \pa \psi^c) Q^b
\nonu \\
&& +5N d^{abc} d^{cde} \psi^a \pa (\psi^d J^e ) J^b
 +5N d^{abc} d^{cde} \psi^a J^b \pa (\psi^d J^e),
\label{2-1} \\
&& \{ \psi^a  Q^a  \; \psi^b  R^b \}_{-1}  = 
-\frac{1}{2} d^{abc} Q^a J^b K^c-2 f^{bce} d^{acd} d^{bfg} ((\psi^a 
\psi^e J^d) J^f)K^g
\nonu \\
&& - N d^{abc} d^{cde} ((\psi^a \pa \psi^b) J^d)K^e
+ 5N d^{abc} d^{cde} \psi^a \pa (\psi^d J^e) K^b,
\label{2-2} \\
&& \{ \psi^a  Q^a \; \psi^b  S^b \}_{-1}  = 
-\frac{1}{2} Q^a S^a -2 f^{bce} d^{acd} d^{bfg} \psi^a \psi^e J^d K^f K^g
\nonu \\
&& - N d^{abc} d^{cde} \psi^a \pa \psi^b K^d K^e,
\label{2-3} \\
&& \{ \psi^a  R^a \; \psi^b  Q^b \}_{-1}  = 
d^{abc} d^{def} \left[-\frac{1}{2} \delta^{cd} (J^a K^b)(J^e J^f)
 - f^{dbg} (\psi^a \psi^g K^c)(J^e J^f) 
\right. \nonu \\
&& +f^{eag} \psi^d (\psi^g J^b K^c) J^f
 - f^{eag} f^{gbh} \psi^d \pa (\psi^h K^c) J^f
+ f^{fag} \psi^d J^e \psi^g J^b K^c
\nonu \\
&&  - f^{fag} f^{gbh} \psi^d J^e \pa (\psi^h K^c)
 + f^{ebg} \psi^d (\psi^a J^g K^c) J^f
 + N \delta^{eb} \psi^d \pa (\psi^a K^c) J^f
\nonu \\
&&  \left. + N \delta^{fb} \psi^d J^e \pa (\psi^a K^c)
+   f^{fbg} \psi^d J^e \psi^a J^g K^c \right], 
\label{2-4} \\
&& \{ \psi^a  R^a  \; \psi^b  R^b \}_{-1}  = 
d^{abc} d^{def} \left[ -\frac{1}{2} \delta^{cd} (J^a K^b)(J^e K^f) 
- f^{dbg} (\psi^a \psi^g K^c)(J^e K^f) 
\right. \nonu \\
&& + f^{eag} \psi^d (\psi^g J^b K^c) K^f
-  f^{eag} f^{gbh} \psi^d \pa (\psi^h K^c)K^f
+ f^{fcg} \psi^d J^e \psi^a J^b K^g
 \nonu \\
&& \left. + \delta^{cd} k \psi^a J^b \pa (\psi^d J^e)
 +f^{ebg} \psi^d (\psi^a J^g K^c)K^f
 + N \delta^{eb} \psi^d \pa (\psi^a K^c) K^f \right],
\label{2-5} \\
&& \{ \psi^a  R^a \; \psi^b  S^b \}_{-1}  = 
d^{abc} d^{def} \left[ 
-\frac{1}{2} \delta^{cd} (J^a K^b)(K^e K^f) 
 - f^{dbg} (\psi^a \psi^g K^c)(K^e K^f)
\right. \nonu \\
&&  + f^{ecg} \psi^d (\psi^a J^b K^g) K^f
 + \delta^{cd} k \psi^a \pa (\psi^d J^e) K^b
+ f^{fcg} \psi^d K^e \psi^a J^b K^g
 \nonu \\
&& \left. + \delta^{cd} k \psi^d K^e \pa (\psi^a J^b) \right],
\label{2-6} \\
&& \{ \psi^a  S^a \; \psi^b  Q^b \}_{-1}  = 
d^{abc} d^{def} \left[ -\frac{1}{2} \delta^{cd} (K^a K^b)(J^e J^f) 
 +    f^{eag} \psi^d (\psi^g K^b K^c) J^f
 \right. \nonu \\
&& \left. +f^{fag} \psi^d J^e \psi^g K^b K^c \right],
\label{2-7} \\
&& \{ \psi^a  S^a \; \psi^b  R^b \}_{-1}  = 
d^{abc} d^{def} \left[ -\frac{1}{2} \delta^{cd} (K^a K^b)(J^e K^f)  
+ f^{fga} J^e \psi^d \psi^g K^b K^c
\right. \nonu \\
&& \left. +(N+2k)\delta^{cd} J^a \psi^b \pa (\psi^e K^f) 
+  f^{eag} \psi^d (\psi^g K^b K^c) K^f  \right],
\label{2-8} \\
&& \{ \psi^a  S^a  \; \psi^b  S^b \}_{-1}  = 
d^{abc} d^{def} \left[ -\frac{1}{2} \delta^{cd} (K^a K^b)(K^e K^f)
+ f^{ega} \psi^d (\psi^g K^b K^c) K^f
\right. \nonu \\
&& 
+(N+2k) \delta^{cd} \psi^a \pa (\psi^ e K^f)K^b 
 +  f^{fga} \psi^d K^e \psi^g K^b K^c \nonu \\
&& \left. +
(N+2k)  \delta^{cd} \psi^a K^b \pa (\psi^e K^f) \right].
\label{2-9}
\eea 
The normal ordered product in third equation of (\ref{2-2})
can be simplified as 
\bea
d^{abc} d^{cde} (\psi^a \pa \psi^b) J^d K^e
& = & 
d^{abc} d^{cde} J^d \psi^a \pa \psi^b K^e +\frac{N^2-4}{2N} \pa^2 J^a K^a.
\label{did1}
\eea
The first and second  terms of (\ref{2-2}) 
contribute to the $c_3$-, $c_{10}$-terms of 
(\ref{UUpole1}), respectively.
The second equation of (\ref{2-2}) should be rearranged
\footnote{That is, 
$f^{abc} d^{cde} d^{bfg} \psi^d (\psi^a J^e)(J^f K^g) =
f^{abc} d^{cde} d^{bfg} \psi^d \psi^a J^e J^f K^g +(N^2-4) f^{abc} \psi^a \pa^2 
(\psi^b K^c)-2(N^2-4) f^{abc} \psi^a \pa (\pa \psi^b K^c) +
N d^{abc} d^{cde} \psi^a \pa (\psi^d J^b K^e) +\frac{N}{2} d^{abc} d^{cde}
\psi^d \pa (\psi^a J^b K^e) -\frac{N}{2} d^{abc} d^{cde} \psi^d \pa (\psi^a 
J^e K^b) -\frac{N}{2} d^{abc} d^{cde} \psi^b \pa (\psi^a J^d K^e)-N
d^{abc} d^{cde} \psi^a \pa \psi^d J^b K^e-N d^{abc} d^{cde} \psi^a \psi^d J^b 
\pa K^e -\frac{N^2-4}{2} J^a \pa^2 K^a -\frac{N^2-4}{2} f^{abc} \psi^a 
\psi^b \pa^2 K^c$.}. 

In (\ref{2-3}), the nonderivative two terms 
contribute to $c_{11}$-, $c_{12}$-terms of (\ref{UUpole1}), respectively. 

In (\ref{2-4}), the following expressions are obtained
\bea
d^{abc} d^{def} f^{eag} \psi^d (\psi^g J^b K^c) J^f
& = & (N^2-4) f^{abc} \psi^a \pa^2 (\psi^b K^c)
+ N d^{abc} d^{cde} \psi^d \pa (\psi^e J^a K^b)
\nonu \\
& + & \frac{N}{2} d^{abc} d^{cde} \psi^d \pa (\psi^a J^b K^e)
-\frac{N}{2} d^{abc} d^{cde} \psi^b \pa (\psi^a J^d K^e)
\nonu \\
&-& \frac{N}{2}
d^{abc} d^{cde} \psi^d \pa (\psi^a J^e K^b)
+ f^{abc} d^{cde} 
\psi^d J^e \psi^b R^a,  
\nonu \\
d^{abc} d^{def} f^{ebg} \psi^d (\psi^a J^g K^c) J^f
& = & 
\frac{N}{2} d^{abc} d^{cde} \psi^d \pa (\psi^b J^a K^e)
-\frac{N}{2}
d^{abc} d^{cde} \psi^b \pa (\psi^d J^a K^e)
\nonu \\
&-& \frac{N}{2} d^{abc} d^{cde} \psi^d \pa (\psi^e J^a K^b)
+ N d^{abc} d^{cde}  \psi^d \pa (\psi^a J^e K^b)
\nonu   \\
& + & N d^{abc} d^{cde} \psi^a J^b \pa \psi^d K^e
+ d^{abc} d^{def} f^{ebg} \psi^d J^f J^g \psi^a K^c, 
\nonu \\
d^{abc} d^{def} f^{eag} f^{gbh} \psi^d \pa (\psi^h K^c) J^f 
& = & -\frac{1}{2} (N^2-4) f^{abc} \psi^a \pa^2 (\psi^b K^c)
- N d^{abc} d^{cde} \psi^d J^e \pa (\psi^b K^a),
\nonu \\
d^{abc} d^{def} \delta^{eb} \psi^d \pa (\psi^a K^c) J^f
&= &
\frac{N^2-4}{2N} f^{abc} \psi^a \pa^2 (\psi^b K^c)
+ d^{abc} d^{cde} \psi^d J^e \pa (\psi^a K^b),
\nonu \\
f^{abc} d^{cde} (\psi^d \psi^a) Q^b K^e
& = &
f^{abc} d^{cde} \psi^d \psi^a Q^b K^e + 
2N d^{abc} d^{cde} \pa \psi^a \psi^d J^e K^b
\nonu \\
&- & N d^{abc} d^{cde} \psi^a \pa \psi^d J^e K^b
-N d^{abc} d^{cde} \psi^a \psi^d \pa J^e K^b
\nonu \\
&-& N d^{abc} d^{cde} \psi^a \psi^d \pa J^b K^e - N d^{abc} d^{cde} 
\pa \psi^b \psi^a J^d K^e 
\nonu \\
&- & N d^{abc} d^{cde} \psi^b \psi^a \pa J^d K^e +
3(N^2-4) f^{abc} \pa \psi^a \pa \psi^b K^c 
\nonu \\
&+ & \frac{9}{2} (N^2-4) f^{abc} \psi^a \pa^2 \psi^b K^c 
-2(N^2-4) \pa^2 J^a K^a 
\nonu \\
&+& N d^{abc} d^{cde} \pa \psi^d \psi^a J^b K^e. 
\label{toto}
\eea
The third, seventh, fourth, eighth and second terms of
(\ref{2-4}) are related to the first, second, third, fourth and fifth 
equation
of (\ref{toto}), respectively.
The $c_3$- and $c_{10}$-terms of (\ref{UUpole1}) can be seen 
from the nonderivative terms of (\ref{toto}).
The product $d^{abc} d^{def} f^{ebg} \psi^d J^f J^g \psi^a K^c(z)$
can be expressed as $c_9$- and $c_{10}$-terms using the Jacobi identity.

The normal ordered products in (\ref{2-5})
can be further simplified as
\bea
d^{abc} d^{def} f^{eag} f^{gbh} \psi^d \pa (\psi^h K^c) K^f 
& = & \frac{k}{3} (N^2-4) \psi^a \pa^3 \psi^a +\frac{N^2-4}{2} f^{abc}
\psi^a \pa^2 (\psi^b K^c)
\nonu \\
&-& N d^{abc} d^{cde} \psi^a K^b \pa (\psi^ d K^e), 
\nonu \\
d^{abc} d^{cde} \psi^d \pa (\psi^a K^b ) K^e 
& = & -\frac{k}{3N}(N^2-4) \psi^a \pa^3 \psi^a -\frac{N^2-4}{2N}
f^{abc} \psi^a \pa^2 (\psi^b K^c)
\nonu \\
& + & d^{abc} d^{cde} \psi^d K^e \pa (\psi^a K^b), 
\nonu \\
d^{abc} d^{def} f^{eag} \psi^d (\psi^g J^b K^c) K^f
& = & 
\frac{N}{2} d^{abc} d^{cde} \psi^d \pa (\psi^a J^e K^b)
-\frac{N}{2} d^{abc} d^{cde} \psi^b \pa (\psi^a J^d K^e)
\nonu \\
&-& \frac{N}{2} d^{abc} d^{cde} \psi^d \pa (\psi^a J^b K^e)
+ f^{abc} d^{cde} \psi^d \psi^b K^e R^a,
\nonu \\
f^{abc} d^{cde}
((\psi^d \psi^a) K^e) R^b
& = &
f^{abc} d^{cde} (\psi^d \psi^a) K^e R^b
+  \frac{1}{2} (N^2-4) \pa^2 (J^a K^a)
\nonu \\
& + & N d^{abc} d^{cde} (\psi^a \psi^d K^e) \pa K^b
+ N d^{abc} d^{cde} \pa (\psi^d \psi^a) (J^e K^b)
\nonu \\
&- & \frac{k}{2N} (N^2-4) J^a \pa^2 J^a -\frac{1}{2}(N^2-4) \pa^2 K^a K^a, 
\nonu \\
f^{abc} d^{cde} K^e (\psi^d \psi^a) R^b
&=& f^{abc} d^{cde} K^e \psi^d \psi^a R^b
+ N d^{abc} d^{cde} K^e \pa \psi^d \psi^a K^b 
\nonu \\
&+& \frac{N}{2} d^{abc} d^{cde} K^d \psi^a  \pa (\psi^b K^e)
-\frac{N}{2} d^{abc} d^{cde} K^d \psi^a \pa (\psi^e K^b)
\nonu \\
&-& \frac{N}{2} d^{abc} d^{cde} K^b \psi^a \pa (\psi^d K^e),
\nonu \\
d^{abc} d^{def} f^{ebg} \psi^d (\psi^a J^g K^c) K^f
& = &
\frac{N}{2} d^{abc} d^{cde} \psi^d \pa (\psi^e J^a K^b)
-\frac{N}{2} d^{abc} d^{cde} \psi^b \pa (\psi^d J^a K^e)
\nonu \\
&- & \frac{N}{2} d^{abc} d^{cde} \psi^d \pa (\psi^b J^a K^e)
\nonu \\
& + & d^{abc} d^{def} f^{ebg} \psi^d \psi^a J^g K^f K^c.
\label{tototo}
\eea
The fourth, eighth, third, second, seventh terms of
(\ref{2-5}) are related to the first, second, third, fourth, and sixth 
equations
of (\ref{tototo}), respectively.
Note that the first term of fourth equation of (\ref{tototo})
is rearranged and represented the fifth equation of (\ref{tototo}).  
In other words,
the fifth equation of (\ref{tototo}) comes from the nonderivative term in 
fourth equation of (\ref{tototo}). 
Furthermore, 
the other two terms appearing in the fourth equation of (\ref{tototo})
should be rearranged.
The $c_4$-, $c_7$- and $c_9$-terms of (\ref{UUpole1}) can be obtained 
from the nonderivative expressions of (\ref{2-5}). 
The normal ordered product 
$f^{abc} d^{cde} \psi^d \psi^b K^e R^a(z)$ can be expressed as
$c_4$- and $c_6$-terms using the Jacobi identity.

The expression in (\ref{2-6}) (the second and third equations) 
leads to the following normal ordered product
\bea
f^{abc} d^{cde} (\psi^d \psi^a K^e) S^b
& = &
f^{abc} d^{cde}
(\psi^d \psi^a) K^e S^b -\frac{N^2-4}{2N}(N+2k) \pa^2 J^a K^a,
\nonu \\
d^{abc} d^{def} f^{ecg} \psi^d (\psi^a J^b K^g) K^f
& = & N d^{abc} d^{cde} \psi^d \pa (\psi^a J^b K^e) \nonu \\
& + &
d^{abc} d^{def} f^{ecg} \psi^d \psi^a J^b K^f K^g.
\label{did2}
\eea
This contributes to the $c_{6}$-term of (\ref{UUpole1}).
Moreover, the first term of (\ref{2-6}) can contribute to
the $c_5$-term of  (\ref{UUpole1}).

For the normal ordered product in (\ref{2-7}),
the relation holds
\bea
d^{abc} d^{def} f^{eag} \psi^d (\psi^g K^b K^c) J^f 
& = & f^{abc} d^{cde} \psi^d \psi^b J^e S^a.
\label{did3} 
\eea
This plays the role of $c_{12}$-term of (\ref{UUpole1}).
The first term of (\ref{2-7}) contributes to $c_{11}$-term of  
(\ref{UUpole1}). The last term of (\ref{2-7}) contributes to $c_{12}$-term of 
(\ref{UUpole1}).

The following identity can be used in (\ref{2-8})
\bea
S^a R^a & = & d^{abc} J^a K^b S^c -\frac{1}{N}(N+2k)(N^2-4) J^a \pa^2 K^a,
\nonu \\
d^{abc} d^{def} f^{eag} \psi^d (\psi^g K^b K^c) K^f & = &
\frac{N+2k}{2N} (N^2-4) f^{abc} \psi^a \pa^2 (\psi^b K^c) -N
d^{abc} \psi^a \pa (\psi^b S^c) \nonu \\
& +& f^{abc} d^{cde} \psi^d \psi^b K^e S^a.
\label{above11}
\eea
The first and last terms of
(\ref{2-8}) are related to the first and second equation
of (\ref{above11}), respectively.
The nonderivative term  of second equation 
in (\ref{above11})  
goes to $c_{16}$-term of (\ref{UUpole1}).

Finally, the normal ordered products in (\ref{2-9})
are summarized as
\bea
f^{eag} d^{def} d^{ghi} \psi^d (\psi^a K^h K^i) K^f
& = & -\frac{1}{2N}(N+2k)(N^2-4) f^{abc} \psi^a \pa^2 (\psi^b K^c)  
+ N d^{abc} \psi^a \pa (\psi^b S^c) \nonu \\
& + & f^{abc} d^{cde} \psi^d \psi^a K^e S^b,
\nonu \\
S^a S^a & = & d^{abc} K^a K^b S^c -\frac{2}{N}(N^2-4)(N+2k) K^a \pa^2 K^a, 
\nonu \\
d^{def} d^{eag} \psi^d \pa (\psi^a K^g) K^f
& = & -\frac{k}{3N} (N^2-4) \psi^a \pa^3 \psi^a -\frac{N^2-4}{2N}
f^{abc} \psi^a \pa^2 (\psi^b K^c)
\nonu \\
& + & d^{abc} d^{cde} \psi^a K^b \pa (\psi^d K^e).
\label{aboveabove}
\eea
The second, first, and third terms of
(\ref{2-9}) are related to the first, second, and third equation
of (\ref{aboveabove}), respectively.
Obviously, the nonderivative terms  of first and second equations 
in (\ref{aboveabove})  
goes to $c_{16}$-, $c_{17}$-terms of (\ref{UUpole1}), respectively.

Collecting all the nontrivial terms in (\ref{2-1})-(\ref{2-9})
using the extra rearrangements for the composite fields with (\ref{did1}),
(\ref{did2}) and (\ref{did3}), 
the final first-order pole  of the OPE $U(z) \; U(w)$
is presented as (\ref{UUpole1}).
The following identities 
can be used to simplify the normal ordered products:
\bea
d^{abc} d^{cde} \pa \psi^a \psi^d J^b J^e & = & 
d^{abc} d^{cde} \pa \psi^a \psi^d J^e J^b +\frac{N^2-4}{N} \left( 
\frac{1}{2} \pa J^a \pa J^a + \frac{N}{3} \psi^a 
\pa^3 \psi^a + N \pa \psi^a \pa^2 
\psi^a \right),
\nonu \\
d^{abc} d^{cde} \psi^a \psi^d \pa J^b J^e & = &
- d^{abc} d^{cde} \psi^d \psi^a J^e \pa J^b +\frac{N^2-4}{2N} \left(
J^a \pa^2 J^a +\frac{4}{3} N \psi^a \pa^3 \psi^a \right),
\nonu \\
d^{abc} d^{cde} \psi^a \psi^e J^b \pa J^d & = &
-d^{abc} d^{cde} \psi^e \psi^a J^b \pa J^d,
\nonu \\
f^{abc} J^a \pa J^b K^c & = & 
-f^{abc} \pa J^a J^b K^c + N \pa^2 J^a K^a,
\nonu \\
f^{abc} \psi^a \pa^2 \psi^b K^c & = & 
\frac{1}{2} \pa^2 J^a K^a -f^{abc} \pa \psi^a \pa \psi^b K^c,
\nonu \\
f^{abc} \psi^a \psi^b \pa^2 J^c & = & J^a \pa^2 J^a +\frac{4}{3} N \psi^a \pa^3 
\psi^a,
\nonu \\
f^{abc} \psi^a \pa \psi^b \pa J^c & = &
\frac{1}{2} \pa J^a \pa J^a +\frac{1}{3} N \psi^a \pa^3 \psi^a -N \pa^2 
\psi^a \pa \psi^a,
\nonu \\
d^{abc} d^{cde} \psi^a \pa \psi^d J^b J^e & = & 
-d^{abc} d^{cde} \pa \psi^d \psi^a J^b J^e.
\nonu
\eea 

The coefficient functions in (\ref{UUpole1}) can be obtained
\bea
c_1 & = & -\frac{1}{2} C^2 k^2 (2 k+N)^2,
\qquad
c_2  =  2 C^2 k^2 (2 k+N)^2,
\nonu \\
c_3 & = & 5 C^2 k N (2 k+N)^2,
\qquad
c_4  =  -25 C^2 N^2 (2 k+N)^2,
\nonu \\
c_5 & = & 
50 C^2 N^3 (2 k+N),
\qquad
c_6  =  -50 C^2 N^2 (2 k+N) (2 k+3 N),
\nonu \\
c_7 & = & -\frac{25}{2} C^2 N^2 (2 k+N)^2,
\qquad
c_9  =   5 C^2 N (2 k+N)^2 (-2 k+5 N),
\nonu \\
c_{10}  & = &  
30 C^2 k N (2 k+N)^2,
\qquad
c_{11}  =  -10 C^2 k N^2 (2 k+N),
\nonu \\
c_{12}  & = &  -10 C^2 N^2 (2 k+N) (4 k+5 N),
\qquad
c_{13}  =  5 C^2 k N (2 k+N)^2,
\nonu \\
c_{16}  & = &  -100 C^2 N^3 (2 k+3 N),
\qquad
c_{17}   =   -50 C^2 N^4,
\nonu \\ 
e_1 & = & -C^2 k N (2 k+N)^2 (18 k+25 N),
\qquad
e_3  =  -C^2 k N (2 k+N)^2 (6 k+25 N),
\nonu \\
\qquad
e_6  & = &  -3 C^2 k^2 N (2 k+N)^2,
\qquad
e_7  =  25 C^2 N^3 (2 k+N) (6 k+11 N),
\nonu \\
e_8  & = &  4 C^2 k^2 (-2+N) (2+N) (2 k+N)^2,
\qquad
e_9  =  -\frac{1}{2} C^2 k (-4+N^2)  (2 k+N)^2 (48 k+25 N),
\nonu \\ 
e_{10} & = &  -\frac{5}{4} C^2 (-4+N^2) N  (2 k+N) 
(28 k^2+124 k N+95 N^2),
\nonu \\
e_{11}  & = &  -C^2 k (6 k - 25 N) (-4 + N^2) (2 k + N)^2,
\nonu \\
e_{12}  & = &  -\frac{5}{2} C^2 (-4+N^2) (2 k+N)^2 (4 k+5 N),
\qquad
e_{14}  =  \frac{5}{2} C^2 N^2 (2 k+N)^2 (8 k+5 N),
\nonu \\
e_{16} & = & 25 C^2 (-4+N^2) N^2  (2 k+N) (2 k+5 N),
\nonu \\
e_{17}  & = &   -C^2 k (-4+N^2) N  (2 k+N)^2 (51 k+25 N),
\nonu \\ 
e_{18} & = & -\frac{1}{3} C^2 k (-4+N^2) N  (2 k+N) (242 k^2+271 
k N+175 N^2),
\nonu \\
e_{20}  & = &  \frac{25}{2} C^2 (-4+N^2) N  (2 k+N)^2,
\qquad
e_{21}  =  -\frac{75}{2} C^2 N^3 (2 k+N)^2,
\nonu \\
e_{22}  & = &  \frac{25}{2} C^2 N^3 (2 k+N) (10 k+21 N),
\nonu \\
e_{23} & = & -\frac{25}{2} C^2 (-4+N^2) N  (2 k+N) (4 k^2+8 k N+11 N^2),
\nonu \\
e_{24} & = &  -25 C^2 (-4+N^2) N  (2 k+N) (2 k^2+9 k N+8 N^2),
\nonu \\
e_{25} & = & -\frac{5}{2} C^2 N^2 (2 k+N)^2 (22 k+35 N),
\qquad 
\qquad
e_{28}  =  5 C^2 N^2 (2 k+N)^2 (8 k+15 N),
\nonu \\
e_{29}  & = &  25 C^2 N^2 (2 k+N)(-2 k^2-3 k N+N^2),
\nonu \\
e_{30} & = & \frac{5}{2} C^2 N^3 (-36 k^2-28 k N+35 N^2),
\qquad
e_{34}  =  5 C^2 (-4+N^2) N  (2 k+N)^2 (2 k+5 N),
\nonu \\
e_{36} & = & -\frac{5}{2} C^2 N^2 (2 k+N)^2 (14 k+15 N),
\label{coeffUU}
\eea
and their large $N$ limit with fixed $\la$ in (\ref{lambda}) reduces to
\bea
c_1 & \rightarrow & \frac{(-2+\lambda ) (-1+\lambda )^2}{100 N^3 
(1+\lambda ) (2+3 \lambda )},
\qquad
c_2 \rightarrow \frac{(-2+\lambda ) (-1+\lambda )^2}{25 N^3 (1+\lambda ) 
(2+3 \lambda )},
\nonu\\
c_3 & \rightarrow & \frac{(-2+\lambda ) (-1+\lambda ) \lambda }{10 N^3 
(1+\lambda ) (2+3 \lambda )},
\qquad
c_4 \rightarrow \frac{(-2+\lambda ) \lambda ^2}{2 N^3 (1+\lambda ) 
(2+3 \lambda )},
\nonu\\
c_5 & \rightarrow & \frac{\lambda ^3}{N^3 (1+\lambda ) (2+3 \lambda )},
\qquad
c_6 \rightarrow 
\frac{\lambda ^2 (2+\lambda )}{N^3 (1+\lambda ) (2+3 \lambda )},
\nonu\\
c_7 & \rightarrow & \frac{(-2+\lambda ) \lambda ^2}{4 N^3 (1+\lambda ) 
(2+3 \lambda )},
\qquad
c_9  \rightarrow  
\frac{(-2+\lambda ) \lambda  (-2+7 \lambda )}{10 N^3 (1+\lambda ) 
(2+3 \lambda )},
\nonu \\
c_{10}  & \rightarrow & 
\frac{3 (-2+\lambda ) (-1+\lambda ) \lambda }{5 N^3 (1+\lambda ) 
(2+3 \lambda )},
\qquad
c_{11}  \rightarrow  \frac{(-1+\lambda ) \lambda ^2}{5 N^3 (1+\lambda ) 
(2+3 \lambda )},
\nonu \\
c_{12} & \rightarrow &   \frac{\lambda ^2 (4+\lambda )}{5 N^3 (1+\lambda ) (2+3 
\lambda )},
\qquad
c_{13}  \rightarrow  \frac{(-2+\lambda ) (-1+\lambda ) \lambda }{10 N^3 
(1+\lambda ) (2+3 \lambda )},
\nonu\\
c_{16} & \rightarrow & 
\frac{2 \lambda ^3 (2+\lambda )}{N^3 (-2+\lambda ) (1+\lambda ) 
(2+3 \lambda )},
\qquad
c_{17} \rightarrow
 \frac{\lambda ^4}{N^3 (-2+\lambda ) (1+\lambda ) (2+3 \lambda )},
\nonu \\
e_1 & \rightarrow & \frac{(-2+\lambda ) (-1+\lambda ) (18+7 \lambda )}{50 N^2 (1+\lambda ) (2+3 \lambda )},
\qquad
e_3  \rightarrow  \frac{(-2+\lambda ) (-1+\lambda ) (6+19 \lambda )}{50 N^2 (1+\lambda ) (2+3 \lambda )},
\nonu \\
e_6 & \rightarrow & \frac{3 (-2+\lambda ) (-1+\lambda )^2}{50 N^2 (1+\lambda ) 
(2+3 \lambda )},
\qquad
e_7  \rightarrow  \frac{\lambda ^2 (6+5 \lambda )}{2 N^2 (1+\lambda ) 
(2+3 \lambda )},
\nonu \\
e_8  & \rightarrow & \frac{2 (-2+\lambda ) (-1+\lambda )^2}{25 N (1+\lambda ) 
(2+3 \lambda )},
\qquad
e_9  \rightarrow  \frac{(-2+\lambda ) (-1+\lambda ) (-48+23 \lambda )}{100 N (1+\lambda ) (2+3 \lambda )},
\nonu \\
e_{10} & \rightarrow & \frac{\lambda  \left(-28-68 \lambda +\lambda ^2\right)}{40 N (1+\lambda ) (2+3 \lambda )},
\qquad
e_{11}  \rightarrow 
\frac{(-2+\lambda ) (-1+\lambda ) (-6+31 \lambda )}{50 N (1+\lambda ) (2+3 \lambda )},
\nonu \\
e_{12} & \rightarrow & 
\frac{(-2+\lambda ) \lambda  (4+\lambda )}{20 N^2 (1+\lambda ) (2+3 \lambda )},
\qquad
e_{14} \rightarrow \frac{(-2+\lambda ) \lambda  (-8+3 \lambda )}{20 N^2 
(1+\lambda ) (2+3 \lambda )},
\nonu \\
e_{16} & \rightarrow & \frac{\lambda ^2}{2 N (1+\lambda )},
\qquad
e_{17}  \rightarrow  
\frac{(-2+\lambda ) (-1+\lambda ) (-32+7 \lambda )}{50 (1+\lambda ) (2+3 \lambda )},
\nonu \\
e_{18} & \rightarrow &
\frac{(-1+\lambda ) \left(242-213 \lambda +146 \lambda ^2\right)}{150 (1+\lambda ) (2+3 \lambda )},
\qquad
e_{20} \rightarrow \frac{(-2+\lambda ) \lambda ^2}{4 N^2 (1+\lambda ) 
(2+3 \lambda )},
\nonu \\
e_{21} & \rightarrow & \frac{3 (-2+\lambda ) \lambda ^2}{4 N^2 (1+\lambda ) 
(2+3 \lambda )}, 
\qquad
e_{22} \rightarrow \frac{\lambda ^2 (10+11 \lambda )}{4 N^2 (1+\lambda ) 
(2+3 \lambda )},
\nonu \\
e_{23} & \rightarrow & 
\frac{\lambda  \left(4+7 \lambda ^2\right)}{4 N (1+\lambda ) (2+3 \lambda )},
\qquad
e_{24} \rightarrow \frac{\lambda  \left(2+5 \lambda +\lambda ^2\right)}
{2 N (1+\lambda ) (2+3 \lambda )},
\nonu \\
e_{25} & \rightarrow & \frac{(-2+\lambda ) \lambda  (22+13 \lambda )}
{20 N^2 (1+\lambda ) (2+3 \lambda )},
\qquad
e_{28}  \rightarrow  \frac{(-2+\lambda ) \lambda  (8+7 \lambda )}
{10 N^2 (1+\lambda ) (2+3 \lambda )},
\nonu \\
e_{29} & \rightarrow & \frac{\lambda  \left(-2+\lambda +2 
\lambda ^2\right)}{2 N^2 (1+\lambda ) (2+3 \lambda )}, 
\qquad
e_{30} \rightarrow \frac{\lambda ^2 \left(-36+44 \lambda +27 
\lambda ^2\right)}{20 N^2 (-2+\lambda ) (1+\lambda ) (2+3 \lambda )},
\nonu \\
\qquad
e_{34} & \rightarrow & \frac{(-2+\lambda ) \lambda }{10 N (1+\lambda )},
\qquad
e_{36} \rightarrow \frac{(-2+\lambda ) \lambda  (14+\lambda )}
{20 N^2 (1+\lambda ) (2+3 \lambda )}.
\label{limitres}
\eea
Therefore, the analysis for the zero mode eigenvalue equation
can be obtained from these limiting values (\ref{limitres}) 
on the coefficient functions 
after the normalization is fixed.

\section{ The OPEs between $\hat{T}(Z)$ and $\hat{W}(Z)$ 
and other OPEs in ${\cal N}=1$ superspace }

The ${\cal N}=1$ 
superconformal algebra is described as the super OPE
\bea
\hat{T}(Z_1) \; \hat{T}(Z_2) & 
= & \frac{1}{z_{12}^3} \,\frac{c}{6} +\frac{\theta_{12}}{z_{12}^2} 
\,\frac{3}{2} \hat{T}(Z_2) +
\frac{1}{z_{12}} \,\frac{1}{2} D \hat{T}(Z_2)
+\frac{\theta_{12}}{z_{12}} \,\pa \hat{T}(Z_2) +\cdots,
\label{tt}
\eea
where 
$z_{12} = z_1 -z_2-\theta_1 \theta_2, 
\theta_{12} = \theta_1 -\theta_2, D = \pa_{\theta} + \theta
\pa_z$ and $\pa =\pa_z$.
The super stress energy tensor is
\bea
\hat{T}(Z) = \frac{1}{2} G(z) + \theta \, T(z), \qquad 
Z=(z, \theta).
\label{that} 
\eea

The primary superfield of dimension-$\frac{5}{2}$, 
\bea
\hat{W}(Z) =
\frac{1}{\sqrt{6}} U(z) + \theta \, W(z),
\label{what}
\eea
satisfies 
\bea
\hat{T}(Z_1) \; \hat{W}(Z_2) & 
= & \frac{\theta_{12}}{z_{12}^2} 
\,\frac{5}{2} \hat{W}(Z_2) +
\frac{1}{z_{12}} \,\frac{1}{2} D \hat{W}(Z_2)
+\frac{\theta_{12}}{z_{12}} \,\pa \hat{W}(Z_2) +\cdots.
\label{superprimary}
\eea

Together with (\ref{tt}) and (\ref{superprimary}), 
the previous OPEs, (\ref{wwexp}), (\ref{wuexp1}) and (\ref{uulast}) 
are summarized as following single ${\cal N}=1$ super OPE
\bea
\hat{W}(Z_1) \; \hat{W}(Z_2) &= &
\frac{1}{z_{12}^5} \, \frac{c}{15} +
\frac{\theta_{12}}{z_{12}^4} \, \hat{T}(Z_2)+
\frac{1}{z_{12}^3} \, \frac{1}{3} 
D \hat{T}(Z_2) +\frac{\theta_{12}}{z_{12}^3} \, \frac{2}{3} \pa \hat{T}(Z_2)
 +  \frac{1}{z_{12}^2} \, \frac{2}{3} \hat{T}^2(Z_2) \nonu \\
& + & \frac{\theta_{12}}{z_{12}^2} \left[ \frac{1}{4} 
\pa^2 \hat{T} + \frac{11}{(4c+21)} \left( 2 \hat{T} D \hat{T}-\frac{1}{4}
\pa^2 \hat{T} \right) +
\frac{1}{\sqrt{6}} \hat{O}_{\frac{7}{2}}\right](Z_2)
\nonu \\
&+& \frac{1}{z_{12}} \left[ \frac{1}{20} D \pa^2 \hat{T} +
\frac{9}{2(22+5c)} \left( D \hat{T} D \hat{T} -\frac{3}{10} D \pa^2 
\hat{T}\right)
\right. \nonu \\
& - & 
\frac{(2c-83)}{2(4c+21)(10c-7)} \left( -\frac{7}{10} D \pa^2 \hat{T} +
\frac{17}{(22+5c)} \left( D \hat{T} D \hat{T} -\frac{3}{10} D \pa^2 
\hat{T}\right) + 4 \hat{T} \pa \hat{T} \right) \nonu \\
& + & \left.
\frac{1}{7\sqrt{6}} D \hat{O}_{\frac{7}{2}} + \frac{2}{\sqrt{6}} 
\hat{O}_4 \right](Z_2)
+ \frac{\theta_{12}}{z_{12}} \left[  \frac{1}{15} 
\pa^3 \hat{T} + \frac{44}{7(4c+21)} \pa \left( 2 \hat{T} D \hat{T}-\frac{1}{4}
\pa^2 \hat{T} \right)  \right. \nonu \\
& + & \left.  \frac{12}{7(10c-7)}
 \left( \frac{8}{3} D \hat{T} \pa \hat{T} -
2 \hat{T} D \pa \hat{T} -\frac{8}{15} \pa^3 \hat{T} \right) 
 + 
\frac{4}{7 \sqrt{6}} \pa \hat{O}_{\frac{7}{2}} + \frac{1}{\sqrt{6}} 
D \hat{O}_4 \right](Z_2) +\cdots
\nonu \\
& = &
\frac{1}{z_{12}^5} \, \frac{c}{15} +
\frac{\theta_{12}}{z_{12}^4} \, \hat{T}(Z_2)+
\frac{1}{z_{12}^3} \, \frac{1}{3} 
D \hat{T}(Z_2) +\frac{\theta_{12}}{z_{12}^3} \, \frac{2}{3} \pa \hat{T}(Z_2)
 +  \frac{1}{z_{12}^2} \, \frac{2}{3} \hat{T}^2(Z_2) \nonu \\
& + & \frac{\theta_{12}}{z_{12}^2} \left[ \frac{(2c+5)}{2(4c+21)} 
\pa^2 \hat{T} + \frac{22}{(4c+21)} \hat{T} D \hat{T} +
\frac{1}{\sqrt{6}} \hat{O}_{\frac{7}{2}}\right](Z_2)
\nonu \\
&+& \frac{1}{z_{12}} \left[ \frac{2(18c+1)}{(4c+21)(10c-7)} 
D \hat{T} D \hat{T}+ \frac{(2c^2-c-37)}{(4c+21)(10c-7)} D \pa^2 \hat{T} 
\right. \nonu \\
& - & \left.  
\frac{2(2c-83)}{(4c+21)(10c-7)} \hat{T} \pa \hat{T} +
\frac{1}{7\sqrt{6}} D \hat{O}_{\frac{7}{2}} + \frac{2}{\sqrt{6}} 
\hat{O}_4 \right](Z_2)
\nonu \\
&+& \frac{\theta_{12}}{z_{12}} \left[  \frac{16(7c-10)}{(4c+21)(10c-7)} 
\hat{T} D \pa \hat{T} +  
\frac{4(2c^2-29c+3)}{3(4c+21)(10c-7)} \pa^3 \hat{T} \right. \nonu \\
& + & \left. \frac{8(18c+1)}{(4c+21)(10c-7)}
 D \hat{T} \pa \hat{T}  + 
\frac{4}{7 \sqrt{6}} \pa \hat{O}_{\frac{7}{2}} + \frac{1}{\sqrt{6}} 
D \hat{O}_4 \right](Z_2) +\cdots,
\label{OPEfirst}
\eea
where the identity $\frac{1}{(z_1-z_2)^n} = \frac{1}{z_{12}^n} -
n \frac{\theta_1 \theta_2}{z_{12}^{n+1}}$ ($n=1, \cdots, 6$) is used
and the following relations for the quasi primaries 
with (\ref{that}) can be used
\bea
T^2 -\frac{3}{10} \pa^2 T & = & 
 \left( D \hat{T} D \hat{T} -\frac{3}{10} D \pa^2 
\hat{T} \right)|_{\theta=0},
\nonu \\
G T -\frac{1}{8} \pa^2 G & = & 
\left( 2 \hat{T} D \hat{T}-\frac{1}{4}
\pa^2 \hat{T} \right)|_{\theta=0}, 
\nonu \\
\frac{4}{3} T \pa G - G \pa T -\frac{4}{15} \pa^3 G & = &
 \left( \frac{8}{3} D \hat{T} \pa \hat{T} -
2 \hat{T} D \pa \hat{T} -\frac{8}{15} \pa^3 \hat{T} \right)|_{\theta=0}.
\label{decom}
\eea
When the higher spin ${\cal N}=1$ 
super currents $\hat{O}_{\frac{7}{2}}(Z)$
and $\hat{O}_4(Z)$ vanish, then the above OPE (\ref{OPEfirst}) becomes
the one for 
the ``minimal'' ${\cal N}=1$ super $W_3$ algebra \cite{IMY} and see also
\cite{BS} where the same convention is used. 
Note that the central charge is given by (\ref{centralc}).

The other ${\cal N}=1$ super OPEs can be obtained from  
the corresponding OPEs with component approach.
Because 
\bea
\hat{O}_{\frac{7}{2}}(Z) = O_{\frac{7}{2}}(z) + \theta \, O_4(z),
\label{7halfhat}
\eea
the ${\cal N}=1$ 
super OPE $ \hat{W}(Z_1) \; \hat{O}_{\frac{7}{2}}(Z_2)$
can be obtained from the OPEs, 
$U(z) \; O_{\frac{7}{2}}(w)$, $U(z) \; O_4(w)$, $W(z) \; O_{\frac{7}{2}}(w)$ and 
$W(z) \; O_4(w)$.
The ${\cal N}=1$ 
super primary field (\ref{7halfhat}) satisfies 
the OPE similar to (\ref{superprimary}).
The $\theta_{12}$ independent terms are originating from 
the first OPE while 
the $\theta_{12}$ dependent terms are originating from 
the third OPE.
Other OPEs can be located at various places appropriately (in ${\cal N}=1$
supersymmetric way).

As done in (\ref{decom}), 
recalling the following relations for the quasi primaries
\bea
G U -\frac{\sqrt{6}}{3} \pa W & = & \sqrt{6} \left( 2 \hat{T} \hat{W} -
\frac{1}{3} \pa D \hat{W} \right)|_{\theta=0}, 
\nonu \\
T W -\frac{3}{14} \pa^2 W & = & \left(
D \hat{T} D \hat{W} -\frac{3}{14} \pa^2 D 
\hat{W} \right)|_{\theta=0},
\nonu \\
G \pa U -\frac{5}{3} \pa G U -\frac{\sqrt{6}}{7} \pa^2 W & = &
\sqrt{6} \left( 2 \hat{T} \pa \hat{W} -\frac{10}{3} \pa \hat{T} \hat{W} -
\frac{1}{7} \pa^2 D \hat{W} \right)|_{\theta=0},
\nonu \\
T U -\frac{1}{4} \pa^2 U & = & \left(2 \hat{T} D \hat{W} -\frac{1}{6} 
\pa^2 \hat{W} \right)|_{\theta=0}, 
\nonu \\
G W -\frac{1}{6\sqrt{6}} \pa^2 U & = &
\sqrt{6} \left( D \hat{T} \hat{W} -\frac{1}{4} \pa^2 \hat{W} 
\right)|_{\theta=0},
\label{decom1}
\eea
where the relations (\ref{that}) and
(\ref{what})
are used,
the second ${\cal N}=1$ 
super OPE with fixed $N=3$ can be expressed as
\bea
&& \hat{W}(Z_1) \; \hat{O}_{\frac{7}{2}}(Z_2) =
\frac{\theta_{12}}{z_{12}^4} \left[
-\frac{(-36+c) (-10+7 c)}{3 (20+3 c) (21+4 c)} \sqrt{6} \, \hat{W}
\right](Z_2)
 \nonu \\
&& +\frac{1}{z_{12}^3} \left[ -\frac{2 (-36+c) (-10+7 c)}{5 (20+3 c) (21+4 c)} 
 \sqrt{6} \, D \hat{W} \right](Z_2) +
\frac{\theta_{12}}{z_{12}^3} \left[ 
-\frac{(-36+c) (-10+7 c)}{3 (20+3 c) (21+4 c)} \frac{2}{5}  \sqrt{6} \,
\pa \hat{W}
\right](Z_2)
\nonu \\
&& +\frac{1}{z_{12}^2} \left[ -\frac{2 (-36+c) (-10+7 c)}{5 (20+3 c) (21+4 c)}
\frac{1}{3}  \sqrt{6} \, D \pa \hat{W} 
\right. \nonu \\
&& \left. 
+\frac{\sqrt{\frac{2}{3}} (-36+c) (-10+7 c)}{(5+2 c) (20+3 c) (21+4 c)}
2 \left( 6 \, \hat{T} \hat{W} - D \pa \hat{W} \right)
+ \sqrt{6} \, \hat{O}_{4'} \right](Z_2) \nonu \\
&& +
\frac{\theta_{12}}{z_{12}^2} \left[
-\frac{(-36+c) (-10+7 c)}{3 (20+3 c) (21+4 c)} \frac{1}{10} 
 \sqrt{6}
\, \pa^2 \hat{W} +\frac{1}{\sqrt{6}} \frac{5}{8} \left( \hat{O}_{\frac{9}{2}} -
D \hat{O}_{4'} \right) +\frac{1}{\sqrt{6}} \, D \hat{O}_{4'}
\right.
\nonu \\
&& 
-\frac{2 (-36+c) (-10+7 c) (47+14 c)}{3 (5+2 c) (37+2 c) (20+3 c) (21+4 c)}
\sqrt{6} \left( D \hat{T} \hat{W} -\frac{1}{4} \, \pa^2 \hat{W} \right)
\nonu \\
&& \left. 
-\frac{2 \sqrt{6} 
(-36+c) (5+6 c) (-10+7 c)}{5 (5+2 c) (37+2 c) (20+3 c) (21+4 c)}
3\left( 
2 \, \hat{T} D \hat{W}-\frac{1}{6} \, \pa^2 \hat{W}
\right)
\right](Z_2)
\nonu \\
&&  +\frac{1}{z_{12}} \left[ -\frac{2 (-36+c) (-10+7 c)}{5 (20+3 c) (21+4 c)}
\frac{1}{14}  \sqrt{6} \, D \pa^2 \hat{W} 
\right. \nonu \\
&& +\frac{\sqrt{\frac{2}{3}} (-36+c) (-10+7 c)}{(5+2 c) (20+3 c) (21+4 c)}
 \frac{3}{8} 2
\, \pa \left( 6 \, \hat{T} \hat{W} - D \pa \hat{W} \right)
 \nonu \\
&&  +\frac{3}{8}  \sqrt{6} \, \pa \hat{O}_{4'}
-\frac{4 (-36+c) (11+45 c)}{5 (37+2 c) (20+3 c) (21+4 c)} 
 \sqrt{6} \left( 
D \hat{T} D \hat{W} -\frac{3}{14} \, D \pa^2 \hat{W}
\right)
\nonu \\
&& \left. 
\frac{\sqrt{\frac{3}{2}} (-36+c) (-618+11 c)}{20 (37+2 c) (20+3 c) 
(21+4 c)} 
6 \left( 2 \, \hat{T} \pa \hat{W} -\frac{10}{3}  \, \pa \hat{T} \hat{W} -
\frac{1}{7} \, D \pa^2 \hat{W} \right)
+  \sqrt{6} \, D \hat{O}_{\frac{9}{2}} \right](Z_2) \nonu \\
&& +
\frac{\theta_{12}}{z_{12}} \left[ 
-\frac{(-36+c) (-10+7 c)}{3 (20+3 c) (21+4 c)} \frac{2}{105} 
 \sqrt{6} \,
\pa^3 \hat{W} +\frac{1}{\sqrt{6}} \frac{5}{8} \frac{4}{9} \, \pa
\left( \hat{O}_{\frac{9}{2}} -
D \hat{O}_{4'} \right) +\frac{1}{\sqrt{6}} \frac{4}{9} \, \pa D \hat{O}_{4'}
\right.
\nonu \\
&& 
-\frac{2 (-36+c) (-10+7 c) (47+14 c)}{3 (5+2 c) (37+2 c) (20+3 c) (21+4 c)}
 \frac{4}{9} \sqrt{6} 
\, \pa \left( D \hat{T} \hat{W} -\frac{1}{4} \, \pa^2 \hat{W} \right)
\nonu \\
&& 
-\frac{2 \sqrt{6} 
(-36+c) (5+6 c) (-10+7 c)}{5 (5+2 c) (37+2 c) (20+3 c) (21+4 c)}
\frac{4}{9} \, \pa \left( 
2 \, \hat{T} D \hat{W}-\frac{1}{6} \, \pa^2 \hat{W}
\right) \nonu \\
&& + 
\frac{2 \sqrt{\frac{2}{3}} (-36+c) (1+18 c)}{15 (5+2 c) (20+3 c) (21+4 c)}
2 \,
\left( \hat{T} \pa D \hat{W} -2 \, \pa \hat{T} D \hat{W} -\frac{1}{21} \, \pa^3
\hat{W} \right)
\nonu \\
&& +  \frac{16 (-36+c) (-41+10 c)}{135 (5+2 c) (20+3 c) (21+4 c)}
\sqrt{6} \, \left( D \hat{T} \pa \hat{W} -\frac{5}{4}\,  \pa D \hat{T}
\hat{W} -\frac{1}{7}\, \pa^3 \hat{W} \right)
\nonu \\
&& \left. -  \frac{2 (-36+c) (-10+7 c)}{5 (20+3 c) (21+4 c)} 
2 \left( \hat{T} \hat{O}_{4'}-\frac{1}{9}\, \pa D \hat{O}_{4'} \right)
\right](Z_2) +\cdots,
\label{expfinal}
\eea
where the last three lines is 
a supersymmetric extension of quasi primary field of spin-$\frac{11}{2}$. 
The ${\cal N}=1$ 
primary super fields appearing in (\ref{expfinal}) are
\bea
\hat{O}_{4'}(Z) =O_{4''}(z) + \theta \, O_{\frac{9}{2}'}(z) , \qquad
\hat{O}_{\frac{9}{2}}(Z) = O_{\frac{9}{2}''}(z) +\theta \, O_5(z). 
\label{othersuper}
\eea
Further simplifications of (\ref{expfinal})
can be made by collecting the coefficient functions in the same 
field content, as in (\ref{OPEfirst}) but the present form is more useful 
eventhough it is rather complicated.
Each ${\cal N}=1$ 
super primary field (\ref{othersuper}) satisfies 
the OPE similar to (\ref{superprimary}).

As done in (\ref{decom}) and (\ref{decom1}), 
from the following identifications between 
the composite fields in the component approach and 
the corresponding super fields at vanishing $\theta$,
\bea
T O_{\frac{7}{2}} -\frac{3}{16} \pa^2 
O_{\frac{7}{2}} & = & \left( D \hat{T} \hat{O}_{\frac{7}{2}} -\frac{3}{16}
\pa^2 \hat{O}_{\frac{7}{2}} \right)|_{\theta=0},
\nonu \\
G P_{4'}^{uu} -\frac{4\sqrt{6}}{9} 
\pa O_{\frac{9}{2}} -\frac{\sqrt{6}}{56}\pa^2 O_{\frac{7}{2}} 
& = & \sqrt{6} \left[ \frac{2}{7} \, (\hat{T} D \hat{O}_{\frac{7}{2}} +
14 \hat{T} \hat{O}_{4})-\frac{4}{9} \, D \pa \hat{O}_4 -\frac{1}{56}
\, \pa^2 \hat{O}_{\frac{7}{2}}\right]|_{\theta=0},
\nonu \\
G P_{4'}^{ww} -\frac{2}{9} \sqrt{\frac{2}{3}} 
\pa O_{\frac{9}{2}} -\frac{1}{14} \sqrt{\frac{2}{3}} \pa^2 O_{\frac{7}{2}}  
& = & \sqrt{\frac{2}{3}} \left[  \frac{2}{7}
\, (4 \hat{T} D \hat{O}_{\frac{7}{2}} +
7 \hat{T} \hat{O}_{4} )-\frac{2}{9} \, D \pa \hat{O}_4 -\frac{1}{14}
\, \pa^2 \hat{O}_{\frac{7}{2}} \right]|_{\theta=0},
\nonu \\
G O_{\frac{7}{2}} +\frac{1}{4\sqrt{6}}  
\pa P_{4'}^{uu} -\frac{\sqrt{6}}{4} \pa P_{4'}^{ww}  
& = & \left[  2 \, \hat{T} 
\hat{O}_{\frac{7}{2}} +\frac{1}{28}\,  \pa \left( D 
\hat{O}_{\frac{7}{2}} + 14 \, \hat{O}_4 \right) -\frac{1}{14}
 \, \pa \left( 4 D 
\hat{O}_{\frac{7}{2}} + 7 \hat{O}_4 \right) \right]|_{\theta=0},
\nonu \\
T P_{4'}^{uu} -\frac{1}{6}  
\pa^2 P_{4'}^{uu} & = &  \frac{\sqrt{6}}{7}\, \left[   D \hat{T} D 
\hat{O}_{\frac{7}{2}} + 14 \, D \hat{T} \hat{O}_4  -\frac{1}{6} 
\, \pa^2  \left( D 
\hat{O}_{\frac{7}{2}} + 14 \, \hat{O}_4 \right)  \right]|_{\theta=0},
\nonu \\
T P_{4'}^{ww} -\frac{1}{6}  
\pa^2 P_{4'}^{ww} & = &  \frac{2}{7}\sqrt{\frac{2}{3}}\, \left[ 
 4 \, D \hat{T} D 
\hat{O}_{\frac{7}{2}} + 7 \, D \hat{T} \hat{O}_4  -\frac{1}{6} 
 \, \pa^2  \left(4 D 
\hat{O}_{\frac{7}{2}} + 7 \, \hat{O}_4  \right) \right]|_{\theta=0},
\nonu 
\eea
and further relations
\bea
G \pa O_{\frac{7}{2}}-\frac{7}{3} \pa G  O_{\frac{7}{2}} +\frac{1}{9\sqrt{6}}  
\pa^2 P_{4'}^{uu} -\frac{1}{3} \sqrt{\frac{2}{3}} \pa^2 P_{4'}^{ww}  
& = &   \left[ 2 
\, \hat{T} \pa \hat{O}_{\frac{7}{2}} -\frac{14}{3} \, \pa \hat{T} 
\hat{O}_{\frac{7}{2}} + \frac{1}{63} \, \pa^2  \left( D 
\hat{O}_{\frac{7}{2}} + 14 \, \hat{O}_4 \right)  
\right.  \nonu \\
& - & \left. \frac{2}{63} \, \pa^2  \left(4 D 
\hat{O}_{\frac{7}{2}} + 7 \hat{O}_4 \right)  \right]|_{\theta=0},
\nonu \\
G  O_{\frac{9}{2}} -\frac{2}{63} \sqrt{\frac{2}{3}}  
\pa^2 P_{4'}^{uu} +\frac{1}{21\sqrt{6}} \pa^2 P_{4'}^{ww}  
& = & \left[ 
 2 \, \hat{T} D \hat{O}_4 -\frac{4}{441} 
\, \pa^2  \left( D 
\hat{O}_{\frac{7}{2}} + 14 \, \hat{O}_4 \right)  \right.
\nonu \\
& + & \left.  \frac{1}{441}\, \pa^2
  \left(4 \, D 
\hat{O}_{\frac{7}{2}} + 7 \, \hat{O}_4 \right)
\right]|_{\theta=0},
\nonu 
\eea
where the relations (\ref{twofour}), (\ref{that}), 
(\ref{7halfhat})
and 
\bea
\hat{O}_4(Z) = O_{4'}(z) +\theta \, O_{\frac{9}{2}}(z)
\label{otherother}
\eea
are used,
the following ${\cal N}=1$ super OPE with fixed $N=3$ can be obtained
\bea
&& \hat{W}(Z_1) \; \hat{O}_{4'}(Z_2)
=\frac{1}{z_{12}^3} \left[ \frac{2 (21+4 c) (16+5 c)}{15 (5+2 c)}
 \sqrt{6} \, \hat{O}_{\frac{7}{2}} \right](Z_2) \nonu \\
&& +
\frac{\theta_{12}}{z_{12}^3}
\left[ \frac{2 (-8+43 c)}{15 (5+2 c)} \frac{\sqrt{6}}{7} \left( D 
\hat{O}_{\frac{7}{2}} + 14 \hat{O}_4 \right)
+
\frac{(156+149 c+10 c^2)}{15 (5+2 c)}
\frac{1}{7} \sqrt{\frac{2}{3}}\left( 4 D 
\hat{O}_{\frac{7}{2}} + 7 \hat{O}_4 \right) \right](Z_2)
\nonu \\
&& +\frac{1}{z_{12}^2} \left[ \frac{2 (21+4 c) (16+5 c)}{15 (5+2 c)}
 \frac{2}{7} \sqrt{6} \, \pa \hat{O}_{\frac{7}{2}}   
+ \frac{(-36+c) (-7+10 c)}{60 (5+2 c)} \sqrt{6} \, D \hat{O}_4
\right](Z_2) \nonu \\
&& +
\frac{\theta_{12}}{z_{12}^2}
\left[ \frac{2 (-8+43 c)}{15 (5+2 c)} \frac{3}{8} 
\frac{\sqrt{6}}{7} \, \pa \left( D 
\hat{O}_{\frac{7}{2}} + 14 \hat{O}_4 \right)
+
\frac{(156+149 c+10 c^2)}{15 (5+2 c)} \frac{3}{8}
\frac{1}{7} \sqrt{\frac{2}{3}} \, \pa \left( 4 D 
\hat{O}_{\frac{7}{2}} + 7 \hat{O}_4 \right) \right. \nonu \\
&& \left. 
+ \frac{11 \sqrt{\frac{2}{3}} (16+5 c)}{5 (5+2 c)} \left( 2 \, \hat{T} 
\hat{O}_{\frac{7}{2}} +\frac{1}{28}\,  \pa \left( D 
\hat{O}_{\frac{7}{2}} + 14 \, \hat{O}_4 \right) -\frac{1}{14}
 \, \pa \left( 4 D 
\hat{O}_{\frac{7}{2}} + 7 \hat{O}_4 \right)
 \right)
\right](Z_2)
 \nonu \\
&& +\frac{1}{z_{12}} \left[  \frac{2 (21+4 c) (16+5 c)}{15 (5+2 c)}
 \frac{3}{56} \sqrt{6} \, \pa^2 \hat{O}_{\frac{7}{2}} 
- \frac{(-36+c) (-7+10 c)}{60 (5+2 c)} \frac{1}{3} \sqrt{6} \,
D \pa \hat{O}_4
\right. \nonu \\
&& 
+ \frac{4 (16+5 c) (94+9 c)}{5 (5+2 c) (53+2 c)} \sqrt{6} \left( D \hat{T} 
\hat{O}_{\frac{7}{2}} -\frac{3}{16} \pa^2 \hat{O}_{\frac{7}{2}} \right) +
\sqrt{6} \, \hat{O}_{\frac{11}{2}}
\nonu \\
&& 
-\frac{\sqrt{\frac{2}{3}} 
\left(-17576-722 c-481 c^2+5 c^3\right)}{5 (5+2 c) (53+2 c) (20+3 c)}
6 \left( \frac{2}{7} \, (\hat{T} D \hat{O}_{\frac{7}{2}} +
14 \hat{T} \hat{O}_{4})-\frac{4}{9} \, D \pa \hat{O}_4 -\frac{1}{56}
\, \pa^2 \hat{O}_{\frac{7}{2}} \right)
\nonu \\
&& \left. 
-\frac{\sqrt{\frac{3}{2}} 
\left(-56948-21835 c-2128 c^2+40 c^3\right)}{5 (5+2 c) (53+2 c) (20+3 c)}
2 \left(  \frac{2}{7}
\, (4 \hat{T} D \hat{O}_{\frac{7}{2}} +
7 \hat{T} \hat{O}_{4} )-\frac{2}{9} \, D \pa \hat{O}_4 -\frac{1}{14}
\, \pa^2 \hat{O}_{\frac{7}{2}} \right)
\right](Z_2) \nonu \\
&& +
\frac{\theta_{12}}{z_{12}} \left[ 
 \frac{2 (-8+43 c)}{15 (5+2 c)} \frac{1}{12} 
\frac{\sqrt{6}}{7} \, \pa^2 \left( D 
\hat{O}_{\frac{7}{2}} + 14 \hat{O}_4 \right)
\right.
\nonu \\
&& +
\frac{(156+149 c+10 c^2)}{15 (5+2 c)} \frac{1}{12}
\frac{1}{7} \sqrt{\frac{2}{3}} \, \pa^2 \left( 4 D 
\hat{O}_{\frac{7}{2}} + 7 \hat{O}_4 \right)
+ \frac{2}{11} \sqrt{\frac{2}{3}} \, \left( D \hat{O}_{\frac{11}{2}} +\hat{O}_6 
\right) 
\nonu \\
&& 
+ \frac{11 \sqrt{\frac{2}{3}} (16+5 c)}{5 (5+2 c)} \frac{2}{5}
\, \pa \left( 2 \, \hat{T} 
\hat{O}_{\frac{7}{2}} +\frac{1}{28} \, \pa \left( D 
\hat{O}_{\frac{7}{2}} + 14 \hat{O}_4 \right) -\frac{1}{14}
 \, \pa \left( 4 D 
\hat{O}_{\frac{7}{2}} + 7 \hat{O}_4 \right)
 \right)
\nonu \\
&&  +
\frac{4 
\left(-1808304-2883416 c+702041 
c^2+180062 c^3+5992 c^4\right)}{
45 (-1+c) (5+2 c) (53+2 c) (61+2 c) (20+3 c)}
\nonu \\
&&  \times  \frac{\sqrt{6}}{7}\, \left(   D \hat{T} D 
\hat{O}_{\frac{7}{2}} + 14 \, D \hat{T} \hat{O}_4  -\frac{1}{6} 
\, \pa^2  \left( D 
\hat{O}_{\frac{7}{2}} + 14 \, \hat{O}_4 \right) \right)
\nonu \\
&&  -
\frac{4 \left(-4858716-2922925 c+48619 c^2+114376 c^3+11156 c^4
+240 c^5\right)}{15 (-1+c) (5+2 c) (53+2 c) (61+2 c) (20+3 c)}
\nonu \\
&&  \times  \frac{1}{7}\sqrt{\frac{2}{3}}\, \left( 
 4 \, D \hat{T} D 
\hat{O}_{\frac{7}{2}} + 7 \, D \hat{T} \hat{O}_4  -\frac{1}{6} 
 \, \pa^2  \left(4 D 
\hat{O}_{\frac{7}{2}} + 7 \, \hat{O}_4 \right) \right) \nonu \\
&& 
+\frac{2 
\sqrt{\frac{2}{3}} (21+4 c) (16+5 c) (164+11 c)}{175 (-1+c) (5+2 c) (53+2 c)}
\left( 2 \, \hat{T} \pa \hat{O}_{\frac{7}{2}} -\frac{14}{3} \, \pa \hat{T} 
\hat{O}_{\frac{7}{2}} + \frac{1}{63} \, \pa^2  \left( D 
\hat{O}_{\frac{7}{2}} + 14 \, \hat{O}_4 \right)  
\right. \nonu \\
&& \left. -\frac{2}{63} \, \pa^2  \left(4 D 
\hat{O}_{\frac{7}{2}} + 7 \hat{O}_4 \right) \right)  -
\frac{(-36+c) (16+5 c) (-7+10 c)}{5 \sqrt{6} (5+2 c) (61+2 c) (20+3 c)}
\nonu \\
&& \left. \times \left( 2 \, \hat{T} D \hat{O}_4 -\frac{4}{441} 
\, \pa^2  \left( D 
\hat{O}_{\frac{7}{2}} + 14 \, \hat{O}_4 \right)  +\frac{1}{441}\, \pa^2
  \left(4 \, D 
\hat{O}_{\frac{7}{2}} + 7 \, \hat{O}_4 \right) \right) 
 \right](Z_2) +\cdots,
\label{finalexp}
\eea
where the ${\cal N}=1$ 
primary super fields appearing in (\ref{finalexp}) are given
\bea
\hat{O}_{\frac{11}{2}}(Z) =O_{\frac{11}{2}}(z) + \theta \, O_{6}(z) , \qquad
\hat{O}_6(Z) = O_{6'}(z) +\theta \, O_{\frac{13}{2}}(z). 
\label{otherotherother}
\eea
Further simplifications of (\ref{finalexp})
can be made by collecting the coefficient functions in the same 
field content, as in (\ref{OPEfirst}).
Each ${\cal N}=1$ 
super primary field in (\ref{otherother}) or 
(\ref{otherotherother}) satisfies 
the OPE similar to (\ref{superprimary}).


\end{document}